\documentclass[journal]{IEEEtran}
\usepackage{enumitem}
\usepackage[lined,boxruled]{algorithm2e}
\usepackage{algorithmic}
\usepackage{booktabs}
\newcommand{\tabitem}{~~\llap{\textbullet}~~}
\usepackage{float}
\usepackage{xspace}
\usepackage{textcomp}
\usepackage{fancyhdr}
\usepackage{amssymb,amsmath, amsfonts}
\usepackage{textcomp} 
\usepackage{bbding}
\usepackage[dvipsnames]{xcolor}
\usepackage{pifont}
\newcommand{\cmark}{\ding{51}}%
\newcommand{\xmark}{\ding{55}}%

\usepackage{comment}
\usepackage[hyphens]{url}
\usepackage{cite}
\usepackage{graphicx}
\usepackage{enumitem}
\usepackage{amssymb}

\usepackage{rotating}
\usepackage{ifthen}
\usepackage{keyval}
\usepackage{trig}

\graphicspath{ {./images/} }
\usepackage{colortbl}
\ifCLASSOPTIONcompsoc
    \usepackage[caption=false, font=normalsize, labelfont=sf, textfont=sf]{subfig}
\else
\usepackage[caption=false, font=footnotesize]{subfig}

\fi
\newcommand{\fakeparagraph}[1]{\vspace{.1mm}\noindent\textbf{#1}}
\newcommand{\fakepar}[1]{\fakeparagraph{#1}}

\usepackage{hyperref}

\begin{document}
\bstctlcite{IEEEexample:BSTcontrol}
\font\myfont=cmr10 at 19 pt



\title{{\myfont Industry 5.0 is Coming: A Survey on Intelligent NextG Wireless Networks as Technological Enablers}}



\author{
Shah~Zeb,~\IEEEmembership{Student~Member,~IEEE,}
	Aamir~Mahmood,~\IEEEmembership{Senior~Member,~IEEE,}
	Sunder~Ali~Khowaja,~\IEEEmembership{Senior~Member,~IEEE,}
	Kapal~Dev,~\IEEEmembership{Senior~Member,~IEEE,}
	Syed~Ali~Hassan,~\IEEEmembership{Senior~Member,~IEEE,}
	Nawab~Muhammad~Faseeh~Qureshi,~\IEEEmembership{Senior~Member,~IEEE,}
	Mikael~Gidlund,~\IEEEmembership{Senior~Member,~IEEE}
	Paolo~Bellavista,~\IEEEmembership{Senior~Member,~IEEE}
	\thanks{
	S.~Zeb and S.~A.~Hassan are with the School of Electrical Engineering and Computer Science (SEECS), National University of Science and Technology (NUST), 44000, Islamabad, Pakistan., (email: \{szeb.dphd19seecs,~ali.hassan\}@seecs.edu.pk).


    A.~Mahmood and M.~Gidlund are with the Department of Information Systems \& Technology, Mid Sweden University, Sweden, (email: \{firstname.lastname\}@miun.se).
    
    S.~A.~Khowaja is with Faculty of Engineering and Technology, University of Sindh, Jamshoro, Pakistan, and Department of Mechatronics Engineering, Korea Polytechnic University, South Korea. Email: sandar.ali@usindh.edu.pk, sunderali@kpu.ac.kr
    
    K.~Dev is associated with the Department of Computer science, Munster Technological University, Ireland and Institute of Intelligent Systems, University of Johannesburg, South Africa, (e-mail: kapal.dev@ieee.org).

    N.~M.~F.~Qureshi is associated with the Department of Computer Education, Sungkyunkwan University, Seoul, South Korea., (e-mail: faseeh@skku.edu.)
    
    Paolo Bellavista is associated with the Department of Computer Science and Engineering, University of Bologna, 40136 Bologna, Italy (e-mail: paolo.bellavista@unibo.it).

    Corresponding authors: Syed~Ali~Hassan and N.~M.~F.~Qureshi}
	\vspace{-20pt}
	}


\maketitle

\begin{abstract}

\textcolor{black}{ 
Industry 5.0 vision, a step toward the next industrial revolution and enhancement to Industry 4.0, envisioned the new goals of resilient, sustainable, and human-centric approaches in diverse emerging applications, e.g., factories-of-the-future, digital society.
The vision seeks to leverage human intelligence and creativity in nexus with intelligent, efficient, and reliable cognitive collaborating robots (cobots) to achieve zero waste, zero-defect, and mass customization-based manufacturing solutions. However, the vision requires the merging of cyber-physical worlds through utilizing Industry 5.0 technological enablers, e.g., cognitive cobots, person-centric artificial intelligence (AI), cyber-physical systems, digital twins, hyperconverged data storage and computing, communication infrastructure, and others. In this regard, the convergence of the emerging computational intelligence (CI) paradigm and next-generation wireless networks (NGWNs) can fulfill the stringent communication and computation requirements of the technological enablers in the Industry 5.0 vision, which is the aim of this survey-based tutorial. 
In this article, we address this issue by reviewing and analyzing current emerging concepts and technologies, e.g., CI tools and frameworks, network-in-box architecture, open radio access networks, softwarized service architectures, potential enabling services, and others, essential for designing the objectives of CI-NGWNs to fulfill the Industry 5.0 vision requirements. 
Finally, we provide a list of lessons learned from our detailed review, research challenges, and open issues that should be addressed in CI-NGWNs to realize Industry 5.0.
}
\end{abstract}
\vspace{-5pt}

\IEEEpeerreviewmaketitle
\section{Introduction}
\label{sec:Introduction}
{\color{black}
Intelligent, resilient, and reliable next-generation--or NextG--wireless networks (NGWNs) are envisioned to form a technological enabler for meeting the diverse communication and computation requirements of Industry 5.0 applications and services~\cite{CI-NGWNs1}.
which are based on three core elements; \textit{resilience}, \textit{sustainability}, and \textit{human-centric industry}~\cite{Industry5.0}. 
Resilience refers to the increase in industrial production while reducing the disruptions and times of crisis. The sustainability mainly addresses environmental factors, such as energy reduction and reuse of the available resources. The human-centric approach places the human interests and needs at the core of the production process, including personalization and machine interaction. Concerning the core elements of Industry 5.0, vital communication and computation paradigm components and their subdivision in NGWNs\footnote{Note that important acronyms are defined in Table.~\ref{table:Acronyms}} are illustrated in Fig.~\ref{fig:core-elements}.


\begin{figure}
\centering
\includegraphics[width=0.85\linewidth]{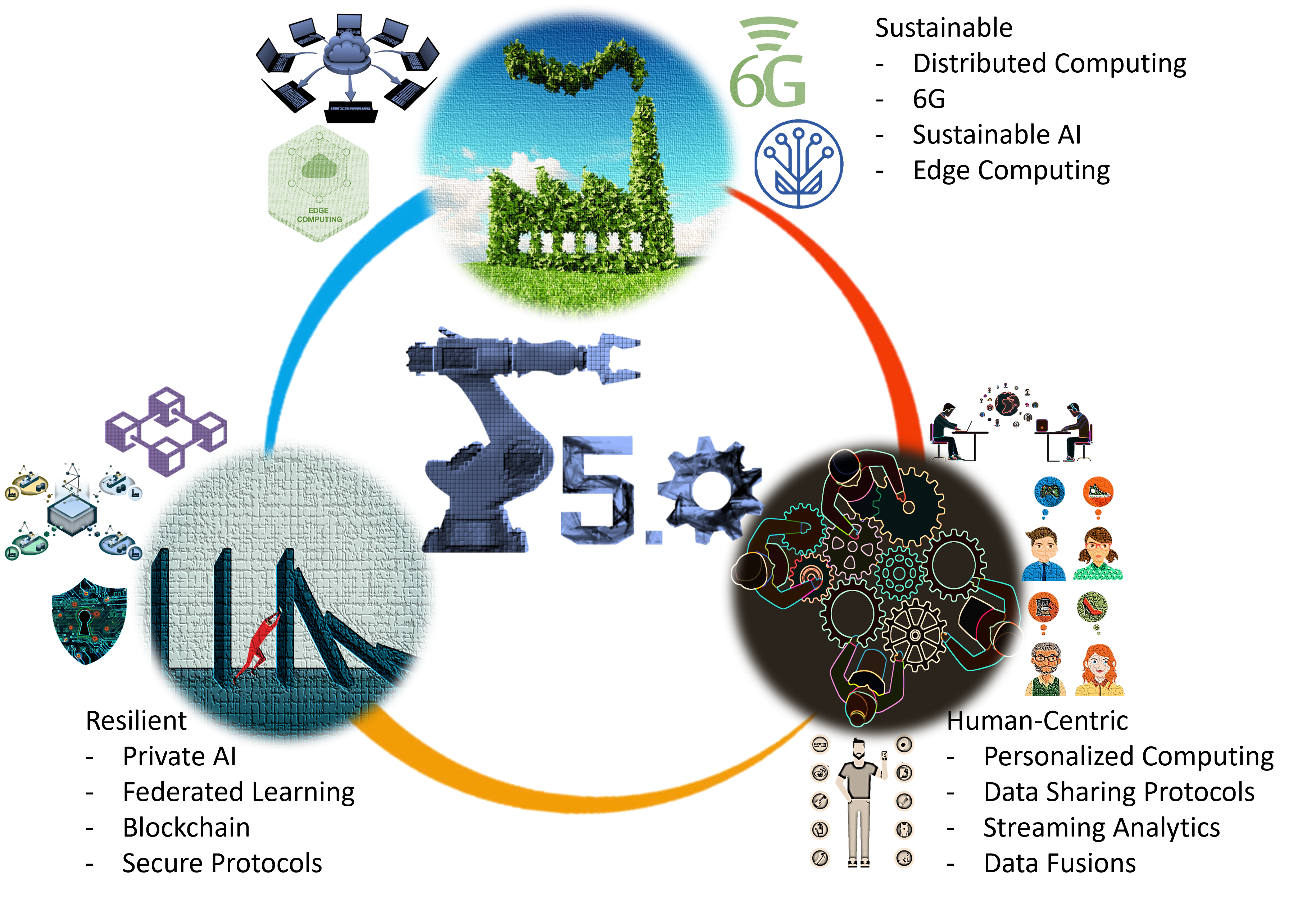}
	\caption{Industry 5.0 core elements and their envisioned technical enablers.}
	\label{fig:core-elements}
\vspace{-10pt}	
\end{figure}

\definecolor{Gray}{gray}{1}
\definecolor{Gray1}{gray}{0.94}
\definecolor{Gray2}{gray}{0.88}
{\renewcommand{\arraystretch}{1.1}
\begin{table}[t!]
\centering
	\caption{List of Important Acronyms} 
	\scalebox{0.75}{
	\setlength\tabcolsep{10pt}
 
 
  

\begin{tabular}{|ll|}
\hline 
 \rowcolor{Gray2} \textbf{Acronym} & \textbf{Definition}  \\ \hline\hline
 
 ICT & \begin{tabular}[c]{@{}c@{}}Information and Communications Technology \end{tabular} \\
 
 CI & \begin{tabular}[c]{@{}c@{}}Computational Intelligence \end{tabular} \\
 
 NGWNs & \begin{tabular}[c]{@{}c@{}}Next-Generation Wireless Networks \end{tabular} \\ 
 
 6G & \begin{tabular}[c]{@{}c@{}}Sixth-Generation \end{tabular} \\ 
 
 B5G & \begin{tabular}[c]{@{}c@{}}Beyond 5G \end{tabular} \\
 
 KPI & \begin{tabular}[c]{@{}c@{}}Key Performance Indicator \end{tabular} \\
 
 NIB& \begin{tabular}[c]{@{}c@{}}Network-in-Box \end{tabular} \\ 
 
 SDN& \begin{tabular}[c]{@{}c@{}}Software-defined Networking \end{tabular} \\
 
 NFV & \begin{tabular}[c]{@{}c@{}}Network Functions Virtualization \end{tabular} \\
 
 O-RAN & \begin{tabular}[c]{@{}c@{}}Open Radio Access Networks \end{tabular} \\
 
 3GPP & \begin{tabular}[c]{@{}c@{}}3rd Generation Partnership Project \end{tabular} \\
 
 AI & \begin{tabular}[c]{@{}c@{}}Artificial Intelligence \end{tabular} \\ 
 
 Cobots & \begin{tabular}[c]{@{}c@{}}Collaborating Robots \end{tabular} \\ 

 CCPS & \begin{tabular}[c]{@{}c@{}}Cognitive Cyber-Physical Systems \end{tabular} \\ 
 
 DT & \begin{tabular}[c]{@{}c@{}}Digital Twin \end{tabular} \\ 
 
 IIoT & \begin{tabular}[c]{@{}c@{}}Industrial Internet-of-Things \end{tabular} \\
 
 UAVs & \begin{tabular}[c]{@{}c@{}}Unmanned Aerial Vehicles \end{tabular} \\
 
 mMIMO & \begin{tabular}[c]{@{}c@{}}Massive Multiple-Input and Multiple-Output \end{tabular} \\
 
 THz and mmWave & \begin{tabular}[c]{@{}c@{}}Terahertz and Millimeter Wave \end{tabular} \\
 
 RIS & \begin{tabular}[c]{@{}c@{}}Reflecting Intelligent Surfaces \end{tabular} \\
 
 NG-IoT & \begin{tabular}[c]{@{}c@{}}Next-Generation IoT \end{tabular} \\
 
 mLLMTC& \begin{tabular}[c]{@{}c@{}}Massive Low Latency Machine Type Communication\end{tabular} \\ 
 
 UmMTC & \begin{tabular}[c]{@{}c@{}}Ultra-massive Machine Type Communication\end{tabular} \\ 
 
 ERLLC & \begin{tabular}[c]{@{}c@{}}Extremely Reliable and Low Latency Communications \end{tabular} \\ 
 
 MBBL & \begin{tabular}[c]{@{}c@{}}Mobile Broadband and Low-Latency \end{tabular} \\ 
 
 FeMBB & \begin{tabular}[c]{@{}c@{}}Further-enhanced Mobile Broadband \end{tabular} \\ 
 
 LDHMC & \begin{tabular}[c]{@{}c@{}}Long Distance and High Mobility Communication \end{tabular} \\ 
 
 ELPC & \begin{tabular}[c]{@{}c@{}}Extremely Low Power Communication \end{tabular} \\ 
 
 AEC & \begin{tabular}[c]{@{}c@{}}AI-assitive Extreme Communications \end{tabular} \\ 
 
 MEC & \begin{tabular}[c]{@{}c@{}}Multi-access Edge Computing \end{tabular} \\
 
 AMRs & \begin{tabular}[c]{@{}c@{}}Autonomous Mobile Robots \end{tabular} \\ 
 
 FoFs & \begin{tabular}[c]{@{}c@{}}Factories-of-the-Future \end{tabular} \\ 
 
 HCI & \begin{tabular}[c]{@{}c@{}}Hyper-converged Infrastructure \end{tabular} \\ 
 
 CFE & \begin{tabular}[c]{@{}c@{}}Cloud-Fog-Edge \end{tabular} \\ 
 
 QoS and QoE & \begin{tabular}[c]{@{}c@{}}Quality-of-Services and Quality-of-Experiences \end{tabular} \\

 \hline

\end{tabular}}
\label{table:Acronyms}
\vspace{-10pt}
\end{table}
}

\fakeparagraph{NextG Wireless Networks (NGWNs).}
Lately, with the deployment of 5G networks, a broad spectrum of research is being carried out to envision next-generation wireless networks or 6G wireless systems--- closely integrating the omnipresent AI into every aspect of networking systems to enable hyper-flexible intelligent end-to-end (E2E) networking architecture~\cite{6G1,6G2,6G3}. Primary goals include the 1) AI-native service-oriented architecture that possesses the inherent features of flexible and enhanced service-based architecture (SBA) to provide hybrid 6G services, 2) high data-transportation and computation networks, and 3) support for knowledgeable and autonomous human-centric systems. In this regard, CI integration can play a big part in realizing the 6G vision. 

\fakeparagraph{Computational Intelligence (CI)} paradigm provides intelligent communication, computation, and functional storage capabilities to numerous applications by utilizing~\cite{CI3,CI2,CI1}, 1) state-of-the-art computational algorithms, e.g., deep neural networks, fuzzy systems, evolutionary AI, and swarm AI, 2) hyperconverged computing infrastructure (HCI), e.g., cloud/edge computing infrastructure, on-premise private data centers, public data lakes, and 3) enabling communication technologies, e.g.,  NG-IoT and IIoT, cognitive CPS systems, B5G/6G wireless systems. The International Data Corporation (IDC) estimates the global spending on AI will increase from 85.3 billion US\$ in 2021 to over 204 US billion US\$ in 2025~\cite{CI4}.

\fakeparagraph{CI-native NextG Wireless Networks (CI-NGWNs).}
A new concept of CI-native NGWNs is emerging following the convergence of CI paradigms, SDN- and NFV-enabled communication networks, coupled with softwarized service architectures (i.e., microservices architectures), which can provide the powerful integrated solution to sustain the needs of Industry 5.0 vision~\cite{CI-NGWNs1,CI-NGWNs2}. It enables fully automated and intelligent network resource management and operations to deploy dynamic network functions and connection services, which provides a high degree of flexibility in numerous demanding scenarios~\cite{CI-NGWNs3,CI-NGWNs4,CI-NGWNs5}, e.g., 1) integrated network softwarization, 2) AI-native service-based architecture, 3) multi-tenancy and service orchestrations, 4) self-sustainable zero-touch networks, 5) O-RAN architecture, 6) DT implementation for human-machine interaction, and 7) time-sensitive networks (TSN).

\fakeparagraph{Industry 5.0 and CI-NGWNs.}
The fifth industrial revolution powered by enabling communication and computation technologies paves the way for new opportunities for firms to create value for their customers. In this respect, CI-NGWNs unlock new business model opportunities for firms to create and capture value from this technology. However, it requires all actors (e.g., leading telecom and vertical industries) to collaborate towards realizing the economic impact that would give everyone involved a fair return on their investments. Furthermore, understanding customer needs (i.e., connectivity requirements) is a prerequisite to understanding successful value creation and capturing. \textit{Therefore, the real value of CI-NGWNs is defined by improving the manufacturing processes in terms of their responsiveness to the demand, efficiency, capacity utilization, and product/service customization and innovation requirements in Industry 5.0. In this sense, it requires the human- and value-centric design of CI-NGWNs through verticals-technology joint design to enable the actual realization and adaptation of Industry 5.0 goals.}

\begin{figure*}[!t]
\centering
\includegraphics[width=0.72\linewidth]{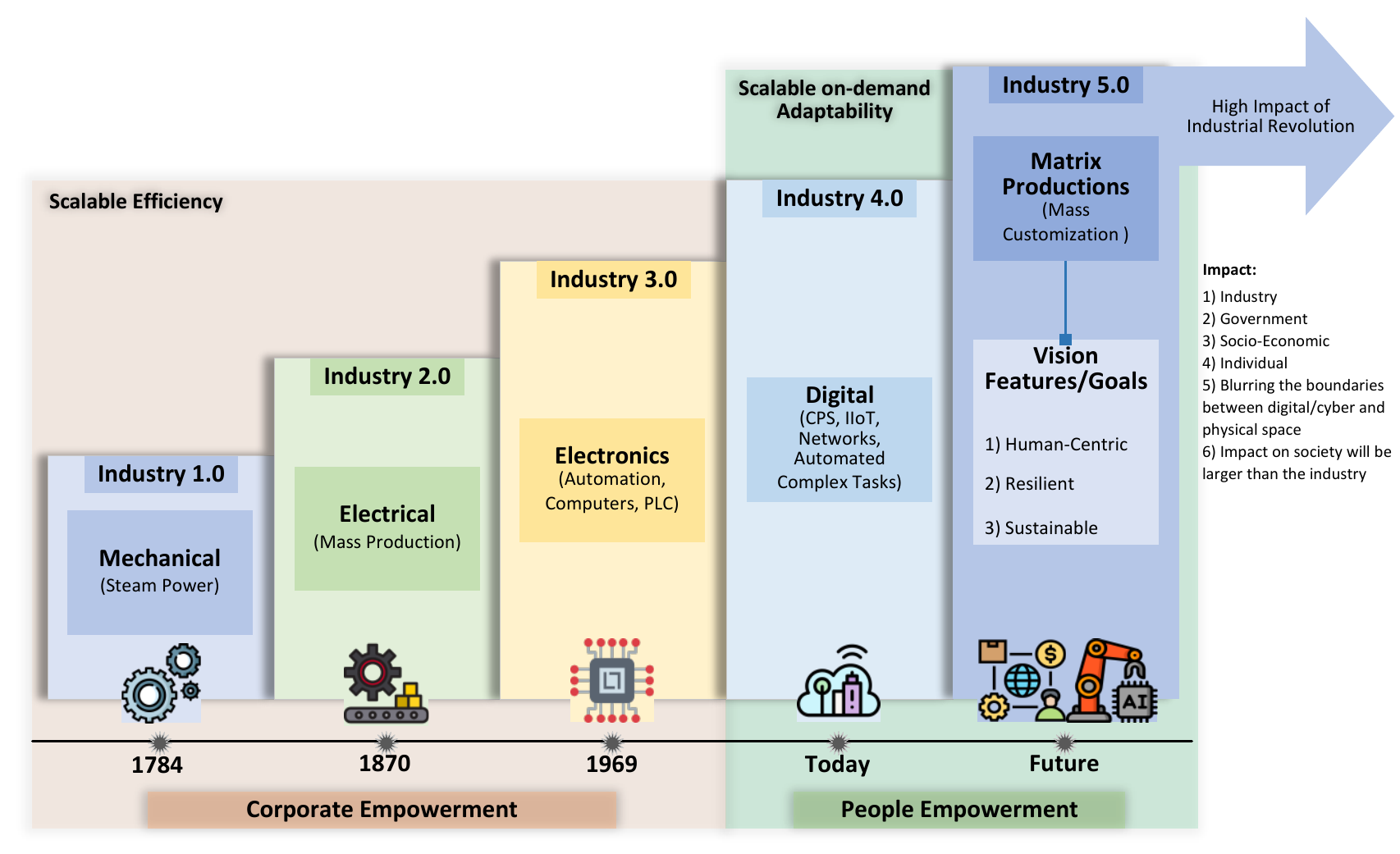}
	\caption{Main enablers and goals for the different industrial revolutions, up to Industry 5.0. }
	\label{fig:I5.0Transition}
 	\vspace{-10pt}
\end{figure*}

\section{The Rise of Industry 5.0 Vision}
\label{Sec:VisionTransition}
\fakepar{Transition towards Industry 5.0}
The world is experiencing the fourth industrial revolution, or Industry 4.0, one of the future projects that the German government outlined in its 2010 "High-Tech Strategy 2020" action plan.~\cite{I2,I1}. 
The end goal was to enhance digitization in smart production and bring automated control under the vision of mass production, like in the preceding three industrial revolutions~\cite{I3}. However, one of the key differences among different revolutions lies in utilizing the state-of-the-art enabling technologies of the respective age (as illustrated in~Fig.~\ref{fig:I5.0Transition}), e.g., the introduction of diverse ICT technologies facilitates the migration from electronics-based Industry 3.0 to the vision of Industry 4.0~\cite{I4}. Hence, the digitized revolution of Industry 4.0 reshapes today's industries by increasing the efficiency and intelligence in manufacturing and automation processes\cite{I5}.
Lately, academia and researchers have started to discuss the futuristic visions for the next industrial revolution or Industry 5.0--- an evolution and logical continuation that complements the existing Industry 4.0 paradigm. 

\fakeparagraph{Industry 5.0 Visions.}
Industry 5.0 aims for mass customized production with zero waste, minimum cost and maximum accuracy while a critical component in Industry 4.0 was mass production with minimum wastage and enhanced efficiency~\cite{I6,Industry5.0}.
However, the Industry 5.0 concept is yet entirely to be evolved. For example,~\cite[Sec.~II]{Industry5.0} discusses the various definitions and observations of leading industry researchers and practitioners regarding the Industry 5.0 vision. 
All relevant Industry 5.0 concepts emphasize humans and robots working in harmony to create an intelligent society where humans perform the creative tasks of innovation while robots (cobots) carry out the rest. 
In this regard,
European Commission (EC) coined the futuristics blueprints to add resiliency, human-centering, and sustainable approaches to the industry 5.0 vision~\cite{EU1,EU2,EU3}. It enables people empowerment (i.e., workforce, digital society) in nexus with the goals of mass-customization in optimized supply chains, e.g., adaptability and scalable product varieties, bringing positive impacts and benefits for different fields of society and realizing the next industrial revolution (c.f.~Fig.~\ref{fig:I5.0Transition}).

\fakeparagraph{Why Industry 5.0.}
The industries of the future need to play a functional part in equipping solutions to imminent societal challenges~\cite{xu2021industry,sharma2022moving,Sindhwani2022}, e.g., 1) preserving environmental and natural resources and climate change (sustainability), 2) circular production models, emerging enabling ICT technologies, and revision of policies for energy consumption for efficient utilization of natural resources in the event of external shocks, e.g., Covid-19 pandemic (resiliency), and 3) digital hyperconnectivity and evolving digital skills for people empowerment and social stability (human-centric value). It is imperative for Industry 5.0 vision components to realize and enable the 17 goals of the Sustainable Development Goals (SDG) or Global Goals outlined in Agenda 2030 of the United Nations (UN)~\cite{United-Nations}. 


\fakeparagraph{Building blocks of Industry 5.0.}
In the face of imminent societal challenges, FoFs in Industry 5.0 can provide innovative solutions via the integration of current and emerging technologies and techniques.
Notably, there have been significant advancements in various emerging technologies, e.g., robotics~\cite{robotic}, IIoT~\cite{aamir2022}, computer vision (CV)~\cite{CV}, big data analytics and computation technologies~\cite{bigdata}, B5G wireless networks~\cite{B5G_intro}, and artificial intelligence (AI)~\cite{AI_intro}, which enable the new vertical and horizontal industrial applications. Some promising examples/applications of an upcoming changing trend are cloud manufacturing, the deployment of artificial-, virtual-, and extended-reality (AR-VR-XR) in mesh-connected digital twins (DTs) of factory, supply chain management, DT-enabled cobots, automated guided vehicles (AGVs), predictive maintenance, intelligent healthcare, and smart education~\cite{EU1}. It is not long before humans will start interacting closely with AI- and DT-enabled cobots systems and machines inside the FoFs.
Additionally, Industry 5.0 brings the other critical features to the existing Industry 4.0 paradigm. For example, \textit{Matrix Production}, based on grid-layout categorized and standardized manufacturing cells, combines the concepts of cell manufacturing~\cite{matrix1}, lean productions~\cite{matrix2}, cell-neutral and smart wireless networked multirobot systems~\cite{I5}, and AGVs~\cite{matrix3}. It enables hyper customization and discards the monolithic rigid production by decoupling the warehouse logistics and manufacturing floor production and using the control software to connect multiple manufacturing cells in matrix grids for sustainable, flexible, resilient and customized productions flows~\cite{matrix4KUKArobotics}. By using individually configurable load handling attachments, these autonomously navigating AGVs can easily pick up and transport different workpieces or tools from warehouses.
Similarly, other enhanced features also exist, e.g., zero-defect and smart additive manufacturing~\cite{zero-defect1,zero-defect2}, monitoring and predictive maintenance~\cite{predictivemaint1}, human-robot collaborative manufacturing~\cite{humanrobotscolab}, renewable energy resources~\cite{renewableFeature}, and green industries~\cite{greenproductline}.
\definecolor{Gray}{gray}{1}
\definecolor{Gray1}{gray}{0.94}
\definecolor{Gray2}{gray}{0.88}
{\renewcommand{\arraystretch}{1.1}
\begin{table}[t!]
\centering
	\caption{Methodology on Screening Papers} 
	\scalebox{0.8}{
\begin{tabular}{ll}
\hline 
 \rowcolor{Gray2} \textbf{Searching Index} & \textbf{Content and Evaluation} \\ \hline
 Search Time-period & From: January 2000, To: March 2022 \\ \hline
 Article Database & \begin{tabular}[c]{@{}c@{}}IEEE Xplore, Scopus, Science Direct, and Google Scholar.    \end{tabular} \\ \hline
 Articles Type & \begin{tabular}[c]{@{}c@{}}Early Access and Published Peer-reviewed Technical Conferences\\ and Journals. \end{tabular}    \\ \hline
 Screening Procedures & \begin{tabular}[c]{@{}c@{}}Each paper's relevance to the research topic is determined by its \\ abstract, introduction, and conclusion.    \end{tabular}  \\ \hline
  Search Strings & \begin{tabular}[c]{@{}c@{}}“Industry 4.0", “Industry 5.0", “resilience", “sustainability", \\“human-centric", “digital society 5.0", “factories-of-the-future",\\ “zero-defect manufacturing", “digital twin", “computational \\intelligence", “industrial IoT", “softwarization", and “6G", etc. \end{tabular}  \\ \hline
\end{tabular}}
\label{methodology}
\vspace{-10pt}
\end{table}
}

\subsection{Research Methodology, Motivations and Contributions}
This subsection discussed the screening method we adopted, motivations, and contributions of our paper, followed by its organization.

\definecolor{Gray}{gray}{1}
\definecolor{Gray1}{gray}{0.94}
\definecolor{Gray2}{gray}{0.88}
\bgroup
{\renewcommand{\arraystretch}{0.9}
\begin{table*}[!ht]
\centering
\caption{Summary and categorization of the reviewed studies concerning Industry 5.0}
\scalebox{0.75}{
\begin{tabular}{|c|c|c|c|c|c|c|c|c|c|c|c|c|} 
\hline
\rowcolor{Gray2}\textbf{Authors \& Year} & {\rotatebox[origin=c]{90}{\textbf{Reference Numbers}}}   & {\rotatebox[origin=c]{90}{\textbf{Cloud-Fog-Edge}}} & {\rotatebox[origin=c]{90}{\textbf{Computational Intelligence}}} & {\rotatebox[origin=c]{90}{\textbf{CFE-CI (Sustainability)}}} & {\rotatebox[origin=c]{90}{\textbf{CFE-CI (Resiliency)}}} & {\rotatebox[origin=c]{90}{\textbf{CFE-CI (Human-Centric)}}} & {\rotatebox[origin=c]{90}{\textbf{Softwarized Networks}}} & {\rotatebox[origin=c]{90}{\textbf{Service Orchestration}}} & {\rotatebox[origin=c]{90}{\textbf{Digital Twin Implementation}}} & {\rotatebox[origin=c]{90}{\textbf{Private 5G Networks \& O-RAN}}} & {\rotatebox[origin=c]{90}{\textbf{Frameworks, Projects \& Demos}}} 
& \begin{tabular}[c]{@{}c@{}}\textbf{Remarks} \\ \textbf{(Compliance with Industry 5.0 Vision)} \end{tabular}
\\ 
\hline\hline
\begin{tabular}[c]{@{}c@{}}Draghici~\textit{et al}., \\ 2022 \end{tabular} & \cite{Draghici2022}         & \cellcolor{green!20} \cmark  & \cellcolor{green!20} \cmark & \cellcolor{green!20} \cmark & \cellcolor{blue!20} M & \cellcolor{red!20} L & \cellcolor{red!35} \xmark  & \cellcolor{red!35} \xmark & \cellcolor{red!35} \xmark & \cellcolor{red!35} \xmark & \cellcolor{red!35} \xmark & Sustainability and Innovation in Industry 5.0    \\ \hline

\rowcolor{Gray1} \begin{tabular}[c]{@{}c@{}}Sindhwani~\textit{et al}., \\ 2022 \end{tabular} & \cite{Sindhwani2022}     & \cellcolor{green!20} \cmark   & \cellcolor{green!20} \cmark  & \cellcolor{blue!20} M  & \cellcolor{green!20} \cmark   & \cellcolor{red!20} L  & \cellcolor{red!35} \xmark & \cellcolor{red!35} \xmark & \cellcolor{red!35} \xmark & \cellcolor{red!35} \xmark & \cellcolor{red!35} \xmark & Social Value Creation in Industry 5.0    \\ \hline

\begin{tabular}[c]{@{}c@{}}Khowaja~\textit{et al}., \\ 2022 \end{tabular} & \cite{Khowaja2021}     & \cellcolor{green!20} \cmark &\cellcolor{green!20} \cmark  &  \cellcolor{red!20} L   &\cellcolor{green!20} \cmark & \cellcolor{red!20} L  & \cellcolor{blue!20} M & \cellcolor{red!35} \xmark & \cellcolor{red!35} \xmark & \cellcolor{red!35} \xmark & \cellcolor{blue!20} M  & Private AI and Industry 5.0   \\ \hline

\rowcolor{Gray1} \begin{tabular}[c]{@{}c@{}}Khowaja~\textit{et al}., \\ 2022 \end{tabular} & \cite{Khowaja2022}   &\cellcolor{green!20} \cmark &\cellcolor{green!20} \cmark  & \cellcolor{red!20} L    &\cellcolor{green!20} \cmark & \cellcolor{blue!20} M  & \cellcolor{red!35} \xmark  & \cellcolor{red!35} \xmark  & \cellcolor{red!35} \xmark & \cellcolor{red!35} \xmark & \cellcolor{blue!20} M  & Model Inversion Attacks concerning Industry 5.0   \\ \hline

\begin{tabular}[c]{@{}c@{}}Farsi~\textit{et al}., \\ 2021 \end{tabular} & \cite{Farsi2021}  &  \cellcolor{green!20} \cmark  &\cellcolor{green!20} \cmark &\cellcolor{green!20} \cmark   &\cellcolor{green!20} \cmark   &   \cellcolor{red!20} L    & \cellcolor{red!20} L   & \cellcolor{red!35} \xmark  & \cellcolor{red!35} \xmark  & \cellcolor{red!35} \xmark & \cellcolor{red!35} \xmark &  Sustainability and Reliability in Industry 5.0   \\ \hline

\rowcolor{Gray1} \begin{tabular}[c]{@{}c@{}}Maddikunta~\textit{et al}., \\ 2021 \end{tabular} & \cite{Maddikunta2021}   &\cellcolor{green!20} \cmark     &\cellcolor{green!20} \cmark   &\cellcolor{green!20} \cmark   &  \cellcolor{red!20} L      &\cellcolor{green!20} \cmark  & \cellcolor{red!35} \xmark  & \cellcolor{red!35} \xmark  & \cellcolor{red!35} \xmark & \cellcolor{red!20} L & \cellcolor{red!35} \xmark & Enabling Technologies for Industry 5.0    \\ \hline

\begin{tabular}[c]{@{}c@{}}Thakur~\textit{et al}., \\ 2021 \end{tabular} & \cite{Thakur2021}   &\cellcolor{green!20} \cmark &\cellcolor{green!20} \cmark   & \cellcolor{blue!20} M    &\cellcolor{green!20} \cmark & \cellcolor{blue!20} M  & \cellcolor{red!35} \xmark  & \cellcolor{red!35} \xmark  & \cellcolor{red!35} \xmark & \cellcolor{red!35} \xmark & \cellcolor{red!35} \xmark  & Cyberphysical systems for Industry 5.0   \\ \hline

\rowcolor{Gray1} \begin{tabular}[c]{@{}c@{}}fraga~\textit{et al}., \\ 2021 \end{tabular} & \cite{fraga_lamas2021}  &\cellcolor{green!20} \cmark  &\cellcolor{green!20} \cmark      &\cellcolor{green!20} \cmark    &\cellcolor{green!20} \cmark      &\cellcolor{green!20} \cmark  & \cellcolor{red!20} L & \cellcolor{red!35} \xmark  & \cellcolor{red!35} \xmark & \cellcolor{red!35} \xmark & \cellcolor{red!20} L    & Sustainability and Circular Economy for Industry 5.0   \\ \hline

\begin{tabular}[c]{@{}c@{}}Du~\textit{et al}., \\ 2021 \end{tabular} & \cite{Du2021}   &\cellcolor{green!20} \cmark   &\cellcolor{green!20} \cmark      &   \cellcolor{red!20} L    &\cellcolor{green!20} \cmark     &  \cellcolor{red!20} L & \cellcolor{red!20} L  & \cellcolor{red!35} \xmark  & \cellcolor{red!35} \xmark & \cellcolor{red!35} \xmark & \cellcolor{red!20} L  & Fault Diagnosis with Edge AI in Industry 5.0   \\ \hline

\rowcolor{Gray1} \begin{tabular}[c]{@{}c@{}}Shahzadi~\textit{et al}., \\ 2021 \end{tabular} & \cite{Shahzadi2021}       &\cellcolor{green!20} \cmark    &\cellcolor{green!20} \cmark &\cellcolor{green!20} \cmark   &  \cellcolor{red!20} L   &  \cellcolor{red!20} L    & \cellcolor{red!20} L  & \cellcolor{red!20} L  & \cellcolor{red!35} \xmark & \cellcolor{red!35} \xmark & \cellcolor{red!35} \xmark   & 6G and Industry 5.0  \\ \hline


 \begin{tabular}[c]{@{}c@{}}Bhat~\textit{et al}., \\ 2021 \end{tabular} & \cite{Bhat2021}     &\cellcolor{green!20} \cmark   &\cellcolor{green!20} \cmark      &   \cellcolor{red!20} L        &  \cellcolor{red!20} L       &\cellcolor{green!20} \cmark & \cellcolor{red!20} L  & \cellcolor{red!35} \xmark & \cellcolor{red!35} \xmark & \cellcolor{red!35} \xmark & \cellcolor{red!35} \xmark & 6G and Human-centric Industry   \\ \hline

\rowcolor{Gray1} \begin{tabular}[c]{@{}c@{}}Wang~\textit{et al}., \\ 2021 \end{tabular} & \cite{Wang2021}      &\cellcolor{green!20} \cmark          &\cellcolor{green!20} \cmark     &  \cellcolor{blue!20} M     &  \cellcolor{blue!20} M     &\cellcolor{green!20} \cmark  & \cellcolor{red!35} \xmark & \cellcolor{red!35} \xmark & \cellcolor{red!35} \xmark & \cellcolor{red!35} \xmark & \cellcolor{red!20} L & Explainable AI and its use case in Industry 5.0   \\\hline

 \begin{tabular}[c]{@{}c@{}}Rupa~\textit{et al}., \\ 2021 \end{tabular} & \cite{Rupa2021}      &\cellcolor{green!20} \cmark           &\cellcolor{green!20} \cmark       &  \cellcolor{red!20} L        &\cellcolor{green!20} \cmark       &  \cellcolor{blue!20} M  & \cellcolor{red!35} \xmark  & \cellcolor{red!35} \xmark & \cellcolor{red!35} \xmark & \cellcolor{red!35} \xmark & \cellcolor{red!35} \xmark  & Medical Data Protection Concerning Industry 5.0    \\ \hline



\rowcolor{Gray1} \begin{tabular}[c]{@{}c@{}}Jain~\textit{et al}., \\ 2021 \end{tabular} & \cite{Jain2021}      &\cellcolor{green!20} \cmark   &\cellcolor{green!20} \cmark   &  \cellcolor{blue!20} M    &\cellcolor{green!20} \cmark    &  \cellcolor{red!20} L  & \cellcolor{red!35} \xmark & \cellcolor{red!35} \xmark & \cellcolor{red!35} \xmark & \cellcolor{red!35} \xmark & \cellcolor{red!35} \xmark   & UAVs and Secure Communication in Industry 5.0     \\ \hline

 \begin{tabular}[c]{@{}c@{}}Chin~\textit{et al}., \\ 2021 \end{tabular} & \cite{Chin2021}      &\cellcolor{green!20} \cmark     &\cellcolor{green!20} \cmark     &  \cellcolor{red!20} L      &   \cellcolor{red!20} L     &\cellcolor{green!20} \cmark    & \cellcolor{red!35} \xmark  & \cellcolor{red!35} \xmark & \cellcolor{red!35} \xmark & \cellcolor{red!35} \xmark & \cellcolor{red!35} \xmark  & Emotional Intelligence in context of Industry 5.0    \\ \hline

\rowcolor{Gray1} \begin{tabular}[c]{@{}c@{}}Elim~\textit{et al}., \\ 2020 \end{tabular} & \cite{Elim2020}    &\cellcolor{green!20} \cmark     &\cellcolor{green!20} \cmark        &\cellcolor{green!20} \cmark      &   \cellcolor{red!20} L     &  \cellcolor{red!20} L  & \cellcolor{red!35} \xmark & \cellcolor{red!35} \xmark & \cellcolor{red!20} L & \cellcolor{red!35} \xmark & \cellcolor{red!35} \xmark & Nexus between Society 5.0 and Industry 5.0     \\ \hline

 \begin{tabular}[c]{@{}c@{}}Nayak~\textit{et al}., \\ 2020 \end{tabular} & \cite{Nayak2020}       &\cellcolor{green!20} \cmark  &\cellcolor{green!20} \cmark   &\cellcolor{green!20} \cmark       &  \cellcolor{red!20} L      & \cellcolor{red!20} L & \cellcolor{red!20} L & \cellcolor{red!35} \xmark & \cellcolor{red!35} \xmark & \cellcolor{red!35} \xmark & \cellcolor{red!35} \xmark   & 6G and Industry 5.0   \\ \hline

\rowcolor{Gray1} \begin{tabular}[c]{@{}c@{}}Longo~\textit{et al}., \\ 2020 \end{tabular} & \cite{Longo2020}        & \cellcolor{blue!20} M    &\cellcolor{green!20} \cmark    &       \cellcolor{red!20} L       &   \cellcolor{red!20} L     &\cellcolor{green!20} \cmark  & \cellcolor{red!35} \xmark  & \cellcolor{red!35} \xmark & \cellcolor{red!20} L & \cellcolor{red!35} \xmark & \cellcolor{red!35} \xmark    & Factory of the Future: Industry 5.0    \\ \hline

 \begin{tabular}[c]{@{}c@{}}Javaid~\textit{et al}., \\ 2020 \end{tabular}     &  \cite{Javaid2020covid} & \cellcolor{blue!20} M    &\cellcolor{green!20} \cmark    &   \cellcolor{red!20} L    &   \cellcolor{red!20} L    &  \cellcolor{red!20} L  & \cellcolor{red!35} \xmark & \cellcolor{red!35} \xmark  & \cellcolor{red!35} \xmark & \cellcolor{red!35} \xmark & \cellcolor{red!35} \xmark  & Healthcare and Industry 5.0     \\ \hline

\rowcolor{Gray1} \begin{tabular}[c]{@{}c@{}}Martynov~\textit{et al}., \\ 2019 \end{tabular} & \cite{Martynov2019}      &\cellcolor{green!20} \cmark         &\cellcolor{green!20} \cmark       & \cellcolor{red!20} L      &    \cellcolor{red!20} L       &   \cellcolor{red!20} L   & \cellcolor{red!35} \xmark & \cellcolor{red!35} \xmark & \cellcolor{red!35} \xmark & \cellcolor{red!35} \xmark &   \cellcolor{red!35} \xmark    & Digitalization with Industry 5.0   \\ \hline

 \begin{tabular}[c]{@{}c@{}}Kent~\textit{et al}., \\ 2019 \end{tabular} & \cite{Kent2019}            &\cellcolor{green!20} \cmark    &\cellcolor{green!20} \cmark        &   \cellcolor{red!20} L       &   \cellcolor{red!20} L         &  \cellcolor{red!20} L  & \cellcolor{red!35} \xmark & \cellcolor{red!35} \xmark & \cellcolor{red!35} \xmark & \cellcolor{red!35} \xmark & \cellcolor{red!35} \xmark   & Digital Society in Industry 5.0     \\ \hline

\rowcolor{Gray1} \begin{tabular}[c]{@{}c@{}}Javaid~\textit{et al}., \\ 2019 \end{tabular} & \cite{Javaid2019}     &  \cellcolor{blue!20} M    &\cellcolor{green!20} \cmark       &  \cellcolor{red!20} L      &    \cellcolor{red!20} L       &   \cellcolor{red!20} L  & \cellcolor{red!35} \xmark  & \cellcolor{red!35} \xmark  & \cellcolor{red!35} \xmark & \cellcolor{red!35} \xmark &   \cellcolor{red!35} \xmark  & Healthcare and Industry 5.0      \\ \hline

 \begin{tabular}[c]{@{}c@{}}Nahavandi~\textit{et al}., \\ 2019 \end{tabular} & \cite{Nahavandi2019}     &  \cellcolor{red!20} L        &\cellcolor{green!20} \cmark         &   \cellcolor{red!20} L     &  \cellcolor{red!20} L       &\cellcolor{green!20} \cmark  & \cellcolor{red!35} \xmark & \cellcolor{red!35} \xmark & \cellcolor{red!35} \xmark & \cellcolor{red!35} \xmark &  \cellcolor{red!35} \xmark  & Human-centric solution for Industry 5.0      \\ \hline

\rowcolor{Gray1} \begin{tabular}[c]{@{}c@{}}Demir~\textit{et al}., \\ 2019 \end{tabular} & \cite{Demir2019}     &  \cellcolor{blue!20} M    &\cellcolor{green!20} \cmark  &\cellcolor{green!20} \cmark       &   \cellcolor{red!20} L   &\cellcolor{green!20} \cmark  & \cellcolor{red!35} \xmark & \cellcolor{red!35} \xmark & \cellcolor{red!35} \xmark & \cellcolor{red!35} \xmark & \cellcolor{red!35} \xmark  & Human-Robot in Industry 5.0      \\ \hline

 \begin{tabular}[c]{@{}c@{}}Ozdemir~\textit{et al}., \\ 2018 \end{tabular} & \cite{Ozdemir2018}      &\cellcolor{green!20} \cmark    &\cellcolor{green!20} \cmark       &  \cellcolor{red!20} L     &  \cellcolor{red!20} L     &   \cellcolor{red!20} L  & \cellcolor{red!35} \xmark & \cellcolor{red!35} \xmark & \cellcolor{red!35} \xmark & \cellcolor{red!35} \xmark &  \cellcolor{red!35} \xmark   & AI and IoT for Industry 5.0      \\ \hline

\rowcolor{Gray1} \begin{tabular}[c]{@{}c@{}}Han~\textit{et al}., \\ 2017 \end{tabular} & \cite{Han2017}    &  \cellcolor{blue!20} M      &\cellcolor{green!20} \cmark         &   \cellcolor{red!20} L      &    \cellcolor{red!20} L        &\cellcolor{green!20} \cmark   & \cellcolor{red!35} \xmark & \cellcolor{red!35} \xmark & \cellcolor{red!35} \xmark & \cellcolor{red!35} \xmark &  \cellcolor{red!35} \xmark   & Human-Robot Collaboration in Industry 5.0     \\ \hline

 \textbf{Our Article} & \textbf{---}        &\cellcolor{green!20} \cmark  &\cellcolor{green!20} \cmark &\cellcolor{green!20} \cmark &\cellcolor{green!20} \cmark &\cellcolor{green!20} \cmark  & \cellcolor{green!20} \cmark & \cellcolor{green!20} \cmark  & \cellcolor{green!20} \cmark & \cellcolor{green!20} \cmark & \cellcolor{green!20} \cmark  &  \begin{tabular}[c]{@{}c@{}}This comprehensive survey presents an overview of NGWNs with CI-enabled\\ capabilities, e.g., CI-native CFE architecture, O-RAN, network intelligence,\\ softwarized service orchestration \& others, which can contribute to\\ meeting the critical communication \& computation needs of Industry 5.0\\\& its components in terms of sustainability, resiliency,\& human-centricity. \end{tabular} 
 \\ \hline
\end{tabular}}

\vspace*{3pt}

\fbox{\begin{tabular}{cccccccc}
\cellcolor{green!20}\cmark & High Coverage &  \cellcolor{blue!20} M & Medium Coverage & \cellcolor{red!20} L & Low Coverage & \cellcolor{red!35} \xmark & Absent/Unavailable
\end{tabular}}

\label{table:reviewedstudiessummary}
\end{table*}
}

\subsubsection{Research Screening Methodology}
\label{sec:screeningmethodology}
The method for selecting the most relevant and authentic literature is summarized in Table.~\ref{methodology}. For the publication screening process to be reliable, the screening process was repeated through three separate independent campaigns. Our findings are compiled from 2000 to 2022 for selected search strings (c.f.~Table.~\ref{methodology}). 
Prior to final processing, it is essential to establish whether the compiled data is relevant to the area of interest, i.e., to determine whether it is based on the papers' abstract, introduction, and conclusion. The search database contained a number of articles with the keyword "digital" or "twin" in the title or abstract, which is not intended to refer to the "digital twin" or "virtual image" of the process as a whole. Similarly, we encountered the same situation while searching for industrial IoT, zero-defect manufacturing, computational intelligence, industry 5.0, and other relevant data. We applied the detailed pre-processing check based on screening procedures and excluded articles of this sort from the final database of literature for our survey paper.

\subsubsection{Our Motivation}
\label{sec:papermotivation}
Industry 5.0 and CI-NGWN have already become critical topics in academia and industry (c.f.~Sec.~\ref{sec:Introduction}~and~\ref{Sec:VisionTransition}). 
We read through all the incorporated papers in our collected literature database, obtained through the aforementioned screening methodology. 

\fakepar{Literature Review and Limitations.}
Many research works are published on Industry 5.0 and CI-NGWNs, vision, enabling technologies, possible applications, and use cases.
In this regard, we categorized and summarized some of the available literature and reviewed works in Table~\ref{table:reviewedstudiessummary}, based on the critical enabling features of CI-NGWNs and its compliance with the Industry 5.0 vision.
We tried to find the common grounds and proposition towards the nexus of Industry 5.0 and CI-NGWNs, especially how CI-NGWNs can act as a technological enabler for realizing the Industry 5.0 vision. From Table~\ref{table:reviewedstudiessummary}, it is observed that all authors have explored Industry 5.0 with a different perspective, for which we gave the summarized remarks and marked the nature of discussion in reviewed literature on the basis of covering the CI-NGWNs feature coverage, i.e., \textit{High Coverage, Medium Coverage, Low Coverage and Unavailable}. From literature review content, it has been revealed that there has been limitation of less coverage of the Industry 5.0 vision concerning numerous features of CI and NextG wireless networks, i.e., \textit{O-RAN, softwarized NIBs, service provisioning, cloud/edge-native computing network architecture and others}. For example, Nayak \textit{et al.,} and Bhat \textit{et al.,} in \cite{Nayak2020} and \cite{Bhat2021}, respectively, reviewed the various features relevant to some of CI techniques and applications in 6G vision, while Shahzadi \textit{et al.,} in \cite{Shahzadi2021} has addressed the human-centric features in futuristic 6G.
Similarly, fewer attempts have been made to identify and cover the KPIs of three pillars of Industry 5.0 vision (i.e., sustainability, resilience, and human-centric) and how various critical features of CI-NWGNs can meet these KPIs requirements.  

Based on the facts mentioned above, CI-NWGNs is expected to open up novel opportunities for Industry 5.0 research and development as a critical technological enabler in the foreseeable future.

\subsubsection{Our Contributions}
As can be seen from Table.~\ref{table:reviewedstudiessummary}, this is the first attempt to survey the path towards Industry 5.0 comprehensively while considering a broad range of CI-NGWNs elements. Hence, the impact of this study lies in following key contributions:
\begin{itemize}[leftmargin=*]
    \item \textbf{Technological Enablers and Requirements:} An important requirements of technological enablers concerning the Industry 5.0 vision is identified and discussed in this paper.
    \item \textbf{CI-NGWNs and NIBs:} The design objectives, vision and NIBs architecture of CI-NGWNs has been categorized and discussed to meet the diverse KPIs of technological enablers in Industry 5.0.
    \item \textbf{Multi-tenant Softwarized Architecture:}
    An industrial automation and control software framework that allows customized and intelligent services to be provided for Industry 5.0 applications has been described and explored in the paper in light of the technological transition in industrial automation.
    \item \textbf{Open and Intelligent RAN:} This paper discusses and shape the O-RAN and Private network for Industry 5.0.
    \item \textbf{CI in Industry 5.0:} In Industry 5.0 framework, sustainable, human-centric, and resilient features are explored and discussed as a result of the combination of CI and cloud-fog-edge computing networking features.
    \item \textbf{Potential Enabling Services:} The paper explores and discusses the potential enabling services for Industry 5.0, including DT-as-a-Service, NFV-as-a-Service, Tracking-as-a-Service, AI-as-a-Service, and others.
    \item \textbf{Summarized ongoing projects and Demos:} Lastly, projects currently underway and completed, publications, demos, and collaborations related to Industry 5.0 are discussed. 
    \item \textbf{Lessons Learned and Research Challenges:} Lessons learned from the reviewed research, recommendations and roadmap for future research challenges were presented to the community based on the reviewed research.
\end{itemize}

\subsubsection{Paper Organization}

The rest of the article is organized as follows. Section~\ref{sec:Industry5.0VisionandRequirements} gives an overview of the critical requirements incurred by the technological enablers of Industry 5.0 vision and NIBs architecture of CI-NGWNs, followed by Section~\ref{sec:computationalintelligenceI5.0} that discusses the role and requirements of CI techniques towards Industry 5.0. Section~\ref{sec:6G_Architecture} discusses the virualization in RAN of NGWNs which brings the new futuristic opportunities for Industry 5.0 applications. Similarly, Section.~\ref{sec:softwarizedservices} provided a discussion on multi-tenant softwarized service provisioning for Industry 5.0 applications, which enables the customizable new potential services which are covered in Section~\ref{sec:Softwarizedexamples}. Section~\ref{sec:Openstandards} covers the ongoing research work, projects and demos relevant to Industry 5.0. Finally, Section~\ref{sec:lessonlearned} covers the lesson learned and recommendations, reserach challenges and the future opportunities, and Section~\ref{sec:conclusion} gives concluding remarks. 

}
\begin{figure*}[!t]
\centering
\includegraphics[width=0.9\linewidth]{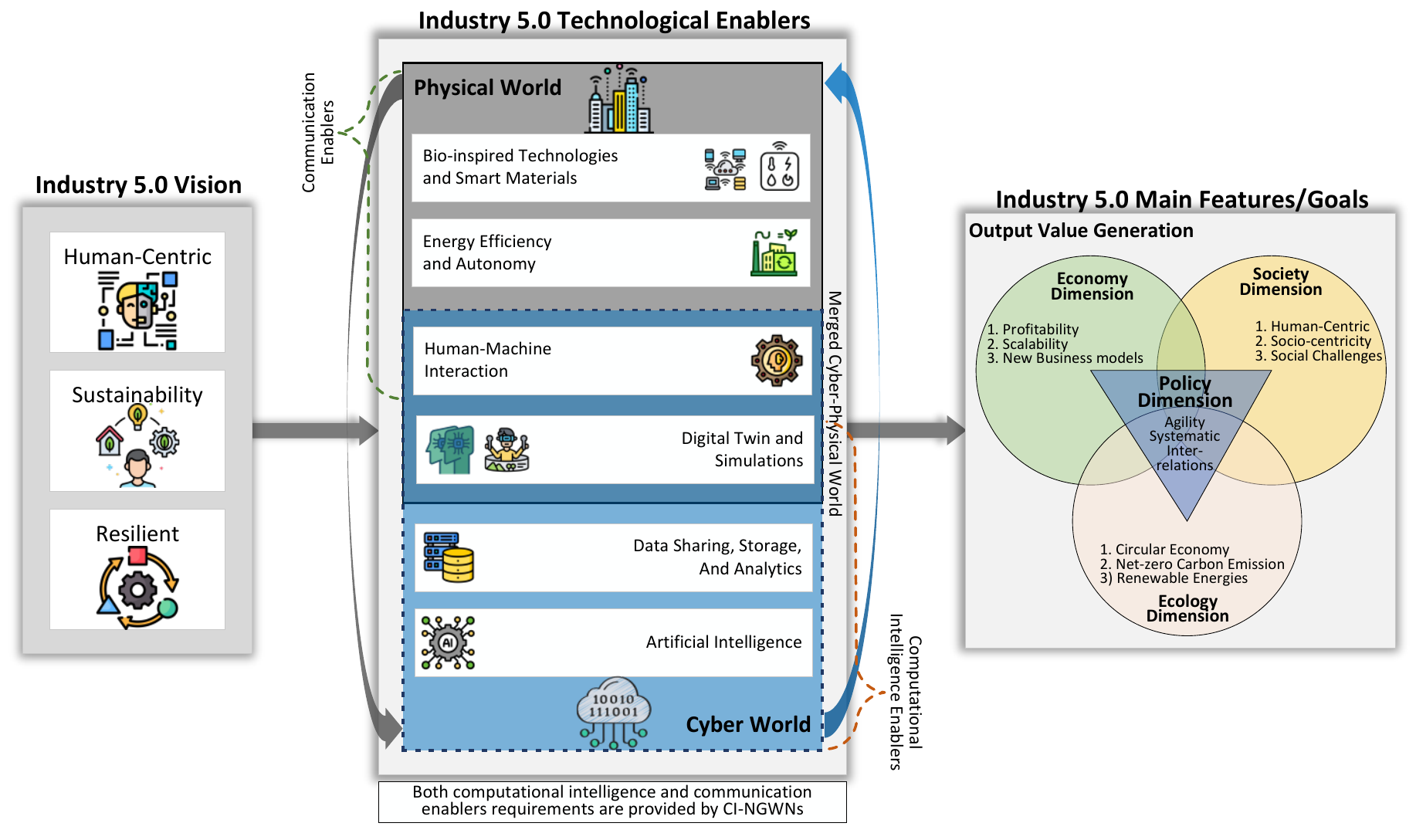}
	\caption{The framework provides a quick overview of the concept of Industry 5.0 as well as broad vision/definition, technological enablers and main goals. }
	\label{fig:I5.0Enablers}
 	\vspace{-10pt}
\end{figure*}
{\color{black}
\section{Technological Enablers of Industry 5.0 Vision and How CI-NGWNs KPIs Support it}
\label{sec:Industry5.0VisionandRequirements}
The optimization and/or expansion of existing Industry 4.0 application requirements (e.g., extended coverage, enhanced features) and new vertical and advanced applications in Industry 5.0 incurs enhanced communication and computation service requirements. In this section, we first discuss the critical technological enablers to realize the Industry 5.0 vision (Sec.~\ref{sec:I5.0Enablers}), followed by the enhanced KPIs of CI-NGWNs that can facilitate the required technological enablers needed for the Industry 5.0 vision (Sec.~\ref{sec:CI-NGWNsupport}).
\subsection{Required Technological Enablers}
\label{sec:I5.0Enablers}
The industry 5.0 enabling technologies are multiple complex systems resulting from seamless convergence of emerging technologies and paradigms that bind physical and cyberspaces to realize the vision (c.f.~Fig.~\ref{fig:I5.0Enablers}). The successful working symbiotic relationship between multiple complex systems and supporting technological frameworks in tandem can only enable the true multi-dimensional potential of the Industry 5.0 features~\cite{EU2,EU3}.  
These industry 5.0 technological enablers are,

\subsubsection{Human-Machine Interaction}
\label{Human-centric}
Human-machine interaction technologies can enable the hyperconnectivity between machines and humans to provide human-centric innovation, solutions, and intelligence, e.g., cognitive and affective computing, digital assistance, visual insights, smart management, and control. Following enabling technologies supports human-centric value design in Industry 5.0.

\fakeparagraph{Cognitive and Cyber-Physical-Social Systems (CCPSS)} is a rapidly growing interdisciplinary technology that combines cognitive computing architecture at the nexus of three critical spaces of machine/cobots to deliver intelligent solutions~\cite{cognitive1}. These three crucial cobots spaces are cyber, physical, and counterpart social (human) components for enabling human-machine interaction. By integrating machine learning and artificial intelligence with brain-inspired Affective computing designs, CCPSS uniquely combines the best of both worlds through human intelligence to aid the Industry 5.0 vision~\cite{CPSS-1}.

\fakeparagraph{Autonomous Mobile Robots (AMR).} 
By combining AMRs with AGVs, multidirectional production lines can be laid out instead of the linear layout possible with conventional conveyors~\cite{AMR-1,AMR-2}. It understands the surrounding environment in matrix productions~\cite{matrix4KUKArobotics}. In the plant and between the warehouses and loading bays, AMRs ensure the movement of components inside and among work cells. Several of the processing tasks related to AMRs, including collision avoidance and visual navigation, are migrated to the on-premises cloud/edge. 

\fakeparagraph{Augmented Reality Technologies}, integrates virtual information with the real world to provide computer-generated perceptual information, e.g., AR-VR-XR~\cite{ARXR1,ARXR2}. The technical uses include multimedia, 3D-modelling, real-time intelligent interaction, sensing, visual insights, and more.

\fakeparagraph{Cognitive Service Provisioning.}
Cognitive service provisioning provides the customized standalone services to meet the service demands of previously discussed human-centric applications~\cite{cognitiveS1,cognitive2,cognitive3}, such as 1) tracking technologies, e.g., real-time monitoring of industrial workforce, navigational support to AMRs, 2) multi-lingual support for natural language processing (NLP) applications, 3) CV for human cognitive abilities, e.g., gesture recognition, pattern recognition and 4) cognitive APIs for CCPSS-based cobots.

\subsubsection{Real-time Virtual Simulation and Digital Twin}
\label{DTsim}
Digital twin technology implementation enables the real-time cyber world through creating a virtual/digital images of counterpart physical assets, which can be devices, process, sensors, etc. DT technology realizes following critical two functional aspects.

\fakeparagraph{DT of Physical Assets and Process Optimization.}
By bridging the physical and cyber world, DT models the entire complex system comprised of deployed industrial assets, enabling them to communicate and interact intelligently to take action and optimize the multi-fold value of the output production process~\cite{DT1,DT4}. Together with IIoT and CPS, DT can facilitate this true digital hyperconnectivity in FoFs~\cite{CI1}. It allows the remote maintenance and predictive maintenance paradigms inside the factory ecosystems~\cite{DT2}. Similarly, massive interconnected DT systems integrated with augmented technologies can efficiently utilize human intelligence for management and control purposes~\cite{DT3}.

\fakeparagraph{Virtual Simulation and Modelling.}
DT provides visualization access to multi-scale and multi-dimensional dynamic modeling of factory processes, machine parts, and other complex systems~\cite{DT5}. Based on the digital images, 3D virtual simulation can be implemented to check the final outputs of factory operations and test new products. Hence, it results in the establishment of operational safety for human-centric purposes before machines or complex factory systems deployments~\cite{DT6}. Similarly, virtual simulations of the green environmental and social impacts of deployed systems can be carried out to take measurements for designing and maintaining value-based production policies.





\subsubsection{Artificial Intelligence-native Smart Systems}
\label{AIusage}
AI-native smart functions can identify and predict the incoming causalities in complex and dynamic industrial assets, e.g., cobots, factory processes, worker's conditions, critical events, which leads to necessary automated actions. For Industry 5.0 vision, AI-native systems need to possess the following characteristics. 

\fakeparagraph{Required AI Systems Features.}
The AI-based innovative system gives the ability to intelligently cater to unexpected critical events fully autonomous and without human intervention~\cite{AIS-1}. However, the problem resides in traditional AI system since they only function and operate with past historical data~\cite{AIS3}. In a real-world scenario, the data is relatively less, and data quality varies. Moreover, the time-series spatial data has a lot of complex statistical cross-correlations among numerous dependent variables, which vary drastically with the heterogeneous data sources, and interrelated dynamic and random systems~\cite{AIS2}. Similarly, state-of-the-art DNN algorithms, i.e., Long Short-Term Memory (LSTM), Bi-LSTM, Gated Recurrent Unit (GRU), Convolutional Neural Network (CNN)-based algorithms, are typically used to capture the complex correlations in past historical time-series data. It is noted that not always past correlation holds out same in the future because of additional causal relationships within the observed data variables, which affects the future prediction. In such cases, causal AI can capture the essence of causality and complex correlations to adopt intelligent choices and actions through integrating human-decision making and reasoning~\cite{AIS4}. 






\fakeparagraph{Individual and Person-centric AI}, a human-centred AI system that focuses on AI algorithms that emerge from continuous human inputs and collaboration learning from human insights~\cite{Human-centric1}. Human-centered AI regards continuous improvements through human input, enables explainable AI paradigms~\cite{Human-centric4}, while allowing humans and cobots to work together in practical ways, e.g., CCPPS systems~\cite{Human-centric2,Human-centric3}.
For example, AI-driven informed algorithms, such as data-driven physics-informed models, are able to learn from incoming data and mathematical physics models. They emulate brain learning, even in unclear, high-dimensional, and partially understood scenarios~\cite{Human-centric4}. 





\fakeparagraph{Secure and Energy-Efficient AI.}
Data has an integral part in realizing the potential of AI systems. Generally, centralized cloud or private on-premise database resources are utilized to perform computations relevant to the AI learning process~\cite{EAIS-1}. In the cloud, distributed computation-based AI models are applied, i.e., data is shared between different geo-distributed computing infrastructures, including data servers (peers)~\cite{EAIS-2}. However, sharing data among different geo-distributed peers increases security risks. Moreover, increasing data computation increases the energy efficiency cost of deployed infrastructure for AI-based computations~\cite{EAIS-3}. Hence, a robust, resilient, and green-aware computing design is critical for realizing the goal of secure and energy-efficient AI models.

\subsubsection{Data Infrastructure, Sharing and Analytics}
\label{cybersafe}
Hyperconverged data infrastructure forms a foundational building block in providing computing and data storage functions, enabling the AI- and DT-based system functionalities. This increases the importance of cyber-safe data transmission, storage, and analysis technologies with the required certain features.

\fakeparagraph{Communication Interfaces.}
Over-the-air, industrial access networks required a communication interface with extreme reliability, ultra-low latency, high data rates, and massive connectivity support to transfer the bulk of collected and sensed informational data to storage infrastructure for further processing~\cite{IndAccess-1,IndAccess-2}. Similarly, data sharing mechanisms and interoperability in heterogeneous systems need to be enhanced, increasing the importance of designing new data-sharing protocols for enabling secure and safe data transmission~\cite{IndAccess-3,IndAccess-4}.

\fakeparagraph{Scalable and Multi-level IT Infrastructure.}
SDN/NFV-assisted scalable (upgradable) and multi-level (cloud/edge) computing IT infrastructure is needed to handle the diverse critical communication and computation requirements that can dynamically provide customized data-related services in every scenario~\cite{ITInfra-1,ITInfra-2}.
For this, softwarization across computing IT infrastructure and emerging service-oriented architectures (e.g., microservices, anything-as-a-service) gain importance for enabling the customized softwarized service provisioning paradigms.
Similarly, blockchain network, AI-based solutions, and private enterprise networks can provide better services for cyber-security in computing infrastructure~\cite{ITInfra-3,ITInfra-4}, e.g., data integrity, secure communication, privacy preservation, detecting intruder attacks. 

\fakeparagraph{Autonomy over Shared Infrastructure Resources.}
Softwarization in IT infrastructures enables autonomy and control over the softwarized shared network resources by adopting the multitenant architecture~\cite{multitenant-1,table_soft3}, where softwarized services orchestration can provide the required secure and service-aware communications over virtual private networks. 

\fakeparagraph{Big Data Analytics and Management.}
Data analytics combines the power of AI frameworks and high data computations in search of extracting critical information from big data to gain knowledge~\cite{bigdata-2,bigdata-3}. This increases the importance of data management, e.g., data traceability, data quality, data features extraction, data storage and backup, data fusion. Similarly, the data preprocess mechanism gains importance for AI applications due to the unlabelled and raw nature of initially collected big data~\cite{bigdata-1}.


\fakeparagraph{Edge Computing and Micro Cloud.}
The edge computing paradigm brings the centralized cloud capabilities closer to the end-users, reducing end-to-end response time, latency, and communication resources in mission-critical applications, e.g., migrating from cloud DT to edge DT. However, to enable an edge computing paradigm, micro clouds are needed~\cite{table_soft10,CI-NGWNs2,DT5}. 
In the edge computing paradigm, a new micro cloud-based infrastructure comprises of compute nodes arranged in small clusters with local storage and secure networking, allowing repeatable, reliable remote deployments and on-demand computing services at the edge of the network~\cite{microcloud-1,microcloud-2}.

\begin{figure*}[!t]
\centering
\includegraphics[width=0.85\linewidth]{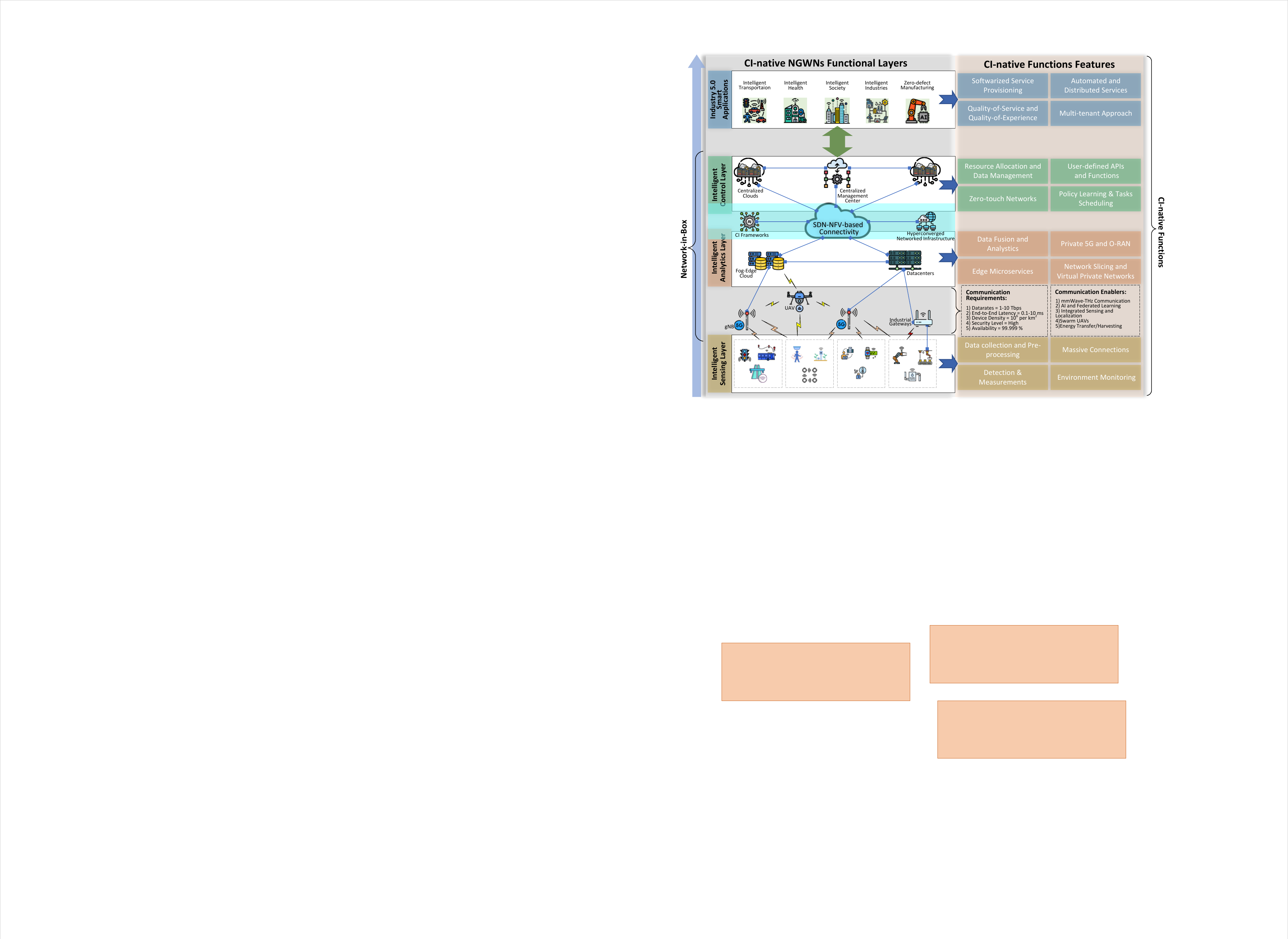}
	\caption{Hyper-level illustration of NIB architecture in CI-native NGWNs to support the technological enablers of Industry 5.0 vision}
	\label{fig:HyperarchitectureCI-NWGNs}
 	\vspace{-10pt}
\end{figure*}

\subsubsection{Bio-inspired Technologies.}
\label{smartmaterials}
It interconnects humans and industrial assets (e.g., cobots) through communication devices or sensors that use the biologically inspired sensing methodology.

\fakeparagraph{Embedded Bio-sensor Network Technologies.}
Any physical environment digitally interacts through the embedded sensor networks (ESNs)--- the mesh network of sensor chips/nodes or computers with many sensors embedded in the physical environment~\cite{ESN-1}. The nodes in ESNs are small and inexpensive chips, each providing specific sensing functions. With wireless Internet connectivity, these embedded sensor networks map the physical world and transmit data information to a centralized cloud, enabling a wide range of applications~\cite{ESN-2,IndAccess-1}, e.g., wireless body area networks (WBANs) for smart health IoT applications, IIoT-based connectivity for enabling CPS systems, etc. One emerging trend in sensor technology is the robust biosensors--- innovative analytical devices that integrate the physiochemical and biological sensing element, i.e., analyte-based sensory, to enable sensing. Biosensors applications are environmental monitoring, toxin detection, drug and pathogen discovery in health applications, and food monitoring~\cite{abid2021biosensors}.

\subsubsection{Energy Efficiency}
\label{EEusage}
Almost technologies mentioned above require many energy resources to operate and provide critical functions. This eventually increases the importance of enabling technologies to adopt green technologies and strategies for achieving goals of emission neutrality.

\fakeparagraph{Integration of Renewable Energy Sources.}
Numerous enabling industries, e.g., manufacturing and automation factories, run on the energies created through fossil fuel-based resources (coal, oil, and natural gas), which causes harm to the environment through greenhouse gases~\cite{renewable-1}. Similarly, dust emissions from industries pollute natural environmental resources. Instead of fossil fuels, renewable energy resources must be adopted for emission-neutral goals~\cite{renewable-1}. Some examples of renewable energy sources are, solar energies, wind energy, hydropower resources, geothermal energies, Hydrogen, and synthetic non-carbon fuels. Similarly, smart dust systems need to be adopted in the factory ecosystem to filter greenhouse gas emissions and dust into the environment.

\fakeparagraph{Energy-autonomous Sensor Systems (EAS)} 
is designed to operate and/or communicate in known or unknown environments that provide data communication, sensing, and storage without connecting to a power source, e.g., grids, batteries~\cite{EAS-1}. 
One example of an EAS system is wireless powered communication networks (WPCNs), where power is transferred over the air to enable data transfer~\cite{WPCN-1}. In WPCNs, energy harvesting paradigms are typically deployed, enabling battery-free information transfer~\cite{backcom-1}, e.g., ambient backscatter communication and cooperative communication or providing low-power data transmissions~\cite{SWIPT-1}, e.g., simultaneous wireless information and power transmission (SWIPT) paradigm.



\subsection{CI-NGWNs Architecture to Support Industry 5.0 Enablers}
\label{sec:CI-NGWNsupport}
The integration of CI-NGWNs and the critical pillars of Industry 4.0 (i.e., IIoT, CPS, and DT), next-generation IoTs (NG-IoT), and SDN/NFV-enabled cloud/edge computing architecture are poised to form critical drivers in supporting a wide range of enabling requirements incurred by the industrial and societal changes in Industry 5.0 scenarios, as illustrated in Fig.~\ref{fig:HyperarchitectureCI-NWGNs}.

\subsubsection{Design Objectives}
\label{sec:NGWNdesignObj}
The design objectives of CI-NGWNs lie in the domain of supporting enhanced communication and networked computation requirements and enabling technological features for Industry 5.0 vision~\cite{CI-NGWNs5} (c.f.~Sec.~\ref{sec:I5.0Enablers}). These design objectives are,
\begin{itemize}[leftmargin=*]

\item Automated and distributed services using multitenancy approaches in CI-NGWNs service architecture to be\textit{ flexible, self-optimized, and self-adapting} and ensure diverse QoS and QoE requirements in numerous industrial application.

\item Capabilities of \textit{maximum and efficient re-usability} of shared networking infrastructure.

\item \textit{Open integration} between numerous industrial applications and CI-NGWN ecosystem to create new and diverse business models for operators and service providers.

\item \textit{Increased performance and lower cost of conventional hardware-defined systems (HWS)} using hyper-converged infrastructure (HCI)--- software-defined IT infrastructure that virtualizes all HWS elements, i.e., computing, storage, networking.

\item Integrates big data analytics and AI/ML-based functionalities across all layers in the network, i.e., core, access network, and edge network.

\item Ensure \textit{security, privacy, and energy efficiency (100x over B5G)} goals over the entire CI-NGWN architecture

\end{itemize}

\subsubsection{Vision and CI-NGWNs NIB Architecture}
\label{sec:NGWNdesignvision}
To support these design objectives, the vision of CI-NGWN has stepped forth towards the \textit{network-in-box (NIB) architecture} where CI features and network softwarization is embedded across all network layers~\cite{CI-NGWNs1,CI-NGWNs2,CI-NGWNs3,CI-NGWNs4}. The hyper-level illustration of CI-enabled NIB architecture for NGWNs is shown in Fig.~\ref{fig:HyperarchitectureCI-NWGNs}, where CI functionalities had roots at each functional layer to provide intelligent features. These enabling CI-native functional features are as follows.


\begin{itemize}[leftmargin=*]
    
    \item The inclusion of CI technologies (swarm AI, evolutionary AI, and others) in conjunction with computing and storage capabilities at cloud-fog-edge level of network hierarchy (e.g., MECs) can bring the required functional edge intelligence closer to the physical layer. With the nexus of blockchain technology, federated learning-based distributed computing paradigms and private networking (e.g., private 5G) reduces the security and privacy concerns in NIB architecture.
    
    \item It integrates multitenancy-enabled softwarized service orchestration and provisioning architecture for providing customized services and modular functions, e.g., DT services, ICAR services, private virtualized network resources, cognitive APIs, AI frameworks, data processing, etc.
    \item It enables the transition from 5G service-based architecture (SBA) to AI-native service architecture for provisioning enhanced services, e.g., mLLMTC, UmMTC, ERLLC, MBBL, FeMBB, LDHMC, ELPC, and AEC. Moreover, it facilitates connectivity-based interaction between softwarized modular functions and AI-native architecture services.
    
    \item The O-RAN ALLIANCE standards are leading to radio access networks (RAN) that are open, intelligent, virtualized, and fully interoperable based on operator-defined specifications.
    
    \item It enables the flexibility in network functions using zero-touch network and service management paradigm, i.e., service-aware networking, new resources allocations, application deployments at edge nodes, dynamic setup of dedicated network resources for service provisioning, network slicing, automated monitoring and control, etc,.
    
    \item It provides the support for integrating diverse access network technologies in industries, i.e., 5G New Radio (NR), evolved LTE, UAV- and satellite-based terrestrial network access, fixed access, WLANs (WiFi-5, WiFi-6), and wired networks (e.g., Industrial Ethernet, Fieldbus communications). Similarly, inclusion of higher mmWave and THz spectrum along with non-orthogonal multiple access schemes and reflecting intelligent surfaces (RIS) increases the high data rates connectivity and spectrum efficiency.
    
\end{itemize}

}

\section{Softwarized Network Control and Service Provisioning for Industry 5.0}
\label{sec:softwarizedservices}
In the following subsections, we discusses the technological transition in various layers of automation pyramid in Industry 5.0, service architecture framework to support softwarized service provisioning, followed by the various customized industrial services scenarios. 

\subsection{Multi-Tenant Service-oriented Industrial Architecture}
\label{sec:servicearchitectural}

\fakepar{Technological Transition in Industrial Automation.}
The 5C smart automation architecture generally consists of five hierarchical levels/layers top to bottom, i.e., configuration, cognition, cyber, conversion, and connection~\cite[CPS 5C architecture]{jiang2018improved}. 
With the convergence of CI-NGWNs along with SDN/NFV-enabled cloud/edge-native computing paradigms and the critical pillars of Industry 4.0 (e.g., IIoT, CPS, and DT) to the 5C smart automation architecture, the technological transition among various layers of pyramid arises with various functions of 5C automation layers now embedded into the new layers, i.e., \textit{infrastructure layer}, \textit{computational intelligence layer} and \textit{service layer} (c.f.~Fig.~\ref{fig:softwarizedarchitecture}). 

\begin{figure*}[!t]
\centering
\includegraphics[width=0.98\linewidth]{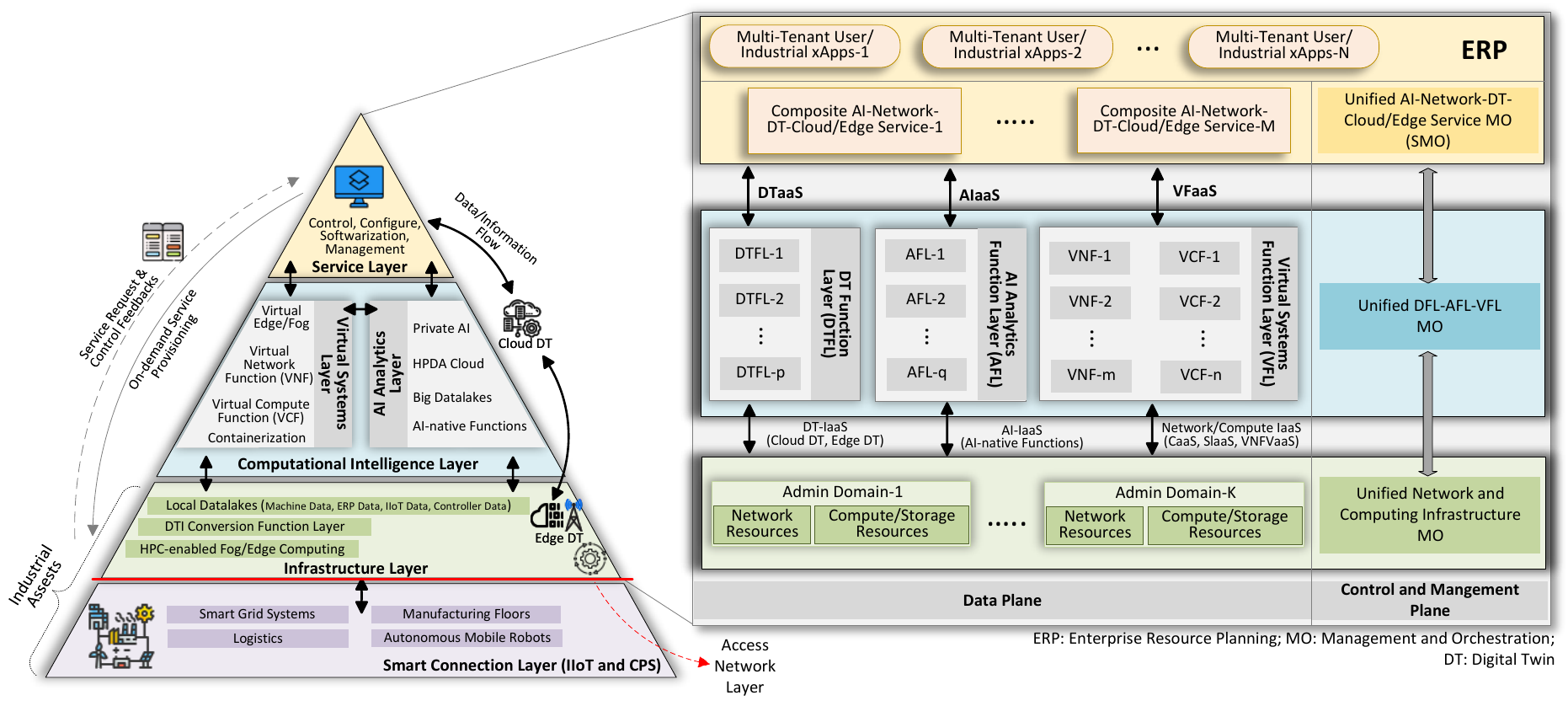}
	\caption{An illustration of multi-tenant softwarized service provisioning framework in Industry 5.0 scenario where CI-enabled NGWNs, CPS, IIoT, DT, and cloud/edge computing converges with the hierarchical automation and control pyramid.}
	\label{fig:softwarizedarchitecture}
 	\vspace{-10pt}
\end{figure*}
\begin{figure}
\centering
\includegraphics[width=0.97\linewidth]{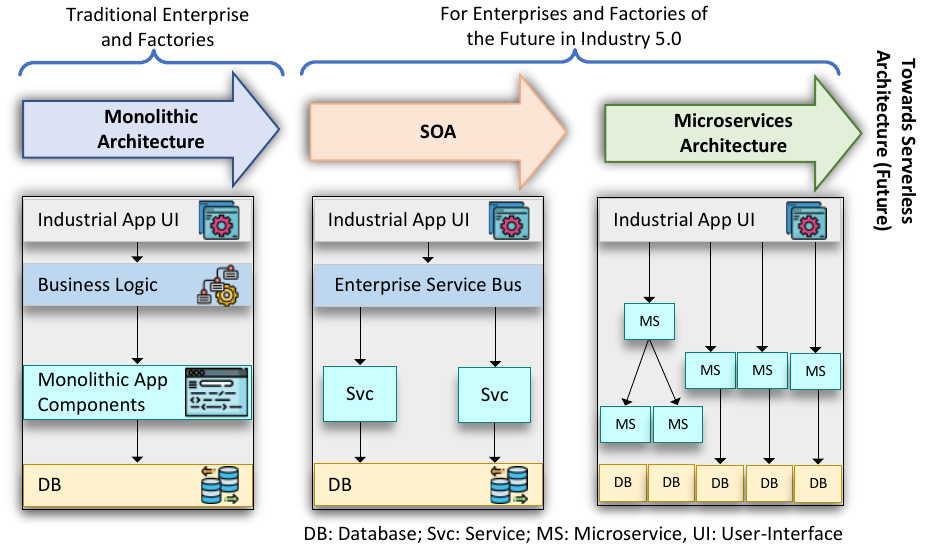}
	\caption{Illustration of evolution of softwarized service provisioning architectures and their compliance with Industry 5.0. }
	\label{fig:evolutionofsoftwarized}
\vspace{-10pt}	
\end{figure}

\fakepar{Softwarized Architecture.}
The multi-tenant softwarized architecture is poised to form key drivers in supporting wide range of new and on-demand service requirements in zero-defect manufacturing and matrix production systems, under the vision of mass production and mass customization in Industry 5.0~\cite{table_soft9}. 
The novel framework can be observed from Fig.~\ref{fig:softwarizedarchitecture} that closely integrates multiple function layers of smart automation pyramid and enables interactions/communication among numerous provisioned services, comprised of multiple customized functions.
Note that each layers of automation pyramid in Fig.~\ref{fig:softwarizedarchitecture} can provides the foundational new function blocks, such as virtualization functions, AI-native functions, DT placements and others, that enables the softwarized service provisioning to industrial applications.
Moreover, as illustrated in Fig.~\ref{fig:evolutionofsoftwarized}, it can closely integrate the principles of \textit{service-oriented architecture (SOA) and microservices architecture} support to enable true digitization in Industry 5.0 vision~\cite{table_soft10,multi-tenant-1,multi-tenant-4}.

\fakepar{Significance.}
The significance of softwarized architecture are, 
\begin{itemize}[leftmargin=*]
    \item Numerous industrial assets or industrial applications at the smart connection layers provides the service requests to the centralized service layer according to the demanding industrial application requirements. 
    \item Centralized management centers at the service layer can orchestrate and provision the required on-demand services needed closer at bottom layer, i.e., smart connection layer, by utilizing cloud/edge-natice computing infrastructural resources efficiently, which can be easily scaled, deployed and monitored.  
    \item Moreover, various monitoring control feedbacks and data transmission for management can be send directly to the service layer in real-time for further resource planning or control purposes. 
\end{itemize}
In this regard, Table.~\ref{table:softarchitectures} provides the taxonomy of various softwarized service architectures that has unique significance and drawbacks concerning numerous industrial use cases. 


\fakepar{Service Provisioning.}
To provision the on-demand incoming service requests, the service layer uses the composite mixture of underlying layers functions to orchestrate and deploy the orchestrated service to the industrial assets of numerous industries~\cite{CI-NGWNs2}. Each layer provides the functional blocks which provides the following functions.


\definecolor{Gray}{gray}{1}
\definecolor{Gray1}{gray}{0.95}
\bgroup
{\renewcommand{\arraystretch}{1.1}
\begin{table*}[t!]
\centering
	\caption{Taxonomy of various softwarized service architectures with their significance, drawbacks and expected usecases scenarios in Industry 5.0 } 
	\scalebox{0.75}{
	\setlength\tabcolsep{2pt}
\begin{tabular}{ccccc}
\toprule
\rowcolor{Gray}
\textbf{References} &\textbf{\begin{tabular}[c]{@{}c@{}}Softwarized \\ Architecture\end{tabular}} &
  \textbf{Significance} &
  \textbf{Drawbacks} &
  \textbf{\begin{tabular}[c]{@{}c@{}}Industrial\\ Usecases\end{tabular}} \\ 
\toprule

\rowcolor{Gray1}
\cite{table_soft1,table_soft2} & Monolithic    & \begin{tabular}[c]{@{}c@{}} \tabitem Simpler design, development and deployment \\ \tabitem Better Performance \end{tabular} & \begin{tabular}[c]{@{}c@{}} \tabitem Lack of adoption to new technologies \\ \tabitem Codebase gets bigger with increase in App scope,\\leading to lower quality of code, increased \\cost and time consuming to manage \end{tabular}  & \begin{tabular}[c]{@{}c@{}}  \tabitem Industrial Apps with small scope \\ \tabitem Small business and factory \\ \tabitem Startups \end{tabular}   \\ \midrule

\rowcolor{Gray}
\cite{table_soft3,table_soft4, table_soft5, table_soft6} & SOA           & \begin{tabular}[c]{@{}c@{}} \tabitem Maintenance, reusability, and reliability of services \\\tabitem Parallel service development \end{tabular} & \begin{tabular}[c]{@{}c@{}} \tabitem Investment costs, i.e., human resources, tools,\\ technology and development, etc,. \\  \tabitem Extra overhead to validate services interaction\\ among each others \end{tabular}  & \begin{tabular}[c]{@{}c@{}} \tabitem Enterprise Apps with complex \\ and medium-scale operations \\ \tabitem Complex service management \end{tabular}  \\ \midrule

\rowcolor{Gray1}
\cite{table_soft7,table_soft8,table_soft9,table_soft10} & Microservices & \begin{tabular}[c]{@{}c@{}} \tabitem Microservices are easy to develop, test and deploy \\ \tabitem Multiple skilled teams and developer design \\parts of softwarized components in services\\ and Apps due to the decoupling of microservices,\\ leading to agile development\\ \tabitem Fully automated vertical scaling (i.e., run same\\ microservices on new bigger machine in future)\\ and horizontal scaling (i.e., add more designed\\ microservices in same service/App. \end{tabular} & \begin{tabular}[c]{@{}c@{}} \tabitem Complexity (service mesh communication) \\ \tabitem Security concerns (API attacks, Service intrusions) \\ \tabitem Various programming languages for \\diverse microservices \end{tabular}  & \begin{tabular}[l]{@{}c@{}} \tabitem Complex large-scale systems \\with numerous industrial Apps\\ \tabitem Zero-defect Manufacturing \\and Matrix Productions \\ \tabitem Society 5.0 \end{tabular}  \\ \midrule

\rowcolor{Gray}
\cite{table_soft14,table_soft11,table_soft12,table_soft13} & Serverless    & \begin{tabular}[c]{@{}c@{}} \tabitem Incorporates function-and/or backend-as-a-service\\ concepts to microservices architecture\\ \tabitem Infrastructure-less dependency (e.g., no database\\ handling, business logic)\\ \tabitem Extremely fast deployment and enhanced scalability\\ of services \\    \end{tabular} & \begin{tabular}[c]{@{}c@{}} \tabitem Vendor lock-in (e.g., AWS Lambda, Azure \\Functions, Kubeless)\\ \tabitem Only for short real-time processes (i.e., one time \\ tasks, auxiliary processes) \end{tabular}  & \begin{tabular}[c]{@{}c@{}} \tabitem Client-heavy Industrial Apps \\ \tabitem Service provisioning in event-\\driven scenario (rapidly changing)\\ \tabitem High-latency background tasks \end{tabular}  \\ \bottomrule

\end{tabular}
}
\label{table:softarchitectures}
\vspace{-10pt}
\end{table*}
}
\egroup

\subsubsection{Service Layer}
 It integrates the functions of top three 5C architecture layers, i.e., cyber layer (twin model for machine components), cognition layer (integrated simulation and synthesis) and configuration layer (self-configure/adjust/optimized for resilience), and customized service provisioning paradigm to enable the true softwarized cloud/edge-native microservices architecture. It comprises of two functional blocks.

\fakeparagraph{Cloud/Edge-native DevOps.}
Service layer supports the cloud/edge-native multi-tenancy in industrial service architecture to orchestrate and deploy multiple diverse services for serving multiple industries, utilizing the underlying shared CI-enabled networking infrastructure. The customized developed industrial apps, such as ERP softwares, self-automated control and management operations, monitoring and visualization, can be orchestrated and deployed leveraging the functionalities of composite service blocks, depending upon the requirements of industrial end-systems on-demand diverse service requests~\cite{ERP-1}. Moreover, the introduction of multi-tenancy also introduces the shift in monolithic software components to cloud/edge-native microservices architecture~\cite{table_soft8}. Thus enabling the integration of agile software development and IT operations (DevOps) strategies to develop smart factory-based microservices closer to the physical layer. The opensource software orchestrator tools, such as Kubernetes (K8s) and OpenStack, can be adopted at the service layer to support the cloud/edge-native scalable service orchestration and deployment.

\fakeparagraph{Service mesh and Composite Service Blocks.}
Depending upon the on-demand service requirement, the unified service management and orchestration (SMO) function block in control and management plane of service layer select the mixture of appropriate softwarized functions, i.e., AFL, VFL, DTFL, and any infrastructure layer-related functions~\cite{table_soft7}. It helps in creating and assigning group of composite service blocks for multi-tenant industrial service application. The service mesh layer enables the required distinct group of softwarized functions in composite service blocks are closely managed and controlled by unified SMO through bi-communicating with the unified MO of underlying layers. Moreover, it controls service-to-service communication over entire network infrastructure. This enables the centralized service management and control center capabilities, helping in orchestrating, deploying and managing industrial service applications remotely from anywhere over Internet infrastructure~\cite{table_soft5}.

\subsubsection{Computational Intelligence Layer}
\label{sec:AFL}
This layer integrates the three critical core function layers, i.e., AI analytics, virtual systems, and DT systems, that work in tandem to enables the paradigm of softwarized intelligent services orchestration and deployment for industrial end-systems. The function of critical function layers are,

\fakeparagraph{AI Analytics Functional Layer (AFL).} 
AI-powered systems can provide diverse and customized AI-native intelligent functions for numerous applications, i.e., business analytics, zero-touch agile network, monitoring and control, etc. Depending upon the application  requirements, these AI-native functions selects and builds the appropriate AI model from available AI models repository to provide the customized intelligent roles, such as data mining and clustering, intelligent decision-making and control~\cite{AI-services-1}. Combining AI-native functions to analytic tools, it can analyze and extract the high value information at the big datalakes that stores arriving bigdata from thousands of heterogeneous data sources in various layers, such as factory processes, industrial assets, core networked devices, DT applications, service application data, data from virtualized industrial network, customer feedbacks, machine data, etc. These analytic operations are summarized in Fig.~\ref{fig:AIanalyticmodels}, which shows the four stages of analytic application types and their description with respect to the value and difficulty of analytic tasks.
By integrating high power computing capability (HPC) with AI analytics, high power data analytic (HPDA) can be applied at the big datalakes with the increase in the value and difficulty of analytic tasks~\cite{bigdata-3}. 
\begin{figure}
\centering
\includegraphics[width=0.9\linewidth]{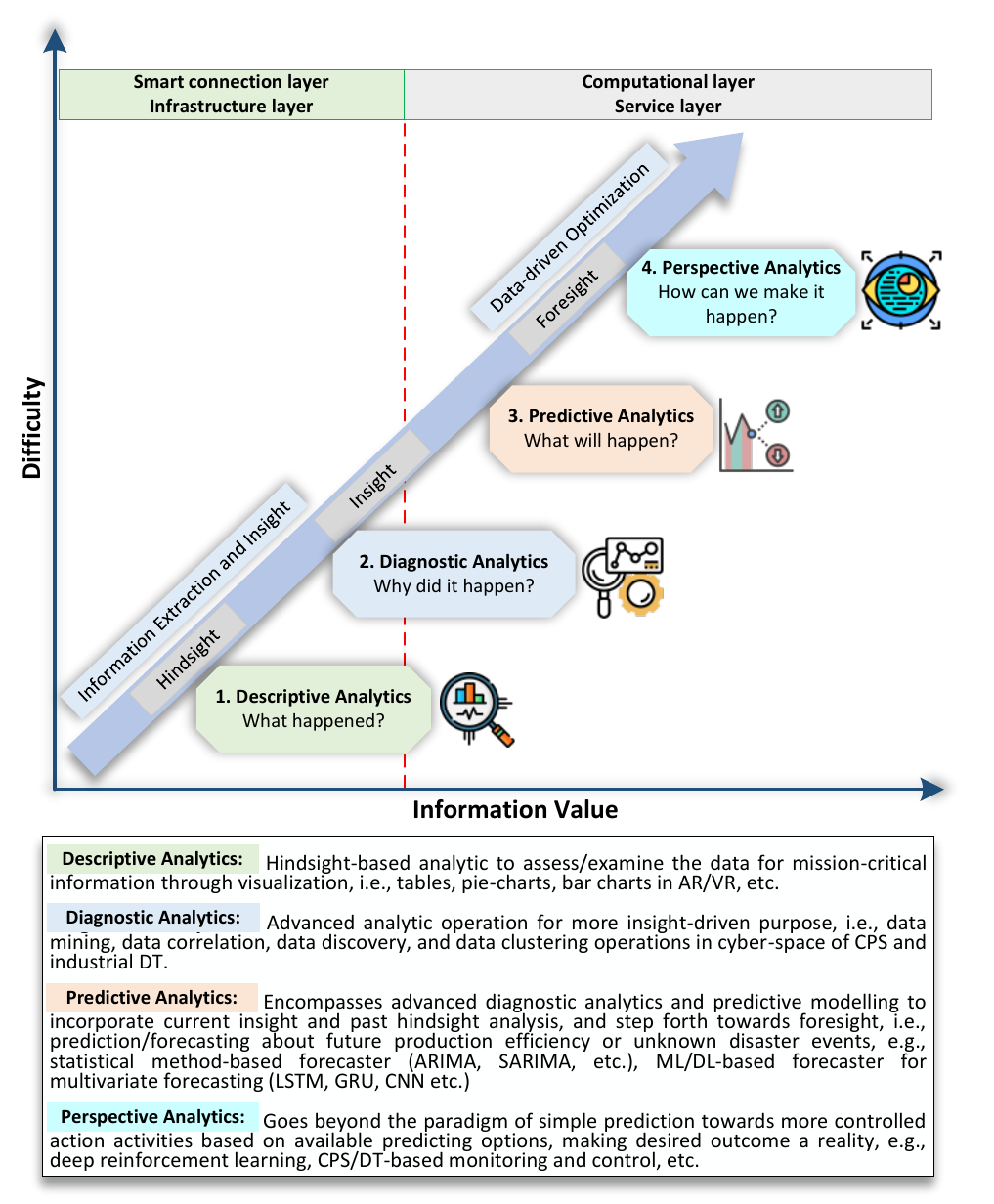}
	\caption{Illustration of AI analytics-based Gartner Analytic Ascendancy model integrated with critical functional layers of industrial automation pyramid and their functional roles description. }
	\label{fig:AIanalyticmodels}
\vspace{-10pt}	
\end{figure}

\fakeparagraph{Virtual Systems Function Layer (VFL).}
Virtualization or virtual system functions offers the dedicated, distributed and isolated physical resources over inter/intra-connected HCI systems (infrastructure layer) based on the on-demand service requirements~\cite{multi-tenant-2}. Following types of virtual functions are provided by VSFL.
\begin{itemize}[leftmargin=*]
    \item \textit{Virtual Storage/Server Functions (VSF)}: Virtual server involves the partitioning of computer server having large memory, GPU, CPU and disk resources into small multiple virtual machines (VMs) with dedicated resources assigned to it~\cite{VSF-1}. Similarly, virtual storage integrates the networked storage devices over multiple machines into single virtual storage device which facilitates the dynamic storage resources allocation as well as data backup and recovery tasks. VSF provides both the required virtual machines and/or virtual storage resources required over the entire enterprise network. 
    \item \textit{Virtual Compute Functions (VCF)}: Virtual compute services are enabled through the usage of virtual servers and virtual storage since compute resources integrates the memory, GPU, CPU and storage resources, which are accessible to the multiple users~\cite{VCF-1}. VCF blocks over infrastructure layer resources provides the virtual compute services to the multiple incoming on-demand service requests.
    \item \textit{Virtual Network Functions (VNF)}: Network virtualization through SDN/NFV functionalities enables the management and control of entire enterprise network (dedicated and shared) through software-defined single or multiple administrative entity~\cite{networkviruali-1,networkviruali-2,barakabitze20205g}. Thus providing the autonomy over enterprise network. Some of the virtual network functions are, 1) SDN routing and switching functions, 2) service-aware routing, 3) provision of virtual network and tunneling over internet, e.g., virtual private network, virtual fog/edge networks 3) traffic analysis, e.g., deep packet inspection, QoS and QoE measurements, 4) optimization of application access requests over network, e.g., load balancers, 5) security functions, e.g., network firewalls, network intrusion detection, etc., and 6) 5G core network components, e.g., radio network controller, user plane function (UPF), data network, network slice selection function (NSSF), core access and mobility management functions (AMF), etc.   
    \item \textit{Container-based Virtualization}: The requirement of scalability, flexibility and agility in DevOps techniques for cloud/edge networked computing paradigm and the dependency of VM over single operating system and hypervisor has led to the
    containerization or container-based virtualization~\cite{VCOF-1}. Container are different from VMs since they do not use hypervisor and easily carries all packaged softwarized components (i.e., runtime libraries and OS dependencies) of microservices/applications~\cite{VCOF-2}. This enables the improved and enhanced integration of virtualization functions that works in tandem with state-of-the-art cloud/edge-native networked computing architecture~\cite{VCOF-3}.
\end{itemize}

\fakeparagraph{DT Function Layer (DTFL).} 
DT function layer uses the functionalities of AFL and VFL to provides the following DT functions, 1) Virtual simulation and performance prediction~\cite{DT5}, 2) Product lifecycle and predictive maintenance, 3) machine failure analysis, and 4) configuration design, motion planning and control development.

\subsubsection{Infrastructure Layer.}
Infrastructure layer forms the backbone between the underlying smart connection layer and the upper topmost layers. Following functions and roles are integrated into this layer.

\fakeparagraph{HCI Resources.} 
Infrastructure layer composed of SDN/NFV-based inter/intra-networked HCI physical systems that provides the fog/edge computing capabilities, Elasticsearch-enabled datalakes, and network resources to industrial end-systems on smart connection layer~\cite{ITInfra-2,ITInfra-1}. 

\fakeparagraph{Industrial Access Networks.}
\textcolor{black}{The industrial access networks, providing a communication interface for the operation technology (OT) domain, enable field-level communication among controllers, machines, and within machines at the routing, control, and sensor level based on the IEC 61784-2 and IEC 61158-1 standards~\cite{TSN-1}. At present, due to diverse industrial connectivity requirements, various wired solutions for the industrial networks exist, such as Ethernet-based (e.g., PROFINET, EtherCAT) and field-buses (e.g., PROFIBUS). However, these variants lead to various issues in realizing the Industry~5.0 vision, including hardware/protocol heterogeneity and multiple solutions for enterprise-wide connectivity while maintaining a strict boundary between OT and IT networks~\cite{TSN-2,TSN-3}. Eliminating or at least thinning this boundary between IT and OT domains is the prerequisite to improving the efficiency and agility of the industrial processes. Time-sensitive networking (TSN), is a step towards this, where the IEEE~802.1 time-sensitive networking (TSN) task group is developing a set of standards (through various TSN profiles) to extend the best-effort Ethernet networking model to provide deterministic streaming services~\cite{aamir2022}. Still, the missing part towards achieving a full-scale network connectivity is the inclusion of wireless access to support massive sensing, mobility, and customizable/agile processes. However, the prominent industrial wireless standards (e.g., ISA100.11a, WirelessHART, Industrial WLAN by Siemens) are designed mainly to support low-rate and/or localized wireless access for industrial monitoring/control applications. These standards operate on unlicensed frequency bands (making them prone to interference issues) and provide short-range connectivity to a limited number of devices. On the other hand, 5G-and-beyond networks with service-based architecture, various RAN and Core enhancements, and 5G-TSN integration(c.f.~Fig.~\ref{fig:softwarizedcommunication}) are becoming attractive to provide robust, scalable, unified, and customizable wired-wireless connectivity for Industry 5.0.}

\fakeparagraph{Autonomy and Control.}
It can provides the autonomy and control over HCI resources and industrial access networks. For example, an enterprise can solely own the access networks and locally installed HCI resources providing total autonomy over enterprise network~\cite{table_soft3}. Similarly, different vendors can also provides the enhanced infrastructure layer resources in the forms of softwarized services to various industrial enterprises, which can also be closely integrated with the locally owned enterprise network. Thus enabling the new digitized and diverse business models for different IT and OT vendors. (c.f.~\cite{Factory5G}) 

\fakeparagraph{Data-to-Information Functions.}
Datalakes closer to access networks layer collect and stores raw machine data, ERP data, IIoT data and controller data from end-systems on smart connection layer~\cite{DTI-2}. That raw data carries an inherent property of 5V's, i.e., volume, velocity, variety, veracity and value. By applying one of the data-to-information (DTI) functions, i.e., pre-process and data fusion, over the collected raw data to extract information and create knowledge that helps in value creation throughout a product life cycle~\cite{DTI-1}. Similarly there are other functions embedded in the DTI functions layers, e.g., data features engineering. Note that based on the obtainable knowledge, useful action can be taken at the manufacturing floor, such as increase production efficiency by redesigning process based on the customer feedbacks, increased resources supply for customized flows from production warehouse in zero-defect manufacturing, etc.

{\color{black}
\section{Open and Intelligent 6G Network Architectures}
\label{sec:6G_Architecture}
\begin{figure*}[!t]
\centering
\includegraphics[width=0.9\linewidth]{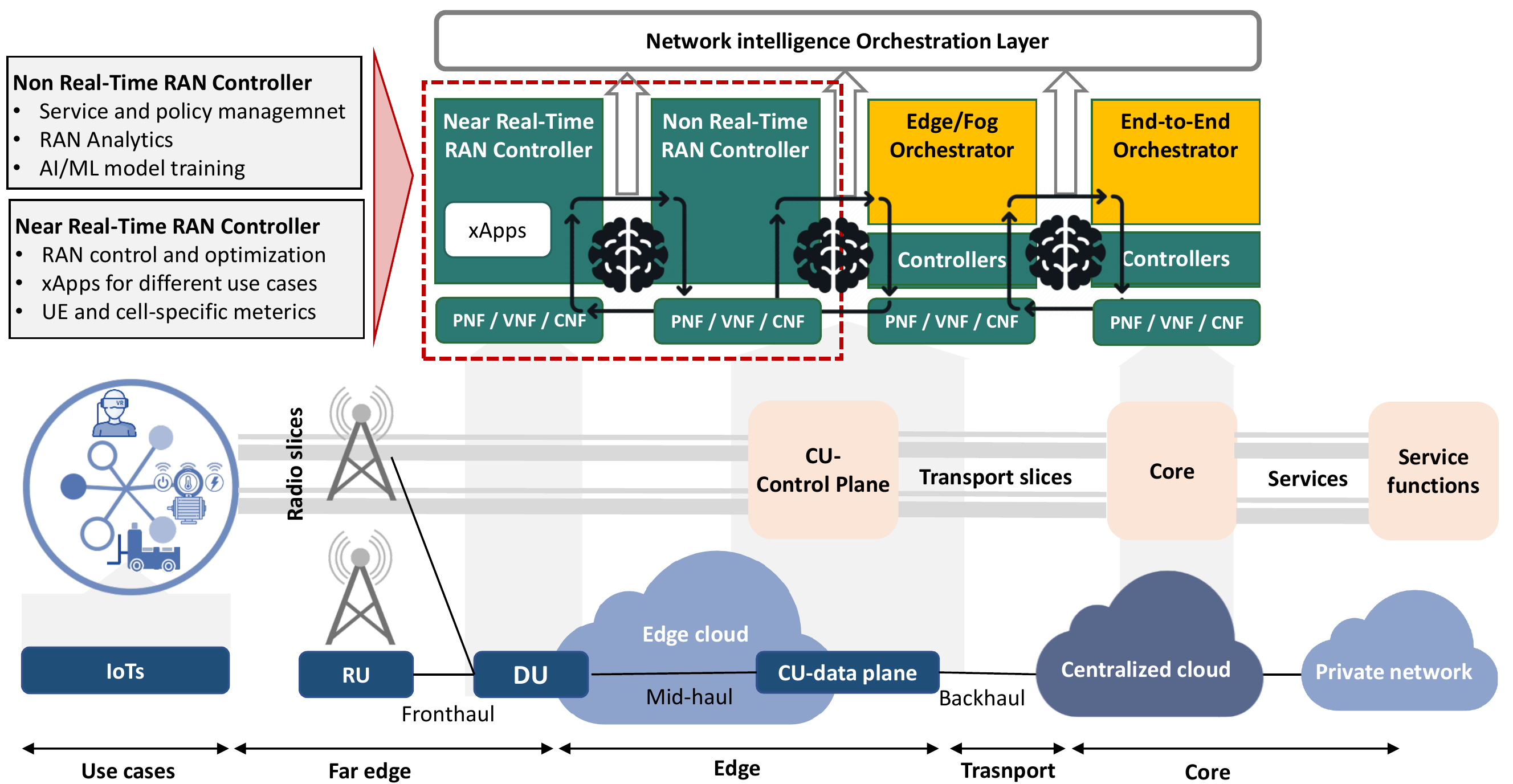}
	\caption{An overview of emerging end-to-end NI-native 6G architecture based on various standardization efforts.}
	\label{fig:6G_Arch}
\vspace{-10pt}	
\end{figure*}
As apparent from previous discussions, Industry~5.0 can be characterized by an increasingly complex ecosystem, requiring an ever-expanding multitude of service classes and their disparate time-varying traffic profiles. 
A forthcoming pathway to address this complexity challenge is the convergence of the firepower of three promising technologies, namely \textit{artificial intelligence, edge computing, and beyond-5G/6G networks}. The convergence is already happening, with an intention of accelerating the automation and optimization of NextG wireless networks to support diverse QoS requirements while capitalizing on edge computing and AI/ML. In this context, \textit{Automation} is required to cope with new scenarios in network/edge planning, operation, and management for vertical industries and enterprise use cases, while also creating the space and opportunity for new infrastructure owners and service providers. Meanwhile, \textit{optimization} is necessary to extend the computing and fast-growing AI/ML applications to the edge so that each network element can provide low latency, high reliability, and pervasive intelligence services.

Be it reducing the data transfer size to the cloud or processing urgency, the data handling must occur closer to the data source, i.e., at the network edge. The edge is defined as a continuum of “edge zones” comprising four key areas: device, premise (on-site, e.g., enterprise campus), access, and metro (upstream aggregation point like service provider) edge~\cite{5GAmerica_WhitePaper}. As evident by names, each edge zone can fulfill different computing/processing requirements well-suited to close the loop (data generation to data processing) according to its local environment without data transmission to a central server. For instance, IIoT devices (e.g., AMRs) can make navigation decisions with autonomous and intelligent local computation considering their local environment. Meanwhile, the AMRs’ integration within the matrix production environment can rely on the on-premise edge for collective intelligence and load training data to the access or metro edge. 


Therefore, the need for intelligent edge automation in NextG networks is vital for Industry~5.0 for automation and optimization for various tasks such as distributed data collection, real-time processing, situational awareness, and network slicing under traffic dynamics. Nevertheless, this new capability comes with its own set of challenges, especially it will require an open and intelligent network infrastructure to support automation, control, and management of network resources and services. Being intelligent implies handling the complex task of efficient management of information flows and, consequently, network infrastructure as every industry and use case will demand different architecture. 


\textit{This section aims to elaborate on various underpinning concepts, activities, and challenges of creating such an intelligent 6G architecture. In particular, we discuss the open and intelligent 6G network architecture, with a focus on Open~RAN  framework for RAN/edge automation, and ensuing end-to-end network intelligence requirements and challenges.}

\subsection{Envisioned 6G Architecture}
\label{subsec:envision_6G_Arch}
To accelerate the needed automation and optimization in NextG networks for on-demand heterogeneous service deployment, beyond-5G and 6G networks are rapidly transitioning from inflexible and monolithic to agile, disaggregated, and cloud-native architectures based on softwarization and virtualization, with the provision network component's reprogrammability. The transition aims to provide multi-tenancy transparent communication network infrastructure for instantly orchestrating end-to-end physical resources and virtual network functions (VNFs) across various network domains. Herein, the separation between the control functions and the hardware fabric as well as the introduction of standardized control interfaces will define custom closed-control loops, enabling edge intelligence and real-time analytics; thus realizing the autonomous, self-organizing networks (SON) effectively.

In this respect, the standardization initiatives are already reflecting the desired architectural redesign, which will further require enhancing NI-nativeness at various network levels (see Sec.~\ref{subsec:NI_6G}). Fig.~\ref{fig:6G_Arch} summarizes envisioned (re)-design of end-to-end network architecture based on the current standardization efforts in 3GPP~\cite{3GPP.TR23.700-91}, ETSI~\cite{D_to_P.ETSI-ENI} and O-RAN~\cite{O-RANAlliance}, where following are the notable points:
\begin{itemize}[leftmargin=*]
    \item The redesign decouples software (network functions) from the hardware to enhance overall network operations, and enable end-to-end softwarization, where the traditional physical network functions (PNFs) are complemented by VNFs and cloud-native network functions (CNFs). 
    \item It includes overarching AI-driven entities (i.e., controllers), responsible for the orchestration and control of PNF, VNFs, and CNFs across all network domains. \textit{Orchestration} relates to lifecycle management of service and network functions, including instantiation, scaling, and termination, as well as control e.g., for context-aware parameter tuning of network functions. 
    
    \item Also, it transitions the classical access-core split into micro-domains or edge zones to increase the management granularity of physical infrastructures in micro-domain. 
\end{itemize}

Fig.~\ref{fig:6G_Arch} shows the expected deployment and operation of controllers/orchestrators in various micro-domains, including Core, Transport, Edge (hosting the RAN's cloud part), and Far Edge (consisting of radio units (RUs) and the fronthaul). Note that the \textit{E2E Orchestrators} operate at a global network level (i.e., in core network deployments and service-related applications in private network deployments~\cite{Factory5G}) while \textit{Edge/Fog Orchestrators} on the mobile edge level. Meanwhile, the \textit{Controllers} introduce enhanced RAN versatility by shifting the network functionalities closer to the user; Non-Real-Time controllers for less time-critical network functionalities at a limited number of sites, and \textit{Real-Time Controllers} for fast-timescale functionalities (e.g., radio resource and interference management) to meet QoS targets.

This novel e2e 6G architecture is expected to enable new functionalities, including ~\cite{ORAN_Learning_NextG}: (i) service differentiation to provide demand-based virtual network slices over a common physical infrastructure, while meeting multi-tenancy and multi-service requirements, (ii) split network functions across multiple software and hardware components, provided by multiple vendors, (iii) capture/expose the KPIs, and network analytics through open interfaces, and (iv) real-time control of the entire network infrastructure via (third party) software applications. Jointly, these functionalities will enable fine-grained network automation and optimization, so that the network functions and their allocated resources can be adapted to traffic dynamics and meeting diverse, heterogeneous QoS requirements.

Over the last few years, multiple independent alliances and forums have initiated research on accelerating this transformation by increasing infrastructure virtualization, combined with embedded intelligence, to deliver more agile services and advanced capabilities to end-users. Open~RAN is such an initiative that enables such transformation via open and intelligent RAN automation, which we discuss in detail in the following section.

\subsection{Evolution towards Open~RAN}
\label{subsec:OpenRAN6G}
Compared to current RAN technology, provided as a hardware and software integrated platform, the ambition for Open RAN initiatives is to create a multi-vendor RAN solution that allows for the separation (or disaggregation) between hardware and software with open interfaces and standard network elements, hosting control and management software in the cloud. 

\fakepar{vRAN vs Open RAN.}
Before going further, the distinction between virtual RAN (vRAN) and Open~RAN is important. Virtualization, which transforms network function into VNFs, decouples VNF software such that it can run on the commercial off-the-shelf (COTS) hardware. However, in vRAN setup, the interfaces between network elements may still be closed or proprietary~\cite{O-RAN_disrupting_vRAN}. In contrast, Open~RAN refers to a set of standards for creating open internal interfaces as well as defines an disaggregated architecture to introduce new capabilities of RAN automation, and service management and orchestration.

\fakepar{RAN Functional Splits.}
RAN has stringent latency requirements, demanding interworking between NFs accordingly. Therefore, some NFs are virtualized at the cloud using central units (CU) while many RAN functions occur at distributed units (DU) (located closer to the cell site), where a DU is split into a baseband unit (BBU) and a radio unit (RU). In this respect, the \textit{Functional splits} refer to the points along the protocol stack where CU and DU signal processing can be separated from the RU. Meanwhile, how CU and DU functions can be distributed in various edge-zones is deployment scenario and use case dependent. 
\definecolor{Gray}{gray}{1}
\definecolor{Gray1}{gray}{0.94}
\definecolor{Gray2}{gray}{0.88}
\bgroup
{\renewcommand{\arraystretch}{1.1}
\begin{table}[t!]
\centering
	\caption{Learning-based closed-control loops in an O-RAN architecture with the challenges and limitations} 
	\scalebox{0.73}{
	\setlength\tabcolsep{2pt}
\begin{tabular}{|c|c|c|c|c|}
\hline
\rowcolor{Gray1} \textbf{\begin{tabular}[c]{@{}c@{}}Objective: Control \\ and Learning\end{tabular}} &
  \textbf{\begin{tabular}[c]{@{}c@{}}Time\\ Scale\end{tabular}} &
  \textbf{\begin{tabular}[c]{@{}c@{}}Devices\\ Scale\end{tabular}} &
  \textbf{Data Input} &
  \textbf{\begin{tabular}[c]{@{}c@{}}Challenges \\ and Limitations\end{tabular}} \\ \hline\hline

 \begin{tabular}[c]{@{}c@{}}AI models, Slices,\\Policies\end{tabular} & \begin{tabular}[c]{@{}c@{}} Non-RT \\($>$1~sec)\end{tabular}  & $>$1000  & \begin{tabular}[c]{@{}c@{}} KPIs at\\ HCI-level\end{tabular} & \begin{tabular}[c]{@{}c@{}} Orchestrating many\\ N-RT RICs and\\ CUs/DUs/RUs \end{tabular}  \\ \hline

\rowcolor{Gray2} \begin{tabular}[c]{@{}c@{}}User Control \& \\Session Management\end{tabular} & \begin{tabular}[c]{@{}c@{}} Near-RT \\($>$10-1000~msec)\end{tabular}  & $>$100  & \begin{tabular}[c]{@{}c@{}} KPIs at CU\\-level, e.g., no\\ of users sessions,\\ PDCP traffic \end{tabular} & \begin{tabular}[c]{@{}c@{}} Processing multiple\\ streams from \\numerous CUs \\\& sessions \end{tabular}  \\ \hline

 \begin{tabular}[c]{@{}c@{}}MAC-level management\\e.g., RAN slicing,\\scheduling policies \end{tabular} & \begin{tabular}[c]{@{}c@{}} Near-RT \\($>$10-1000~msec)\end{tabular}  & $>$100  & \begin{tabular}[c]{@{}c@{}} KPIs at MAC-\\level, e.g., buffering,\\PRB utilization\end{tabular} & \begin{tabular}[c]{@{}c@{}} Mission-critical \\operations \& decisions\\ for several DUs/UEs \end{tabular}  \\ \hline

\rowcolor{Gray2} \begin{tabular}[c]{@{}c@{}}NR management\\e.g., beamforming, \\scheduling resources \end{tabular} & \begin{tabular}[c]{@{}c@{}} RT \\($<$10~msec)\end{tabular}  & 10-100  & \begin{tabular}[c]{@{}c@{}} KPIs at PHY-\\level, e.g., CSI\\ estimation\end{tabular} & \begin{tabular}[c]{@{}c@{}} Support and \\deployment of AI\\ models at DUs \end{tabular}  \\ \hline

\begin{tabular}[c]{@{}c@{}}Device management\\e.g., DL/UL management,\\modulation, blockage\\detection, ranging \end{tabular} & \begin{tabular}[c]{@{}c@{}} RT \\($<$1~msec)\end{tabular}  & 1  & \begin{tabular}[c]{@{}c@{}} KPIs at Device-\\level, e.g., I/Q \\samples, RSSI, ToA\end{tabular} & \begin{tabular}[c]{@{}c@{}} Requirement of\\ standardization\\ at DU-and/or\\RU-level \end{tabular}  \\ \hline

\end{tabular}

}
\label{table:ORANarchitectures}
\vspace{-10pt}
\end{table}
}
\egroup

\fakepar{Fronthaul Interface: CPRI \& eCPRI.}
The fronthaul interface in 4G architecture, called as common public radio interface (CPRI) connects the DU signal processing to the RU. In 4G, the DU-RU split was designed such that the daterate of a maximum of 600~Mbit/s of fronthaul interface was needed for 10~MHz of radio channel bandwidth. 
On the other hand, in 5G systems, with a radio bandwidth of 100~MHz and above and massive multiple-input multiple-output (mMIMO) antennas, fronthaul bandwidth of hundreds of Gbit/s is required. To support this, the 3GPP has introduced an enhanced-CPRI (eCPRI) interface while to reduce the overall fronthaul bandwidth an increased number of functional splits are possible. In fact, 3GPP defines eight functional splits, where parts of the DU’s functions can be combined with the RU. Which split to employ leads to several ramifications, such as 1) additional DU electronics and energy-consumption costs if combined with the RU, 2) the distance (affecting latency) between the two elements, and 3) high-performance computing needs to process signals at the DU. 
Importantly, eCPRI still contains vendor-specific elements. 

\textit{Figure~\ref{fig:I5.RAN} summarises the various stages of evolution of RAN architectures, from traditional to open and disaggregated RAN.} 

\begin{figure*}[!t]
\centering
\includegraphics[width=0.8\linewidth]{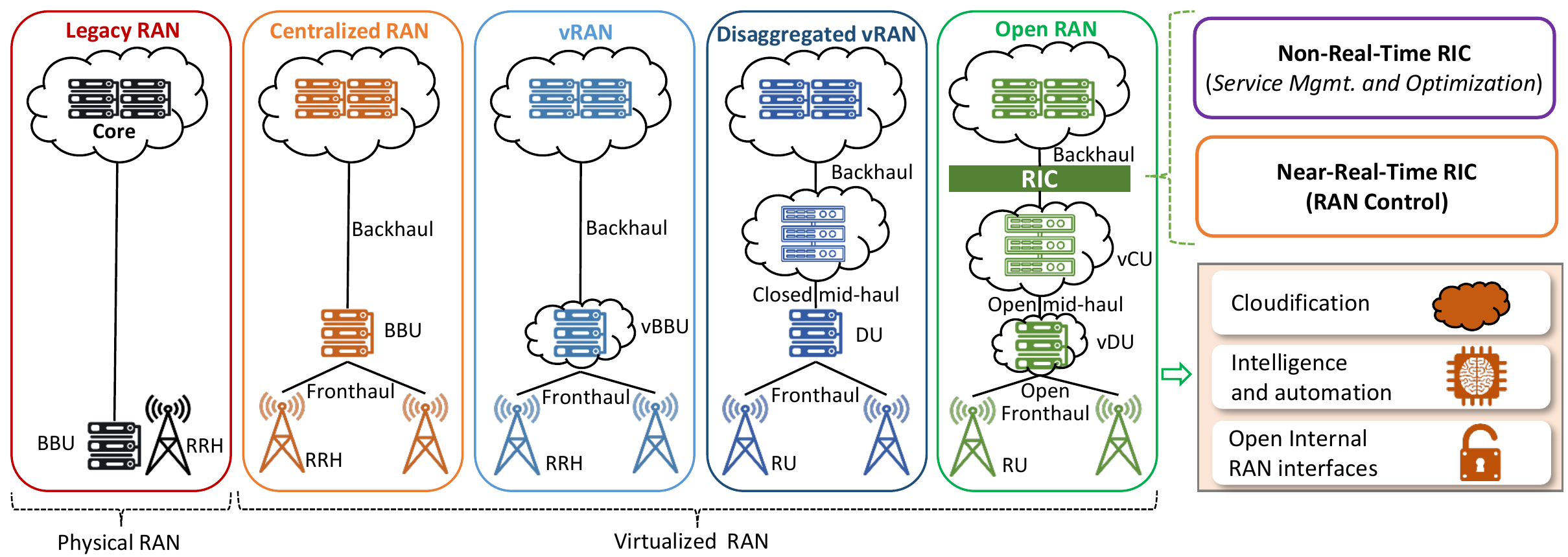}
	\caption{RAN transformation: Evolution from traditional RAN to open and disaggregated RAN architecture.}
	\label{fig:I5.RAN}
\vspace{-10pt}	
\end{figure*}

\fakepar{Disaggregation and Intelligence in O-RAN.} 
The O-RAN Alliance~\cite{O-RANAlliance}, which is a consortium of network operators created in 2018, is working to realize NextG networks by developing specifications, reference architecture, and open interfaces to intelligently manage/control multi-vendor infrastructures to delivering critical, high-performance services. To achieve this goal, O-RAN proposes an architectural changes using two core principles. \textit{First}, O-RAN embraces and promotes the 3GPP functional splits, where a typical Base Station's (BS) functionalities are virtualized as network functions and distributed across various network nodes, i.e., CU, DU, and RU. This facilitates the instantiation and execution of diverse networking processes at different network points. Therefore, one of O-RAN’s work groups is focused on creating an open, interoperable fronthaul interface to enable the integration of DUs and RUs from different vendors; thus, enabling a true multi-vendor ecosystem.
\textit{Second}, which is likely to be an crucial architectural component of 6G systems, is the introduction of the RAN Intelligent Controller (RIC) that allows operators to deploy customizable control plane functions. RIC comes in two variants: 
\begin{itemize}[leftmargin=*]
    \item \textbf{\textit{Non-real time RIC (non-RT RIC)}} provides an open platform for policy-based RAN control using \textit{rApps}, including mechanisms for enabling artificial intelligence and machine learning for more efficient network management and orchestration. The non-RT RIC is expected to perform operations with higher than one second time granularity.
    \item \textbf{\textit{Near-real time RIC (near-RT RIC)}} is aimed at abstracting the control processes, typically embedded in BS, and executing those processes as third-party applications \textit{xApps} in the cloud. The near-RT RIC handles procedures at timescales of tens of milliseconds and implement intelligence through data-driven control loops.
\end{itemize}
In both variants, RIC facilitates RAN optimization through closed-control (action to feedback) loops, between RAN components and their controllers. O-RAN envisions different loops operating at timescales that range from 1~ms (e.g., for real-time control of transmission schemes) to thousands of milliseconds (e.g., for network slicing, traffic forecasting). \textit{Table~\ref{table:ORANarchitectures} summarizes the timescale-related information and challenges in RICs.}



Although the O-RAN architecture is gaining momentum among operators, vendors and researchers, the challenges related to its implementation for true  NextG networks (i.e., built as data-driven, open, programmable, and virtualized) are largely to be handled. In the literature, there are various ongoing efforts towards answering such important concerns, including~\cite{ORAN_Learning_NextG} (i) the functionalities/parameters to be controlled by each network entity, (ii) distribution of network intelligence, (iii) validation and training of data-driven control loop solutions, and (iv) provision of access to data and analytics from the RAN to the AI agents while minimizing the overhead.

Towards these concerns a few notable studies are \cite{ORAN_Learning_NextG, niknam2020_ORAN, Dynamic_CU_DU_ORAN, Elastic_ORAN} (for an extensive review see~\cite{polese2022understanding}). In~\cite{ORAN_Learning_NextG, niknam2020_ORAN}, the potential and challenges of data-driven optimization approaches to network control over different timescales are discussed for disaggregated O-RAN architecture. In \cite{ORAN_Learning_NextG}, the authors optimized the scheduling policies of co-existing network slices (NS). In this respect, while leveraging O-RAN interfaces to collect edge data, the authors demonstrated closed-loop integration of real-time analytics and control through DRL agenets for RAN control through xApps at near-RT RIC. In~\cite{niknam2020_ORAN}, an intelligent radio resource management scheme with LSTM-based traffic congestion handling is demonstrated on a real-world dataset. The LSTM model, trained at non-RT RIC, is used for inference at near-RT RIC for cell splitting of congested cells to improve the related KPIs. In \cite{Elastic_ORAN}, the authors developed O-RAN control and management layer intelligent functions (i.e., near-RT RIC and non-RT RIC) to solve the O-RAN slicing problem for the industrial IoT. In particular, for \emph{O-RAN control}, this study performs the radio connectivity management in the near-RT RIC, where the stable IIoT and small base stations (SBSs) associations distributedly using the matching game are determined. Using the outcomes of the matching game, the IIoT slice requests’ demand are measured to generate IIoT service-specific slice templates at non-RT RIC while considering the slicing constraints at the \emph{QoS management module} in the near-RT RIC. Meanwhile, in \cite{Dynamic_CU_DU_ORAN}, the authors studied how to exploit the O-RAN architecture to improve an RL-based resource block allocation model’s performance in a dynamic network while dynamically relocating an NF at different layers of O-RAN (either at DU or CU) to improve its performance.

\subsection{Network Intelligence in 6G Networks}
\label{subsec:NI_6G}
The vision of 6G networks is set to meeting the requirements of ever-expanding multitude of service classes and their disparate time-varying traffic profiles. This vision is leading towards the need and ability to instantly orchestrate end-to-end physical resource and virtual network functions (VNFs) across different network domains through multi-tenancy transparent communication network infrastructure, which grows the management complexity of already complex network architectures beyond the grasp of human-in-the-loop approaches. Therefore, the success of 6G networks will greatly depend on the Network intelligence (NI), and its quality, to fully automate the network operation and management. Herein, AI models, deployed at network controllers and network orchestrators will be cornerstone in NI design in solving complex and dynamic problems of massive and heterogeneous network traffic.  Specifically, AI-driven NI will be deployed in 6G networks for human-intervention free automated decision making for adapting network resource and functions to the use case demands and requirements. In end-to-end NI-native 6G architecture, AI models needs to be tailored according to the network functional needs at each specific network level. Despite the multiple examples of NI-driven handling of network operation, they are limited in receiving native integration support from the network architecture.
In~\cite{NI_ORAN}, the challenges and directions towards integration of NI-driven functionalities for network capacity and forecasting 6G architecture are presented. The two key identified challenges for incorporating NI in 6G networks are:

\begin{itemize}[leftmargin=*]
    \item \textbf{End-to-end NI-native architecture:} The NextG network architecture should facilitate hosting and integration of NI instances explicitly at NF levels and their coordination across network domains. Meanwhile, the current standardization trends towards updating the existing network design must reflect that.  
    
    \item \textbf{Customized AI for NI:} AI models should facilitate network automation and management, and thereby adapted (w.r.t. latency guarantees training and computational complexity) for NI-specific design aspects according to the needs of network functionalities.  
\end{itemize}

\subsubsection{NI Role in 6G Systems}
To support diverse services, tenants, and slices, NI will be fundamental for the optimal operation of this softwarized, cloudified and split network architecture and managing the multi-level network functions and associated resources.  Considering the variety of network management tasks and the entailing functions, each orchestrator/controller will need to run multiple NI instances. Meanwhile, an NI instance must first  swiftly detect or predict the new requests or their fluctuations. Then, it should react by automously instantiating, relocating, or re-configuring network resources/functions, such that a) traffic demands are met with desired QoS requirements, b) reuse maximization of infrastructure and resources (across multiple tenants and resources), and c) ultimately  completely eliminating human intervention to achieve zero-touch network and service management vision~\cite{Zero_Touch_Talib}. These aspects are all highly critical, and the success of 6G systems will greatly depend on the quality and appropriate integration of NI solutions in the complete network.
The NI across network can be divided at three scales based on the timescale of the associated management and automation, as described below:
\begin{itemize}[leftmargin=*]
    \item \textbf{NI at Orchestrators:} The NI integrated at E2E and edge orchestrators has a broader network scope to manage functions/resources and take decisions, with an effect on large sets of RAN entities. Since, at orchestrator level, the network statistics their dynamics vary relatively slowly, automation and management functions are performed at timescales of minutes to hours. Consequently, NI decisions must endure large time windows.
    
    \item \textbf{NI at non-RT RIC:} As the non-RT controllers must handle reduced time-criticality functionalities of a limited number of radio sites, the associated NI will operate at a local scope, with decisions in the order of tens of seconds.
    
    \item \textbf{NI at near-RT RIC:} the NI deployed within near-RT RIC will ensure fast-timescale functionalities (as expected within O-RAN~\cite{O-RANAlliance} or FOG-05~\cite{Eclipse_fog05} frameworks), such as radio scheduling, interference management, which typically require data-driven decision-making at timescales of milliseconds or even lower. The other proposal is to integrate NI within NFs themselves, e.g., to enable functionalities operating beyond-edge micro-domain, for instance, realizing programmable radio environments (e.g., using ) with localized, fast decision making.
\end{itemize}

\begin{figure*}[!t]
\centering
\includegraphics[width=0.8\linewidth]{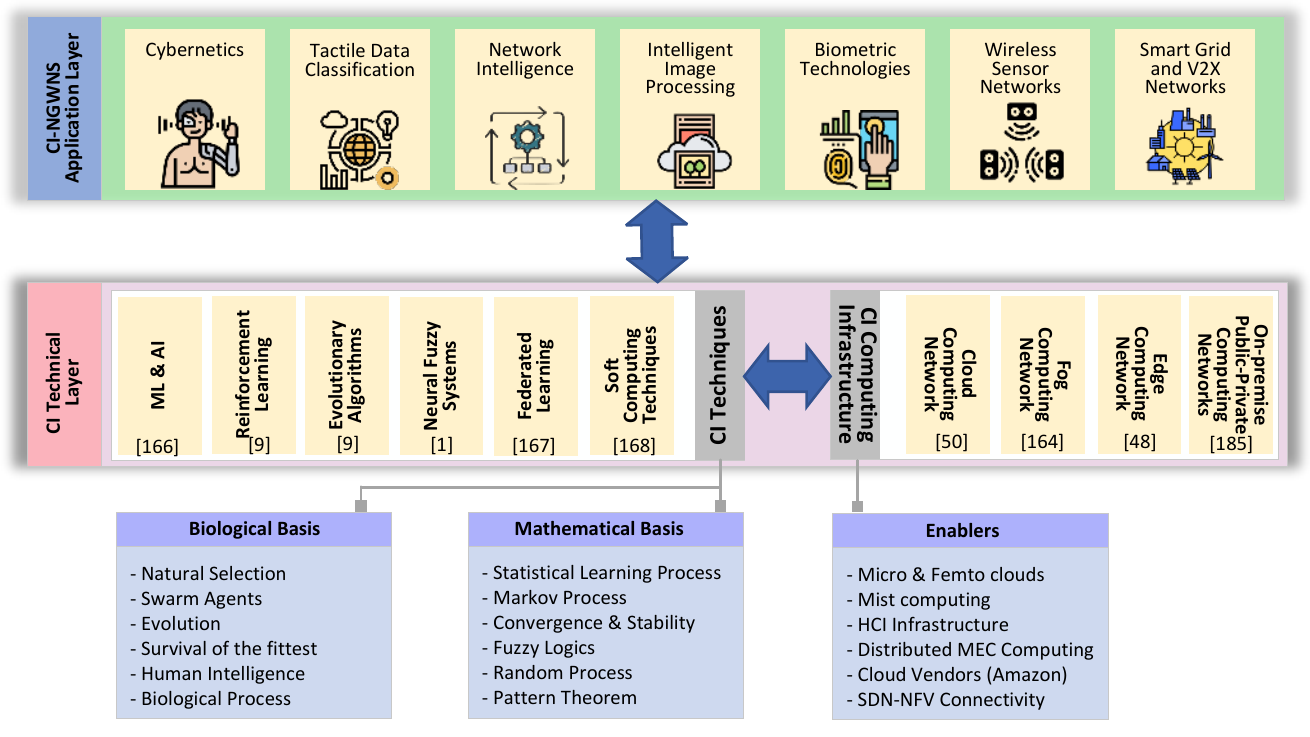}
	\caption{CI techniques and CFE infrastructures, theoretical basis and enablers for CI-NGWNs applications.}
	\label{fig:CItechniquesandCFE}
 	\vspace{-10pt}
\end{figure*}

\subsubsection{NI Integration Challenges}
The current trends in NI for NextG network orchestration, promoted by major standardization bodies, are based on the closed-loop AI~\cite{5GAmerica_WhitePaper}; that is, the NI instances at the orchestrators or controllers work in closed-control loops. These loops abide modern AI’s learning principles; recording the context of management decisions, continuous monitoring for collecting observations on the decision quality, and finally using feedback to improve future decisions. The close-loop models allow NI to apprehend to learn over time to automate optimal decision making towards the expected targets. Currently, the prevailing vision for closed-loop NI expects instances located at orchestrators or controllers in the control plane that interact with NFs deployed in the data plane~\cite{NI_ORAN}. This vision results into two major limitations, as follows. 

\begin{enumerate}[leftmargin=*]
\item \textit{Strictly centralized closed-loop models hinder the effectiveness of network intelligence;} there is a risk in data-to-decision communication delay for critical management scenarios. To be more effective, many network functionalities must operate at timescales that are not compatible with the current lengthy data gathering procedures at NFs, data processing at control plane, and final restitution of actions to be executed in the data plane. As a results, the feedback loop between the control and data planes hinders NI to continuous system monitoring.

\item \textit{The interaction among NI instances at different levels is not promoted in the current architectural trends}; thus, the reciprocal mutual impact of their individual actions cannot be captured~\cite{NI_ORAN}. Therefore, to enable optimal NI operation at each level, cooperation between NI instances is vital to allow for global optimality and convergence of the zero-touch network management process. Therefore, To tackle these limitations, architectural design of 6G networks must necessarily take into account the NI-native choices.
\end{enumerate}

}

{\color{black}
\section{Computational Intelligence for Industry 5.0}
\label{sec:computationalintelligenceI5.0}



Modern technologies rely heavily on CI techniques and CFE computing platform not only to automate the process in a dynamic manner but also to help achieve the desired efficiency. In this regard, we reviewed some of the CI techniques and related work for different targeted domains in Sec.~\ref{sec:CITechniques}, which is followed by the Sec.~\ref{sec:CFETechniques} that focuses discussion on related work on critical CFE computing. Similarly, onwards Sec.~\ref{sec:NexusHuman},~\ref{sec:NexusResiliency}~and~\ref{sec:NexusSustainability} explores the integrated capabilities of CI techniques and CFE computing for three pillars of Industry 5.0 vision.

\subsection{Computational Intelligence}
\label{sec:CITechniques}
 \textit{Along with CFE, computational intelligence is considered one of the key enabling technologies for Industry 5.0.} The CI methods are quite a wide area, thus includes a wide variety of methods including but not limited to neural fuzzy systems, soft computing techniques, evolutionary algorithms, reinforcement learning, unsupervised learning, supervised learning, self-supervised learning, federated learning, and more~(c.f.~Fig.~\ref{fig:CItechniquesandCFE}). The growth, success rate, and evolution of intelligent applications has been drastic in the past decade, thanks to the deep learning architectures and computing advancements. Therefore, the fifth iteration of the revolution takes advantage of the characteristics of CI to achieve sustainable, efficient, and effective industrial processes.

\subsubsection{Supervised Learning}
The supervised learning is the most commonly used learning method in the field of CI~\cite{sharma2021comprehensive}. It requires the set of associated labels with the data to learn the patterns or trends to perform tasks like prediction, forecasting, classification, detection, segmentation, and others.

\subsubsection{Unsupervised Learning}
Unsupervised learning methods are those that do not have a reference to start with. Algorithms in this category identify patterns from acquired data to make sense of them~\cite{li2020systematic}. The learned patterns can then be used for various applications such as data clustering, market segmentation, and other related applications. 

\subsubsection{Self-Supervised Learning}
The self-supervised learning is a new paradigm and can be considered as an intermediate algorithm between unsupervised and supervised learning, respectively. The self-supervised methods assumes that the data is not of the best quality in terms of labels (opposed to the supervised learning), thus it starts with the training of pseudo-labels associated with data and then can carry on to employ supervised or unsupervised learning for improving the desired task~\cite{liu2021self}. Self-supervised learning is relatively new and has been proven to be effective specifically in audio processing, as is evident from its usage for speech recognition on Meta (formerly Facebook).

\subsubsection{Evolutionary Algorithms}
Evolutionary algorithms encompass optimization and meta-heuristic algorithms to solve the task at hand. These algorithms are certainly good when there is no assumption about the optimized values to begin with. Although, the evolutionary algorithms can do the job but at most times they are computationally complex, therefore, they can either be used with hard constraints to lower the computational complexity or with prior approximations that are near to the optimal values~\cite{CI1}. There are various and popular methods that can be categorized as evolutionary algorithms, such as particle swarm optimization, cuckoo search, artificial bee colony, ant colony optimization, gray-wolf optimization, Harris-hawk optimization, sailfish optimization, firefly optimization, harmony search, and others.  

\subsubsection{Neural Fuzzy Systems}
Neural Fuzzy systems use fuzzy theory as its underlying technique to represent the probabilistic values between true and false states in conjunction with neural networks~\cite{CI-NGWNs1}. Neural Fuzzy theory has been extensively applied over the years in the field of home appliance control, power system control, and control systems, in general. 

\subsubsection{Reinforcement Learning}
Reinforcement learning, also known as learning-by-experience, trains an agent by taking actions with perceived inputs and interprets the environment to maximize cumulative rewards~\cite{CI1}. The method heavily relies on the notion of exploration versus exploitation of the prior and acquired knowledge to achieve the desired task. It has gained a lot of interest from the research community in the field of robotics, autonomous driving, resource optimization, and financial technology (FinTech) applications. 

\subsubsection{Federated Learning}
Federated learning on the other hand is also a relatively new technique, but is proposed to address privacy and security concerns while improving the task at hand~\cite{khan2021federated}. The method was first proposed by Google, which does not send the data to the server, in contrast to the other learning algorithms. Federated learning requires only the trained model, which is collected by the number of users, aggregated on the server, and then sent to the user for model update. Federated learning has become very popular due to its secure characteristics and has been extensively applied in the field of communication systems. 

\subsubsection{Soft Computing Techniques}
The last in this discussion are the soft computing techniques that are quite analogous to numerical computation. It kind of combines the characteristics of the aforementioned learning algorithm, such as supervised/unsupervised learning, fuzzy logic, evolutionary techniques, and probabilistic reasoning, to approximate the solution rather than provide an optimized one~\cite{soltanali2021comparative,sharma2021comprehensive}. These techniques are imprecise, but can be effectively tolerant to uncertainty. For real world applications that can cater to higher risk for robust solutions at reasonable cost, the soft computing techniques are ideal and natural fit. Existing studies suggest that without the use of computational intelligence techniques, Industry 5.0 could not be realized in an effective way~\cite{fraga_lamas2021, Maddikunta2021}. 
\subsubsection{CI for Holistic Industry 5.0 Framework}
Some studies have also emphasized on the use of CI techniques for solving sustainability and resiliency concerning Industry 5.0 \cite{fraga_lamas2021, Ozdemir2018,Du2021,Thakur2021} while others related the use of CI to achieve or realize specific characteristics of Industry 5.0 such as human-machine collaboration \cite{Demir2019, Longo2020, Nahavandi2019}, pandemic situations (COVID-19) \cite{Javaid2020covid}, education \cite{Rachmawati2021}, medical and health \cite{Javaid2019}, robotics \cite{Han2017}, Blockchain \cite{Rupa2021}, Digital Society \cite{Kent2019, Martynov2019}, Society 5.0 \cite{Elim2020}, Energy \cite{Carayannis2021, ElFar2021}, UAVs \cite{Jain2021}, Mass Customization \cite{ElFar2021}, and Emotional Intelligence \cite{Chin2021}. A summary of the applications targeted by the summarized studies in Industry 5.0 is shown in Fig.~\ref{fig:I5.0applicationdomains}. Below we reviewed some of the works carried out towards exploring the CI in targeted application domains concerning Industry 5.0. 
\begin{figure}
\centering
\includegraphics[width=\linewidth]{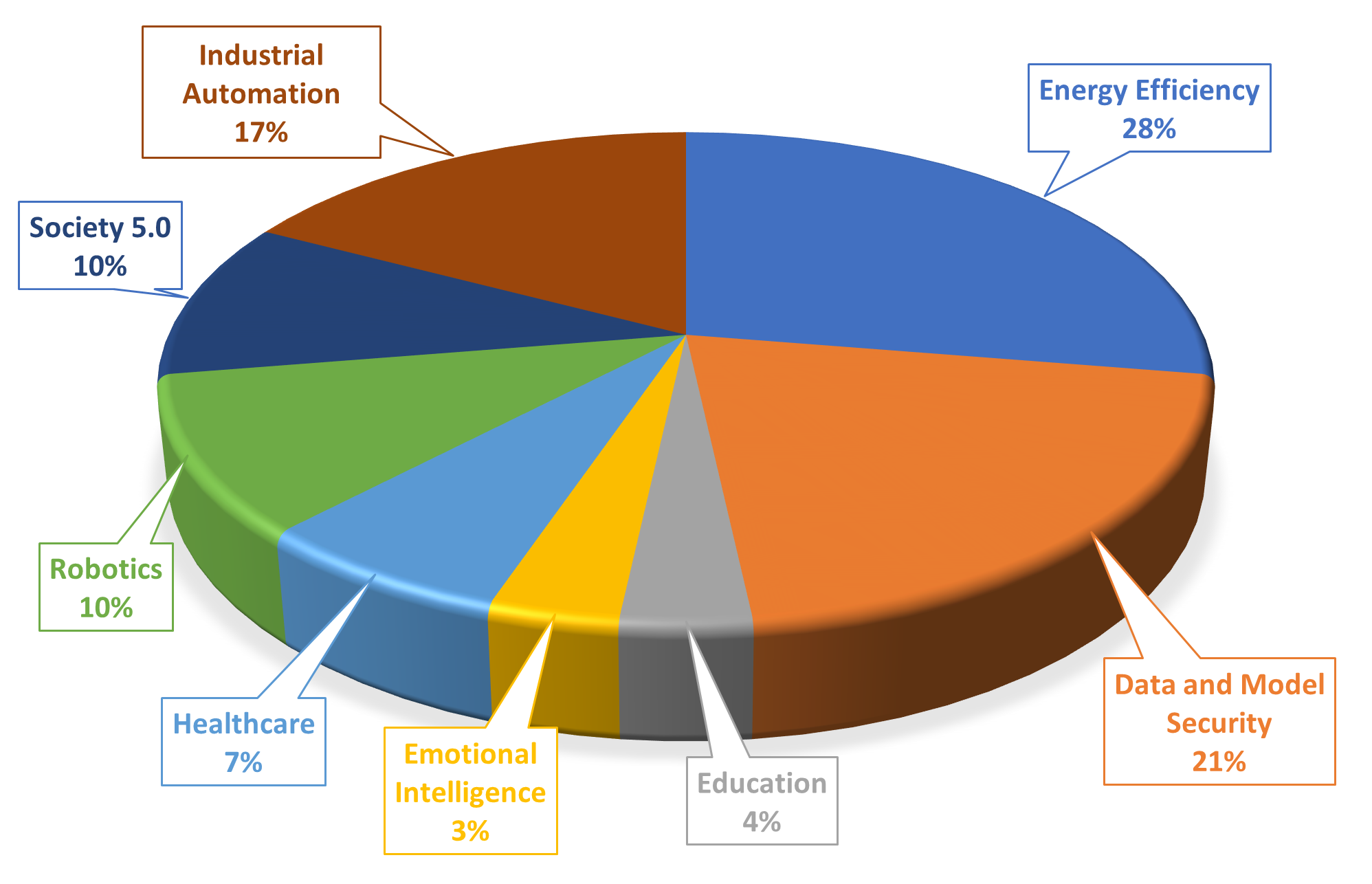}
	\caption{Distribution of the targeted application domains considered by the summarized studies using CI techniques. }
	\label{fig:I5.0applicationdomains}
\vspace{-10pt}	
\end{figure}

\fakepar{Industrial Automation and FoFs.}
Jain et al. \cite{Jain2021} focused mainly on an application of Industry 5.0, that is, allowing communication through UAVs in future industries. The UAVs are assumed to be equipped with CI techniques to recognize, classify, and secure the data contents, specifically the image modality. Han et al. \cite{Han2017} emphasized human-robot collaboration for increased productivity in industries associated with the fifth iteration. However, the study deviated more towards the design of robots that can exhibit not only autonomous intelligence, but also integration of human supervision. Their study proposed the development of an intelligent swarm robotic system that can be used as an intelligent automation system, as well as for functions controlling humanoid organ. To achieve the said task, the study heavily used CI techniques to control the robotic system with human commands and in an autonomous way, accordingly.

\fakepar{Robotics.}
In \cite{Nahavandi2019}, authors emphasized the human-machine collaboration for the realization of Industry 5.0. The study considers sensor data interoperability, digital twins, virtual training, intelligent autonomous systems, and machine cognition to be the driving forces for human-machine collaboration. The study suggested that CI techniques are analogous to threads that connect the aforementioned driving forces in the context of Industry 5.0. 
Similarly, Maddikunta et al. \cite{Maddikunta2021} suggested that CI is one of the key enabling technologies for the realization of Industry 5.0. They suggested that CI can be used to bring sustainability and automation in applications such as smart additive manufacturing, predictive maintenance, hypercustomization, and cyberphysical cognitive systems. Furthermore, they also suggested that the CI plays a key role in realizing cobots, which is one of the key characteristics of Industry 5.0. On the other hand, Javaid and Haleem \cite{Javaid2020} emphasized on the use of CI for realizing Industry 5.0, specifically in the field of manufacturing. The study listed 17 critical components need to be undertaken for adoption of Industry 5.0 in manufacturing. Amongst them, cobots, emergent artificial intelligence, multiagent systems and technologies, complex adaptive systems, and virtual reality heavily rely on the CI for effective realization. Although components, holography, 4D printing, 5D printing, smart manufacturing, Internet of Everything, and Big data are complimentary domains that can be used with CI for various applications.


\fakepar{Healthcare.}
The authors in \cite{Javaid2019,Javaid2020covid} proposed the use of Industry 5.0 for healthcare in the context of the current pandemic (COVID-19), as well as in general. These studies explore various paradigms, issues, and challenges associated with the realization of Industry 5.0 in the healthcare industry to facilitate physicians, practitioners, and patients. Among all enabling technologies, both studies focus on the use of CI to make the smarter, accurate, sustainable, and resilient.

\fakepar{Digital Society.}
The studies \cite{Kent2019},~\cite{Martynov2019},~and~\cite{Elim2020} considered the aspect of digitization and digital society in the industry 5.0 ecosystem. The former two studies evaluated the technologies that are essential for the digitization of Industry 5.0 components, while the latter represented the description of Industry 5.0 as a complex mathematical problem. The study suggested that giving it a formal description would not only help the enterprises to transition from the fourth to the fifth industrial revolution, but also help in measuring the cost, thus, by extension, making the business information support more efficient. Both of the studies highly emphasize on the use of CI as an underlying technology to achieve the digitization process.

\subsection{Cloud/Edge-native Computing Network Infrastructure}
\label{sec:CFETechniques}
The availability of HCI computing infrastructure and communication resources at the core and edge of NIBs (c.f.~Fig.~\ref{fig:softwarizedarchitecture}), i.e., cloud-fog-edge computing network infrastructure, can provide the necessary communication and computation platform for enabling CI techniques in different Industry 5.0 applications. Numerous critical enablers for CFE computing is shown in Fig.~\ref{fig:CItechniquesandCFE}.

\subsubsection{Cloud-Fog-Edge (CFE) Computing}
In order to achieve green communication and sustainability within the Industry 5.0 ecosystem, key components,e.g., IoT, CFE computing in cloud-/edge-native network architecture, and AI are the key enabling technologies \cite{fraga_lamas2021}. 
The amalgamation of IoT and cloud-edge-fog computing for sustainable communication is referred to as \textit{Green IoT (G-IoT) system}, which aims to minimize the energy consumption in the communication system. In general, the main task of the IoT system is to collect data from sensors and send the response to the IoT actuators. The integration of cloud in the general IoT system acts as a liaison for managing the implicit services and interacting with remote users as well as third party services, accordingly. Since the cloud-based IoT systems are centralized, they suffer from long response latency and bottlenecks \cite{Du2021}. To overcome the aforementioned issues, mist, fog, and edge computing were explored for the IoT ecosystem, which helps to offload some communication tasks from the cloud in order to improve latency response and node requests \cite{fraga_lamas2021}. Fog computing gateways and cloudlets were added between IoT devices and the cloud to improve response time. Mist computing devices can perform complex tasks locally as well in collaboration with other IoT nodes, therefore, reducing the communication overhead with the cloud \cite{Markakis2017}. Cloudlets use high-end PCs to perform computing-intensive services and real-time rendering in the local network \cite{fraga_lamas2021}. The fog computing gateways can provide QoS-aware, low-latency, and physically distributed services, accordingly \cite{Markakis2017}. The use of these devices in communication architectures help to achieve greener and sustainable solutions.

\fakepar{CFE-based Smart Circular Economy.}
Fraga-Lamas et al. \cite{fraga_lamas2021} discussed the sustainability issue in the context of smart circular economy and Industry 5.0. The paper focuses mainly on the reduction of carbon footprint issue concerning recycling, operating, and manufacturing processes. The study considers CFE as one of the enabling technologies to achieve sustainability in the Industry 5.0 ecosystem. Furthermore, the study takes into account edge nodes in integrated with mist computing for analyzing energy consumption and its effect on carbon footprint. Furthermore, the impact of sustainability on different countries has also been discussed, accordingly.

\subsubsection{Three-level CFE-based IoT Architecture}

Recently in an Annual Automation Fair, Rockwell Automation (world’s largest company for industrial automation and digital transformation) have also listed cloud-fog-edge architectures as one of the driving forces for Industry 5.0\footnote{http://www.koreaherald.com/common/newsprint.php?ud=20211202000559}. For the sake of generality, we can assume a three-level cloud-fog-edge-based IoT architecture for Industry 5.0. The first layer comprises edge devices or IoT sensors used in the context of Industry 5.0, including high-end vehicles, smartphones, smartwatches, smart pendants, wearable devices, and sensor nodes. The second layer consists of fog gateway, base stations, cloudlets, road-side units, routers, edge servers, and proxy servers to form edge-fog computing layer. This layer provides the essential computation, communication, and storage capabilities for the nodes in the first layer. The last is the cloud computing layer that consists of cloud server(s) and datacenter(s), respectively. As Industry 5.0 is still in its infancy, some studies have been published to cover a few or a single aspect of CFE architectures in the context of Industry 5.0. Subsequently, we consolidate a summary of those works in this subsection.

\fakepar{Edge Computing as Supportive Technology.}
Maddikunta et al. \cite{Maddikunta2021} reviewed the enabling technologies for Industry 5.0 and considered edge computing as one of the supportive technologies. The integration of edge computing was discussed in the context of latency reduction rather than a sustainable solution. According to their review, Industry 5.0 needs to be faster in response time to achieve industrial needs, therefore, edge computing techniques are considered a necessary building block. Similarly, Özdemir and Hekim~\cite{Ozdemir2018} were one of the earliest ones to discuss the implications and diversity of Industry 5.0 through a scientific study. Although the study discusses mostly regarding the technology policy and adaptation, it also emphasizes greatly on the use of edge devices to maintain a symmetry in industrial automation, which is one of the reasons for transition to Industry 5.0. It also suggests that the use of edge-computing and devices is not only limited to implement automated networks, but also to generate pathways and linkages with the human factors in case of hyperconnectivity failure. Since our study follows the model proposed by EC~\cite{EC2021}, we consider CFE as one of the building blocks to achieve sustainability concerning Industry 5.0 systems.
\begin{figure*}[!t]
\centering
\includegraphics[width=0.8\linewidth]{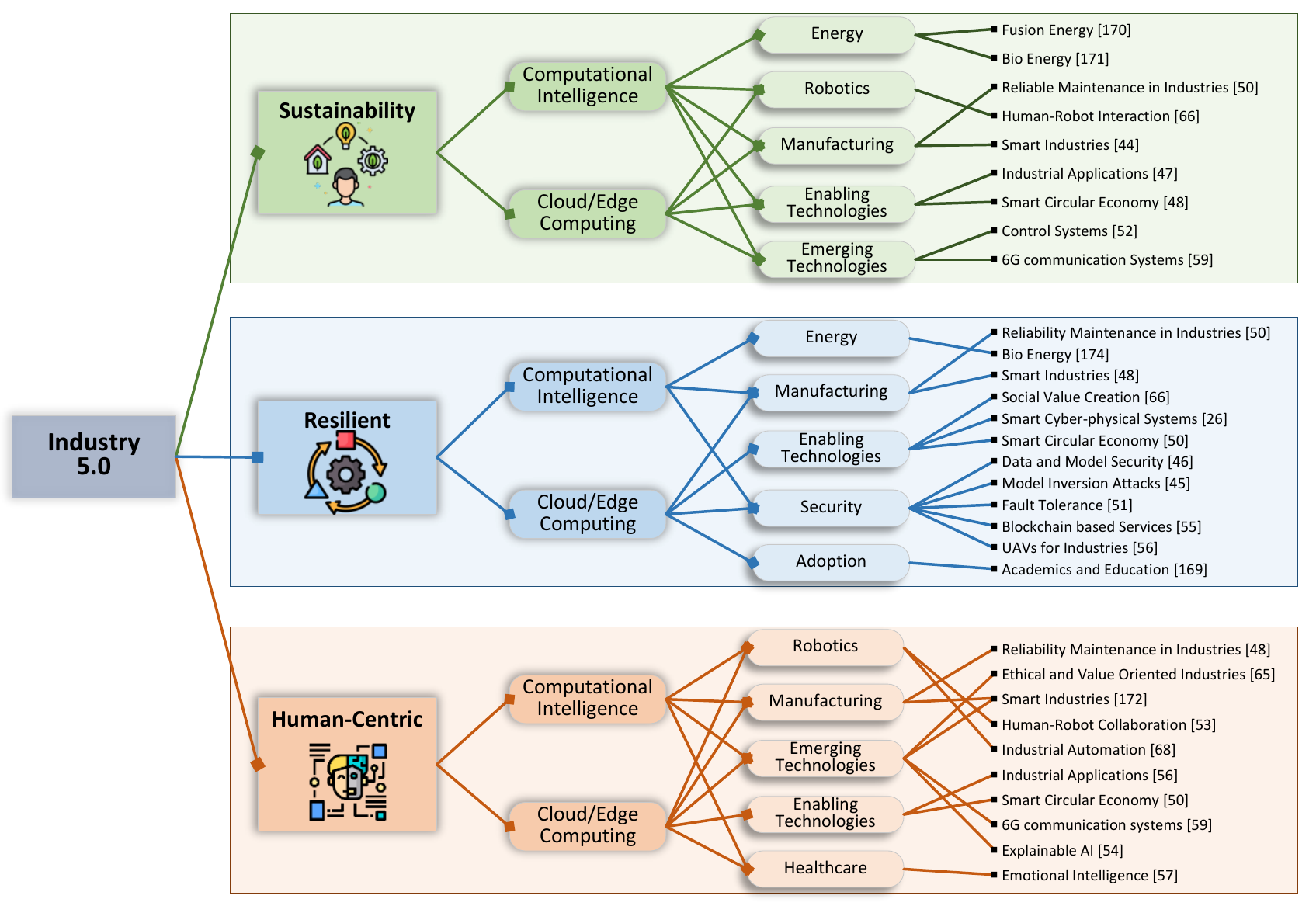}
	\caption{Taxonomy and applications of the reviewed studies that uses the combination of both computational intelligence and cloud-fog-edge network computing resources concerning three components of Industry 5.0 vision }
	\label{fig:I5.0Taxonomy}
\vspace{-10pt}	
\end{figure*}

\fakepar{Industrial IoT at the Edge.}
Thakur and Sehgal \cite{Thakur2021} proposed IIoT and edge-computing based architecture that caters to several industrial processes and applications for heterogeneous smart cyber-physical systems. The processes include hydraulic, pneumatic, and electrical, while the application comprises dynamic voltage and frequency scaling, and voltage frequency islands. The authors emphasized that their edge-computing based heterogeneous architectures is in compliance with Industry 5.0 standards as it requires minimal human intervention. Similarly, Du et al. \cite{Du2021} proposed the use of the industrial edge computing framework for fault diagnosis in the context of Industry 5.0. The three layered architecture consist of edge devices, edge/fog/mist servers, and cloud servers. The study through analysis showed that the edge computing-based framework not only reduces the computation and communication bottleneck, but also reduces the latency and helps shift the industrial process paradigm from centralized to distributed, respectively.



\subsection{Nexus of CI and Cloud-Fog-Edge Computing for Human-centric Solutions}
\label{sec:NexusHuman}
As one of the driving forces for Industry 5.0 is mass customization and personalization, it attracts human eccentricity into the mix that prioritizes human interest and needs in the production process. \textit{Industry 5.0 is also considered to be a society-centric approach, suggesting that the technology would adapt to the diversity and needs of workers.} For example, prioritizing the mental and physical well-being of workers can be performed to preserve the fundamental rights of workers while focusing on the digital privacy of acquired data.\\
\textit{One of the ways of well-being, relevancy, and human-machine collaboration can be achieved is by using the combination of CFE computing and CI.} The edge devices can help in acquiring the data from human workers in a ubiquitous way, while the CI can analysis the data and activate relevant services concerning the desired issue or application. Furthermore, Edge AI (a mergence of CI and CFE computing) can also help to improve human-machine collaboration by understanding action perception in a pervasive manner. Some studies have highlighted the human-centricity issues in their studies (c.f.~\ref{fig:I5.0Taxonomy}), while others focused on a particular scenario and proposed a way to realize it in current industrial environment.

\subsubsection{Enabling Technologies}
Maddikunta et al. \cite{Maddikunta2021} highlighted the term hyper customization that is essential to Industry 5.0. The study also suggests that CI and data from edge devices, such as, specific content, service, or product details, are crucial to move towards the hyper customization process. Furthermore, the study also highlights the specific preferences of the user that can be incorporated in the manufacturing process by utilizing user preference data from edge devices and decision/analysis using CI. The study reveals that the worker is a crucial aspect for achieving personalization, as the manufacturing process must transition to an agile supply chain and a process that requires human intervention to customize the product accordingly. Similarly, the study \cite{fraga_lamas2021} undertakes an industrial scenario where humans and machines work in collaboration. The study focuses on the safety of human workers in a industrial environment by leveraging Edge AI characteristics. For instance, continuous visual monitoring could be performed in order to detect close proximity of robots with humans. If the proximity condition is violated, an alarm could be generated and the machine could be shut down in a remote manner for protecting the human worker. 

\subsubsection{Emerging Technologies}
Bhat and Alqahtani \cite{Bhat2021} linked the vision and key performance indicators of 6G with Industry 5.0 to achieve hyper personalization. It should be noted that both the aforementioned technologies heavily rely on edge computing and CI. According to the study, the use of Internet-of-Everything (IoE), extended Reality, 4D imaging, brain-computer interface (BCI), haptic communication, holography, and telepresence would heavily contribute to personalized systems and automation. The importance of this study can be taken into account by recent developments in the aforementioned technologies from Facebook (Meta)\footnote{https://www.wired.com/story/facebook-haptic-gloves-vr/}.\\
Some studies \cite{Wang2021, Shahzadi2021, Nayak2020} have hypothesized that EdgeAI that drives 6G communication systems would contribute a lot to Industry 5.0 in terms of personalization, human-centric services, and real-time management systems. For instance, Wang et al. \cite{Wang2021}, proposed an AI-powered 6G layered architecture that uses EdgeAI and ensures the delivery of intelligent services to Industry 5.0. The architecture is heavily based on explainable AI (XAI) for not only understanding the logic behind activation of differentiated services, but also how they related to the end-users, which will be helpful for legal auditors and service providers to customize their products and equipment, accordingly. Shahzadi et al. \cite{Shahzadi2021} proposed the use of 6G services in manufacturers based on Industry 5.0 to not only leverage EdgeAI analytics, but also 6G communication systems to increase collaboration and cognitive communication between humans and machines to provide personalized user experience and improved service quality. Similarly, Nayak and Patgiri \cite{Nayak2020} corresponded to the transition of Industry 4.0 (Digitalization) to Industry 5.0 (Personalization) with Edge AI and 6G communication systems. The study highlights several issues concerning data heterogeneity, security, privacy, and mobility in the context of Industry 5.0. The study highlighted that the convergence of Industry 5.0 and 6G communication systems would result in an improved Industry 5.0 (Intelligentization) that could be drive a paradigm shift in the technological industry.

\subsubsection{Ethical and Value Oriented Industries}
Saeid Nahavandi \cite{Nahavandi2019} highlighted several aspects and issues concerning human-centricity in Industry 5.0. For instance, the study highlights that new jobs and positions such as chief robotic office (CRO) could be created in every industry as human-machine collaboration will have a high significance in Industry 5.0 ecosystem. Furthermore, the study also highlights the importance of virtual training to update the skills of the workforce in a simulated environment or remotely, that does not affect the real-time manufacturing process. The study suggests that the role of Edge AI in developing such kind of systems would be highly crucial. The study emphasized on machine cognition and sensing technologies to build a cobots-based systems for improving the quality and accuracy of assistive tasks in industrial environment.

\subsubsection{Emotional Intelligence}
The study in \cite{Chin2021} explored the concept of emotional intelligence in the context of Industry 5.0. The study considered emotional intelligence from two different perspectives. The first is the use of emotional intelligence as a soft skill to prepare the human workforce for machine collaboration in the Industry 5.0 ecosystem. The second aspect is to analyze the emotions and expressions of humans while working with robots to avoid and deal with stressful situation. The study suggested that CI with wearable devices can be used to recognize emotions and expressions to address both of the aforementioned aspects. Similarly, Longo et al. \cite{Longo2020} used the term "\textit{Age of Augmentation}" interchangeably with Industry 5.0. They proposed a value-sensitive design framework to elicit human values in the factory of the future concerning the human-machine symbiosis. Moreover, studies such as \cite{Demir2019, Chin2021} focused specifically on the human-robot systems and highlighted several issues concerning administrative, organizational, regulatory, legal, psychological, social, and ethical issues within Industry 5.0 ecosystem. The studies dive into the consequences of adopting human-robot collaboration and suggest some possible solutions to avoid mishaps in terms of physiological and mental well-being of workers.

\subsection{Nexus of CI and Cloud-Fog-Edge Computing for Resiliency}
\label{sec:NexusResiliency}
\textit{The importance of resilience was highlighted during the COVID-19 pandemic that challenged the economic resilience concerning industry perspective.} In its report, the European Commission \cite{EC2021, EC2022} highly emphasized the resilience to social, economic, and production stabilization in Industry 5.0. For example, the issue of industrial process recovery and manufacturing recovery has been discussed and touched on in an abstract manner in \cite{Maddikunta2021, fraga_lamas2021} while highlighting the use of CI and edge computing as its key enablers. 
In this regard, We consolidate the studies addressing resilience in Industry 5.0 below. 

\subsubsection{Energy}
The study \cite{ElFar2021} discusses resilience in Industry 5.0 in a very brief manner in the context of SDG goals proposed by the UN\footnote{https://www.undp.org/sustainable-development-goals}. This study corresponds mainly to resilience in Industry 5.0 to the 11th (SDG), that is, sustainable cities and communities, which are compliant with both reports of the European Commission \cite{EC2021, EC2022}, accordingly.


\subsubsection{Enabling Technologies}
In \cite{Sindhwani2022}, the authors considered the technology enablers for Industry 5.0 and carried out an extensive review for the evaluation and ranking of the enablers, accordingly. The study developed a criterion for inclusion and selection of technology enablers, followed by its evaluation via Pythagorean fuzzy Delphi method. The ranking of technology enablers was then carried out using a Pythagorean fuzzy analytical hierarchical process by calculating and assigning weights accordingly. The study also provided implications in terms of case studies and several prepositions to improve the Industry 5.0 adoption process.

\subsubsection{Security}
The following studies refer mainly to security and privacy concerns when addressing resilience in Industry 5.0. 
\begin{itemize}[leftmargin=*]
    \item In \cite{Du2021}, authors proposed a collect-reduce-aggregate-collect-aggregate-update (CRACAU) method that use edge computing and Byzantine machine learning algorithm to meet industrial automation requirements concerning Industry 5.0. The study highlighted that fault tolerance is one of the issues that will always be addressed irrespective of the industrial generation being followed; however, due to the employment of heterogeneous devices, the security aspect becomes crucial and can affect the overall production of the industry. In this regard, they proposed the CRACAU framework to not only improve automation processes but also enhance the resiliency and privacy through Byzantine-tolerant algorithm.
    
    \item In \cite{Khowaja2021}, the study highlighted the problem of data and model security when it comes to NGWNs and its associated advancements such as Industry 5.0. The study highlighted several security concerns that are dealt when performing data acquisition and data analysis on edge devices in industrial environment. The study proposed a Private AI-based framework that not only addresses data but also model security that has not been taken into account concerning fourth and fifth industrial revolutions. The study also proposes a data security mechanism for industrial environments to secure the data flow, accordingly.
    
\end{itemize}

While the aforementioned study proposed data and model security but performed evaluation on the data modality, only, the study in \cite{Khowaja2022} dealt with the model security aspect. Although the study focuses mainly on IoMT data, the authors suggest that the model security framework can be applied to the Industry 5.0 environment, as model security is one of the emerging issues and can affect industries heavily when humans and edge devices are hyperconnected. The study proposed a proximal gradient based learning networks that could help to cope with model inversion attacks, respectively. Similarly,
Rupa et al. \cite{Rupa2021} also emphasized on the medical industries concerning Industry 5.0 and proposed the use of blockchain networks for data protection. The study employed \textit{remix Ethereum blockchain} to protect and manage medical certificates on edge computing devices. Furthermore, the study tested their method on distributed applications using the RPC-based Ethereum blockchain and agents in a simulated environment.

\subsection{Nexus of CI and Cloud-Fog-Edge Computing for Sustainability}
\label{sec:NexusSustainability}
\textit{Existing studies have highly focused on the use of Edge AI, i.e., a fusion of CI and CFE computing to achieve the required level of sustainability.} Below is a consolidation of some studies that focus mainly on sustainability using Edge AI in Industry 5.0. We have divided the studies concerning Industry 5.0 sustainability in few categories such as energy, robotics, manufacturing, enabling technologies, emerging technologies, and generalized studies (as illustrated in ~\ref{fig:I5.0Taxonomy}).
\subsubsection{Energy}
The studies \cite{Carayannis2021} and \cite{ElFar2021} focused on the energy crisis, energy requirement, and sustainable development goals in the context of Industry 5.0. The former study emphasized geopolitical involvement to support fusion of energy sources to control global emissions and achieve a sustainable Industry 5.0 solution. The latter study focused on the use of bioenergy through byproduct generation and algae culture for achieving a sustainable solution. Both studies emphasize the use of CI techniques to achieve their respective goals. For instance, the study \cite{Carayannis2021} emphasizes the use of CI not only for the automation process, but also to explore the possible combination for the fusion of energy sources. On the other hand, the study \cite{ElFar2021} utilized CI techniques in conjunction with the prototype of the electro-optical systems bioreactor to reduce carbon emissions for manufacturers in Industry 5.0. The study suggested that Algae Industry could provide alternate solutions for renewable energy sources in the context of Industry 5.0.

\subsubsection{Robotics}
In \cite{Demir2019}, Demir et al. specified that the main purpose of Industry 5.0 is to achieve sustainability and pursue the bioeconomy through the means of renewable resources, bionics, robotics, and CI. The study suggests that the CI is the key to enable the co-work between human and robots as well as develop a sustainable working ecosystem in the context of Industry 5.0. The study also highlights the issues that can occur due to the involvement of human and robots.

\subsubsection{Manufacturing}
The authors in \cite{Draghici2022} explored different aspects of Industry 5.0 for manufacturing systems that include green manufacturing, smart manufacturing, intelligent monitoring, process automation, warehouse ergonomics, product development, and economic impact. Most of the chapters, focused on the feasibility and impact of Industry 5.0 based manufacturing systems on the sustainability. Meanwhile, some chapters also emphasized on using CI and edge computing to move towards green computation and circular economy that could help in reducing the carbon emissions, accordingly.\\
Farsi et al. \cite{Farsi2021} evaluated Industry 5.0 indicators for their feasibility in reliable centered maintenance. They suggested that a high correlation between maintenance and operation phases could achieve sustainability for high-value equipment from the industry 5.0 perspective. The study uses fuzzy logic and edge equipment to measure the impact of reliable, centered maintenance on sustainability. Furthermore, the study also suggests recommendations to improve sustainability through the said indicators.

\subsubsection{Enabling Technologies}
Fraga-Lamas et al. \cite{fraga_lamas2021} proposed the use of Edge AI and G-IoT systems to address issues of carbon footprint. The study proposed hypothetical architecture based on Edge AI and green IoT and performs evaluation in terms of energy efficiency to show that such a fusion can be helpful in reducing carbon emissions. The evaluated scenario is considered in the context of Industry 5.0 and it was suggested that the use of CI and edge computing can be helpful in mitigating sustainability issues. Maddikunta et al. \cite{Maddikunta2021} also emphasized sustainability in Industry 5.0 and highlighted the use of CFE computing and the CI technique to reduce the carbon footprints from manufacturing industries.
\begin{figure*}[!t]
\centering
\includegraphics[width=0.7\linewidth]{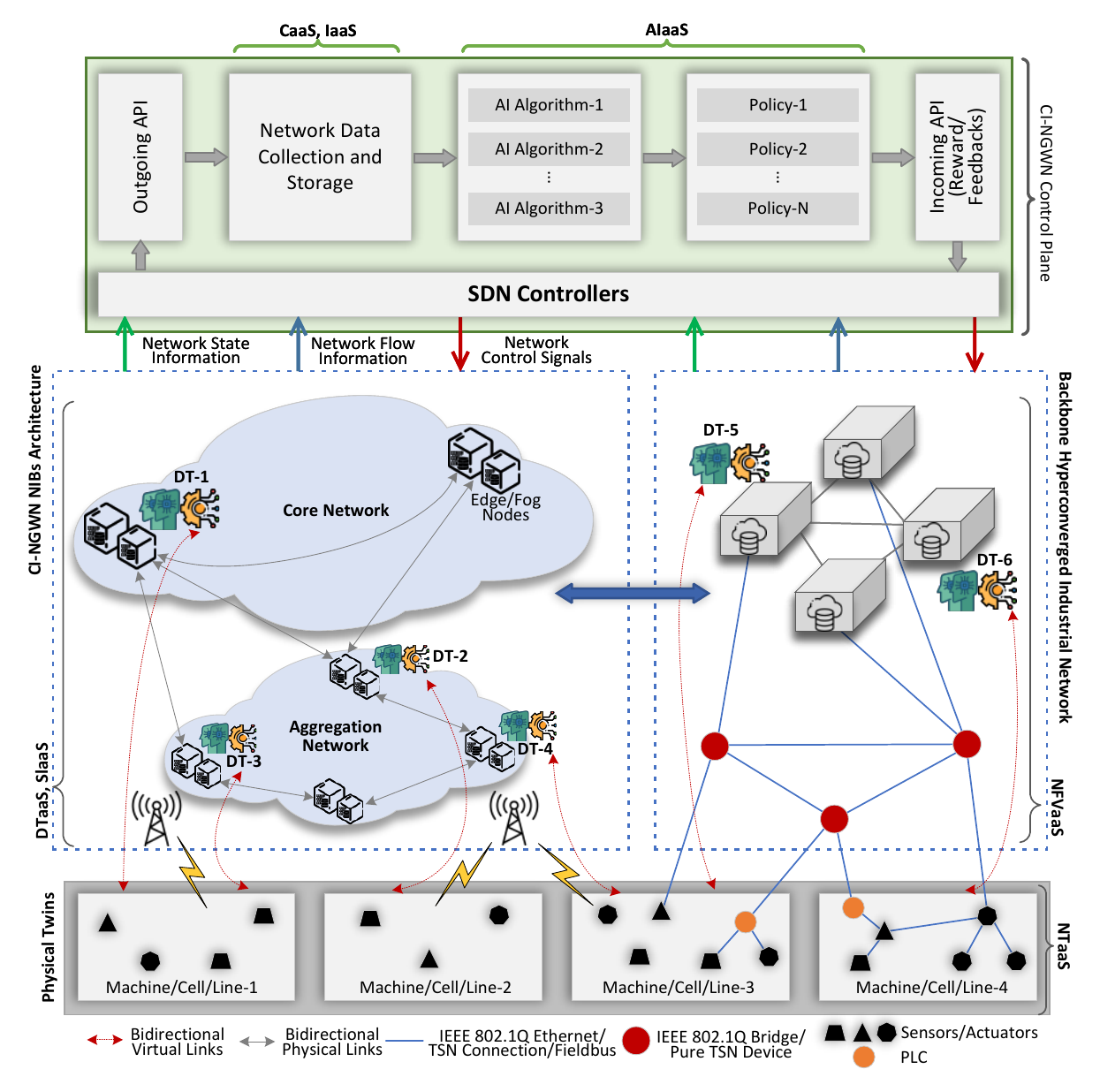}
	\caption{Softwarized communication of physical twins and deployed digital twins over a zero-touch TSN-enabled hybrid NIBs architecture of CI-NGWNs.}
	\label{fig:softwarizedcommunication}
 	\vspace{-10pt}
\end{figure*}

\subsubsection{Emerging Technologies}
Although some studies have not directly proposed sustainability in Industry 5.0 using Edge AI, they have indirectly addressed the use cases of techniques that use both CI and edge computing, accordingly. For instance, Nayak and Patgiri \cite{Nayak2020} proposed a hypothetical architecture for 6G communication that is based primarily on computational intelligence and edge computing techniques. 6G is essentially known for a larger footprint and sustainable communication, therefore, some of the use cases highlight its correspondence with Industry 5.0, assuming that 6G could carry on its sustainable traits to the manufacturing domain as well. Similarly, Shahzadi et al. \cite{Shahzadi2021} proposed a collaborative cognitive communication system that uses CI and CFE computing as its core to enable intelligent services. Furthermore, the main aspect of the system was to leverage the contextual information to activate a particular service. The study shows that one of the use cases of their proposed system is sustainable energy for Industry 5.0.

}

{\color{black}


\section{Potential Enabling Services For Industry 5.0}
\label{sec:Softwarizedexamples}
This section covers the potential examples of enabling services that are supported by multi-tenant softwarized service framework (as shown in Fig.~\ref{fig:softwarizedarchitecture}) to address the challenges of Industry 5.0 vision.

\subsection{Services for TSN-enabled Hybrid NIBs Architecture}
\label{sec:hbridNIBsarchitecture}
The multi-tenant softwarized architecture can combine the diverse enabling service functions (discussed in Sec.~\ref{sec:softwarizedservices}) to provision composite and unified services for diverse applications. For example, Fig.~\ref{fig:softwarizedcommunication} illustrates the TSN-enabled hybrid NIBs infrastructure that combines the CI-NGWNs infrastructure and backhaul hyperconverged industrial network for providing reliable and secure communication and computation services to the interconnected machine cells on the physical manufacturing floor\cite{TSN-4,TSN-5,TSN-6}. The convergence of softwarized architecture over TSN-enabled hybrid NIBs can benefit the CI-NGWNs in providing scalable, agile, and flexible services for the required technological enablers of the Industry 5.0 vision. Some of the other significances are,

\begin{enumerate}[leftmargin=*]
    \item It can provide the significance of seamless integration between both public cloud (e.g., B5G/6G, third party vendors) and private cloud (on-premise) in hybrid NIBs. More enterprises are moving towards the options of both public and private cloud according to the business ecosystem.
    \item With the integration of TSN standards, a promising enabler standardized by 3GPP, hybrid NIBs can enable holistic softwarized communications for industrial automation and control applications with satisfying critical requirements, e.g., time synchronization, bounded latency, ultra-reliability, and dedicated resource allocations.
\end{enumerate}
Additionally, other potential required services can be orchestrated and deployed to enable critical applications, e.g., CaaS, IaaS, and AI-as-a-Service can be deployed at CI-NGWNs control plane (as shown in Fig.~\ref{fig:softwarizedcommunication}) for zero-touch intelligent network management and control over the entire TSN-enabled NIBs architecture. Some of the potential enabling services are explored in the following subsections.

\subsection{Everything-as-a-Service (XaaS)}
\label{sec:XaaS}
Also called anything-as-a-service that combines traditional on-premises, software-as-a-service (SaaS), platform-as-a-service (PaaS), and infrastructure-as-a-service (IaaS) model according to the hierarchy of control and management over computational and communication resources, identified in Fig.~\ref{fig:XaaS}. The XaaS paradigms has led to the massive shift in leveraging subscription-based business models closely integrated with the “\textit{servitization}” in emerging enterprises, i.e., nexus of softwarized products with the on-demand service models in a single package~\cite{XaaS-1,XaaS-2,XaaS-3}.
Hence depending upon the type of XaaS servitization model and subscription, the CI-enabled NGWN can provide the wide range of service advantages, e.g., 

\begin{enumerate}[leftmargin=*]
    \item Communication and computation services to the numerous human-centric robotic industrial processes
    \item Financial savings, and improved productivity for small enterprises
    \item High efficiency in terms of hybrid cloud-edge management, i.e., service request automation, cost management and resource optimization, secure identity and compliance during product lifecycle, governance in package and deliverance procedure
    \item System backup and cybersecurity without exorbitant costs
    \item Agile, scalable and growth-oriented enterprise with integrated evolving technologies and softwares.
\end{enumerate}

Similarly, the emerging trend in the softwarization of networks is paving the way for new hybrid XaaS paradigms~\cite{XaaS-4,XaaS-5}, such as integration of disaster recovery and unified communication-as-a-service, zero-touch network monitoring services, container-as-a-service, health-care-as-a-service, etc., This can be beneficial for the new vision provided for core elements in Industry 5.0. 

\begin{figure}
\centering
\includegraphics[width=0.95\linewidth]{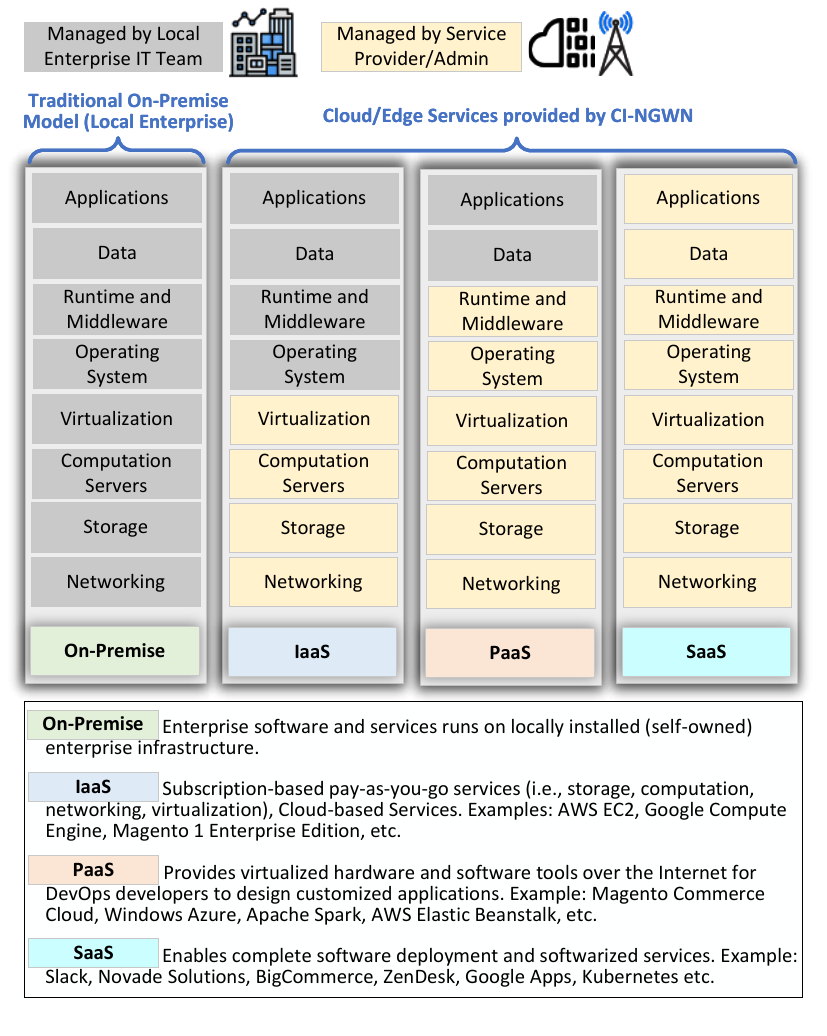}
	\caption{Various servitization frameworks that can be adopted in TSN-enabled hybrid NIBs architecture  }
	\label{fig:XaaS}
\vspace{-10pt}	
\end{figure}

\subsection{Network Function-as-a-Service (NFaaS)}
\label{sec:SlaaS}
The symbiosis of NFV and SDN technologies within the TSN-enabled hybrid NIB architecture and their nexus with the enabling networking techniques, e.g., network programmability, virtualization/containerization, and emerging AI-native SBA paradigm, facilitates the leverageable zero-touch network functionalities, i.e., service-aware communications, network management, and control, and visual services~\cite{NFaaS-1,NFaaS-2}. For this purpose (as illustrated in Fig.~\ref{fig:softwarizedcommunication}), the control plane of NGWN collects the network state information and network flow information from the underlying core of hybrid NIBs and trains the AI models in run time to intelligently manage the softwarized network using network control signals.
Similarly, utilizing the composite enabling functions of the CI and infrastructure layers (c.f.~Fig.~\ref{fig:softwarizedarchitecture}) also forms the building blocks for providing network functions-as-a-services (NFaaS) to aid the service-based provisioning architecture in industry 5.0 scenarios. Some examples of the NFaaS are NFV Infrastructure-as-a-Service (NFVIaaS), Virtual Network Function-as-a-Service (VNFaaS), and Virtual Network-as-a-Service (VNaaS) paradigms.

\fakeparagraph{Slice-as-a-Service (SlaaS).}
NFaaS also forms an building block for network slicing in the hybrid NIB architecture, enabling the network vendor/operator to tailor their network infrastructure and aims to provide a portion of network resources as a slices for diverse tenants~\cite{SlaaS-1,SlaaS-2}, e.g., hybrid eMBB-URLLC-mMTC traffics, e-health, private enterprise networks connected vehicles, etc. This introduces a new business models where long-term agreements can be made between network operators and diverse tenants, and provides a sustainable network slice-as-a-service (SlaaS) to meet the on-demand diverse requirements and network traffic characteristics.

\subsection{AI-as-a-Service (AIaaS)}
\label{sec:AIaaS}
 
Earlier numerous enterprises were not considering AI due to, 1) expensive cost on massive hardware/machines (infrastructure), 2) short supply of technical programmers who can operate over available infrastructure (human resources), and 3) lack of sufficient data~\cite{AIS2,AIaaS-1}. However, the exponential growth in hybrid cloud services has led to the adoption trend towards AI as it becomes more feasible to use because of infinite data collection and service availability.
AI-as-a-Service (AIaaS) can provide remote AI tools enabling emerging enterprises to implement and scale AI techniques~\cite{AIaaS-2}. Moreover, it provides various on-demand AI-native functions (ACF) to various industrial processes (c.f.~Sec.~\ref{sec:softwarizedservices}). This brings numerous advantages, e.g.,

\begin{itemize}[leftmargin=*]
    \item Minimizing the investment risks with cost flexibility
    \item Less dependency over expertise in field of data analytics and ML while staying focused on the core of enterprise business
    \item Gain benefits from the information extracted from critical data, resulting in transparency and manoeuvrability improvement over future course of strategic actions in digital enterprise to increase productivity. 
\end{itemize}
\begin{figure}
\centering
\includegraphics[width=1\linewidth]{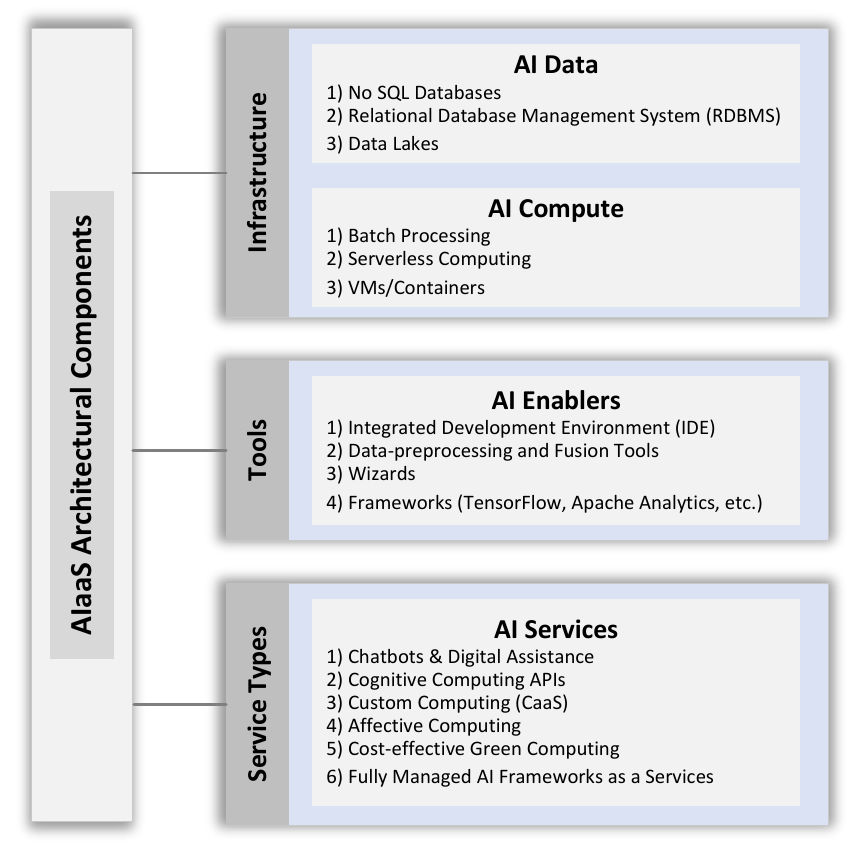}
	\caption{AIaaS Components }
	\label{fig:AIaaSComponents}
\vspace{-10pt}	
\end{figure}

According to the Flexera report on the wide-scale AI adoption in field companies, almost 28\% of enterprises are reported to test-trial with AI/ML techniques while 46\% of enterprises are planning to go for the test-trial experiments in their enterprises ecosystem~\cite{AIaaS-3}.
To enable real time AIaaS architecture, there are generally three essential components which is illustrated in Fig.~\ref{fig:AIaaSComponents}. Further details to the essential components are as follow.  

\subsubsection{AI Infrastructure}
AI infrastructure forms the backbone structure to keep implementing and training multiple AI-native models and functions (c.f.~Sec.\ref{sec:AFL}). Data storages and computations form the critical pillar of AI infrastructure. 

\fakeparagraph{Data.}
AI data from multiple sources has utmost importance at the AFL due to its reliance on trained AI models. These data are fed as input to the AI models, which can come from numerous types of databases, i.e., 1) relational databases management systems (RDBMS) that offer data storage functions possessing security, accuracy, integrity, and consistency~\cite{AIaaS-4}, 2) stored raw data in data lakes\cite{AIaaS-6}, and 3) unstructured large binary objects and stored annotation in NoSQL databases~\cite{AIaaS-5}. Similarly real-time data processing and streaming techniques, e.g., Apache Spark and Kafka, has a scalable ML library that works in parallel with databases to store collected raw data~\cite{AIS2}.  

\fakeparagraph{Computations.}
Advanced AI techniques such as Deep RL, Swarm AI, etc., needs high computation to train on data and converge towards optimal solutions~\cite{CI1}. As a part of an extension to the XaaS paradigm (c.f.~Fig.~\ref{fig:XaaS}), numerous cloud, telecom, and private vendors in AIaaS setup can offer a combination of GPU-CPU-based VMs and containers (IaaS) which is based on a pay-as-you-go model to offer infrastructure-based high computations tasks, such as, zero-touch self-automated AI tasks, perform serial/parallel processing in batch and streaming analytic, serverless computations, etc,. Together they form a compute-as-a-service (CaaS) in hybrid NIBs (c.f.~Fig.~\ref{fig:softwarizedcommunication}), which can easily be availed on-demand and scaled by enterprises since maintaining and owning a dedicated infrastructure (on-premises) can be cost expensive for an emerging factory ecosystem. Under the Industry 5.0 vision, it allows the space for investing spare capital on other goals of human sustainability~\cite{fraga_lamas2021}, e.g., health, human-robots collaboration, digital skills among workers.

\subsubsection{AI Tools}
AI tools help the enterprise developers team to develop and access the various on-demand AI services and other associated services (i.e., IaaS, CaaS, NFVaaS), which are entirely in sync with data and compute platforms.

\fakeparagraph{Wizards \& IDE.}
The nexus of wizards (software) together with the integrated development environment (IDEs) and browser-based notebooks 
enables real-time back/front-end development in a multi-tenant softwarized environment and provides ease by reducing complexity in training AI frameworks for an enterprise management team~\cite{AIaaS-7}. Such AI tools facilitate the smooth DevOps processes in smart industrial applications.

\fakeparagraph{Data Engineering Tools.}
The quality of AI data determines the performance of any trained AI model~\cite{AIaaS-8}. Hence, the cleansing and pre-processing of collected raw data are essential for which data preparation tools have the utmost importance as advanced data tools help extract-transform-load (ETL) jobs, sophisticated data automation, increase the data value, perform feature engineering, and fuse specific critical data, i.e., data fusion~\cite{AIaaS-9}. The extracted data are then fed into the ML pipelines for model training, provided by third-party orchestrated and deployed applications, e.g., Kubeflow. Kubeflow supports wizards, IDEs, and deployable and scalable ML/DL workflows over the softwarized microservice architecture, easily integrated with the CaaS model~\cite{AIS2}.

\fakeparagraph{AI Libraries.}
In the data science environment, the configuration and setting up of ML and DL libraries, such as TensorFlow, Apache MXNet, Torch, Keras, Scikit-learn, etc., has become quite complicated, especially with the emerging updates and patches updates over time~\cite{AIaaS-10}. Numerous ready-to-go VMs and containers templates are set and configured with necessary ML/DL libraries with GPU-supported AI computations, which can be provided to train and implement complex neural networks.

\subsubsection{AI Service APIs and Types}
Numerous cloud vendors provide different types of standalone customer service APIs that can be readily deployed since they do not require dedicated custom-designed AI models as benefits are extracted from the underlying AI infrastructure. Some of the emerging AI service types are given below.

\fakeparagraph{Chatbots \& Digital Assistance.}
Chatbots and digital/virtual assistance in integration with collaborative robots (cobots) can perform the tasks automatically and intelligently in collaboration with human workers~\cite{chatbots3}. They generally use natural language processing (NLP) and perceptual image processing algorithms on voice, text, and image data to learn from human beings and imitate the language and human behavior pattern to provide answers and assistance~\cite{chatbots2}. Enterprises can leverage and integrate AI-based bots and assistance services over the whole factory ecosystem to fulfill the vision of Industry 5.0 by bridging the gap between humans and robots towards the digital revolution~\cite{chatbots1}. Moreover, developers can use these types of AI services in orchestrated industrial xApps to assist in monitoring and controlling applications.

\fakeparagraph{Cognitive Computing APIs.}
Cognitive computing is a technology that imitates human thought processes, personality, and sentiments, which helps in real-time analysis of the environment and making the best-possible appropriate decision~\cite{APIs-1}. The concept of chatbots and digital assistants is part of the cognitive computing products. Various services use cognitive computing APIs as representational state transfer (REST) endpoints--- a way for services to communicate with each other~\cite{APIs-3}. Without developing the application code from scratch, the REST endpoints are accessed by enterprise developers and integrated into the custom industrial application/services they are building~\cite{APIs-2}. Some examples of the cognitive computing APIs are 1) NLP, 2) speech recognition, 3) image recognition, 4) pattern recognition, 4) knowledge extraction, 5) searching, 5) emotion detection.

\fakeparagraph{Fully Managed AI Frameworks as a Services.}
Various ML and AI frameworks (e.g., Deep DL, neural networks, etc.) can be provided on-demand to the enterprise developers that automatically learn and build on the incoming data and evolve with time~\cite{john2020architecting}. 
For example, at the CI-enabled NGWNs control plane, various AI models are provided as disjoint services that take the underlying network's state information to learn and build and select an appropriate control policy to provide zero-touch network functions. 
Similarly, enterprise developers can utilize various intelligent functions using various AI model templates and drag-and-drop tools to assist with more customized and fully automated AI-based applications.

\fakeparagraph{Custom Computing.}
With the rise of different industrial applications and scenarios having diverse communication and computation requirements, various cloud vendors are moving towards providing custom APIs that serve the purpose of specialized custom services~\cite{customcomputing-1}. This strategy results in lessening the overhead of selecting the suitable algorithm and networked computation resources that suit the on-demand QoS and QoE performance.

\fakeparagraph{Cost-effective Sustainable and Green Computing Design.}
Cost-effective green computing or environment-friendly computing design has taken the utmost importance for a sustainable future~\cite{greencomputing-1}. It introduces novel solutions to protect the environment and achieve the goal of net-zero carbon emission. The aims of green computing are~\cite{greencomputing2}, 1) introduction of management and policies to reduce the energy consumption and carbon footprint of the enabling ICT technologies to perform computations while reducing the costs, 2) promote energy-efficient and eco-friendly networking products (hardware and software), 3) reduce the power consumption in machines, and 4) use minimum computation and communication resources in all types of M2M communications. 
By adopting green computing approaches, using AI models in the DevOps operations for the factory softwarized environment can provide green-computing-enabled cost-effective services to achieve various SDG goals in Industry 5.0 applications~\cite{vinuesa2020role}.

\fakeparagraph{Affective Computing.}
According to the author, Rosalind Picard, of "Affective computing"~\cite{picard2000affective}, the computers must be provided the capability to have \textit{recognition}, \textit{understanding}, and \textit{tenancy of emotions and expression}, in case we want computers to possess high intelligence and interact naturally with us. The recent advancements in AI algorithms (cognitive computing, conversational AI, emotional intelligence) and robotics (i.e., cobots for zero-defect manufacturing) and convergence of multidisciplinary fields, i.e., engineering, psychology, education, cognitive science, sociology, and others, have brought forth the new demanding vertical applications in human-computer interactions and affective computing~\cite{wang2022systematic}. These new vertical applications are human health, computer-assisted AR-VR-XR technologies, entertainment, factories-of-the-future, etc. Similarly, the integration of affective mediation technologies and the assistive tools using AR-VR-XR into the FoFs ecosystem can significantly help assist disabled workers during their daily experience.

\subsection{Connected Digital Twin Services and Metaverse}
\label{sec:DTandMeta}

\begin{figure}[!t]
\centering
\includegraphics[width=0.65\linewidth]{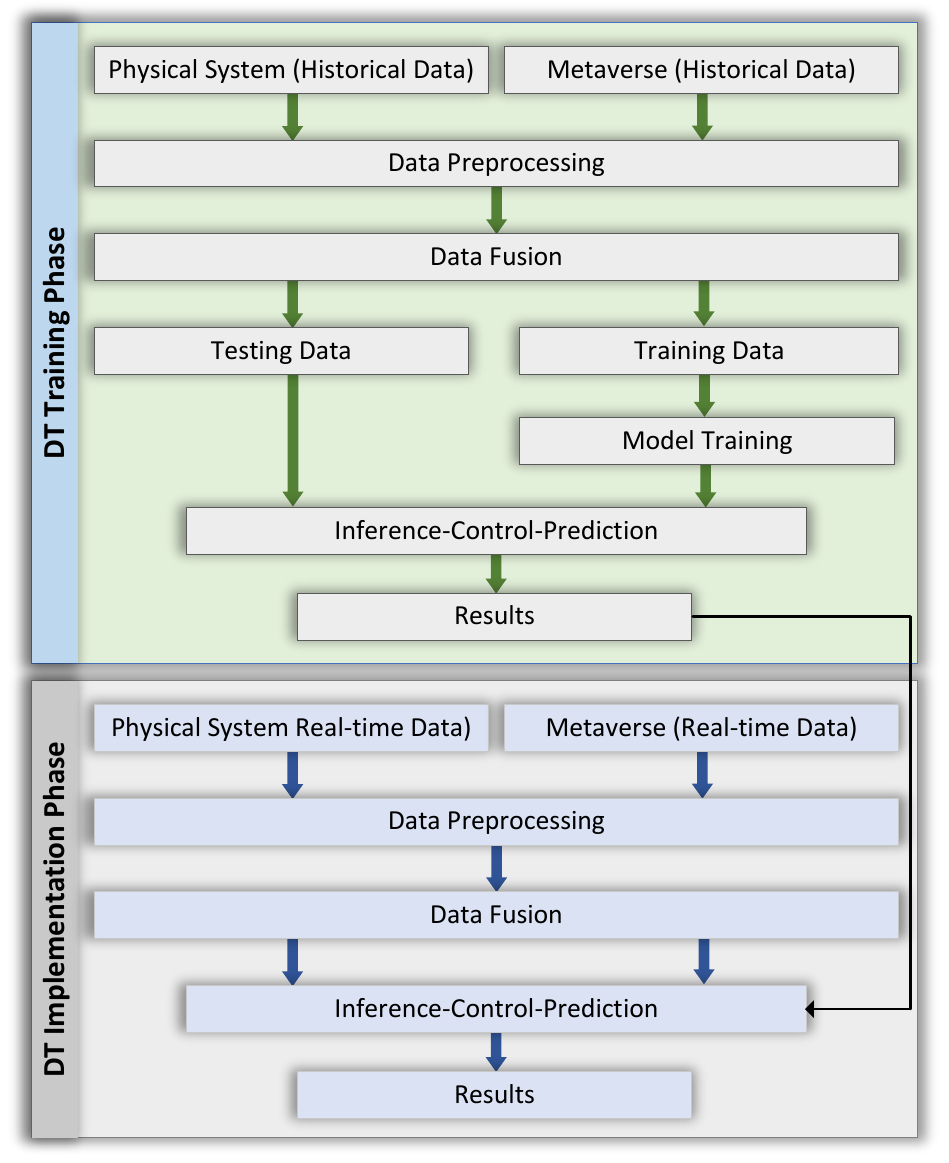}
	\caption{Illustration of autonomous digital twin with deep learning (DTaaS Implementation). }
	\label{fig:DTaaSImplementation}
\vspace{-10pt}	
\end{figure}
\begin{figure*}[!t]
\centering
\includegraphics[width=0.8\linewidth]{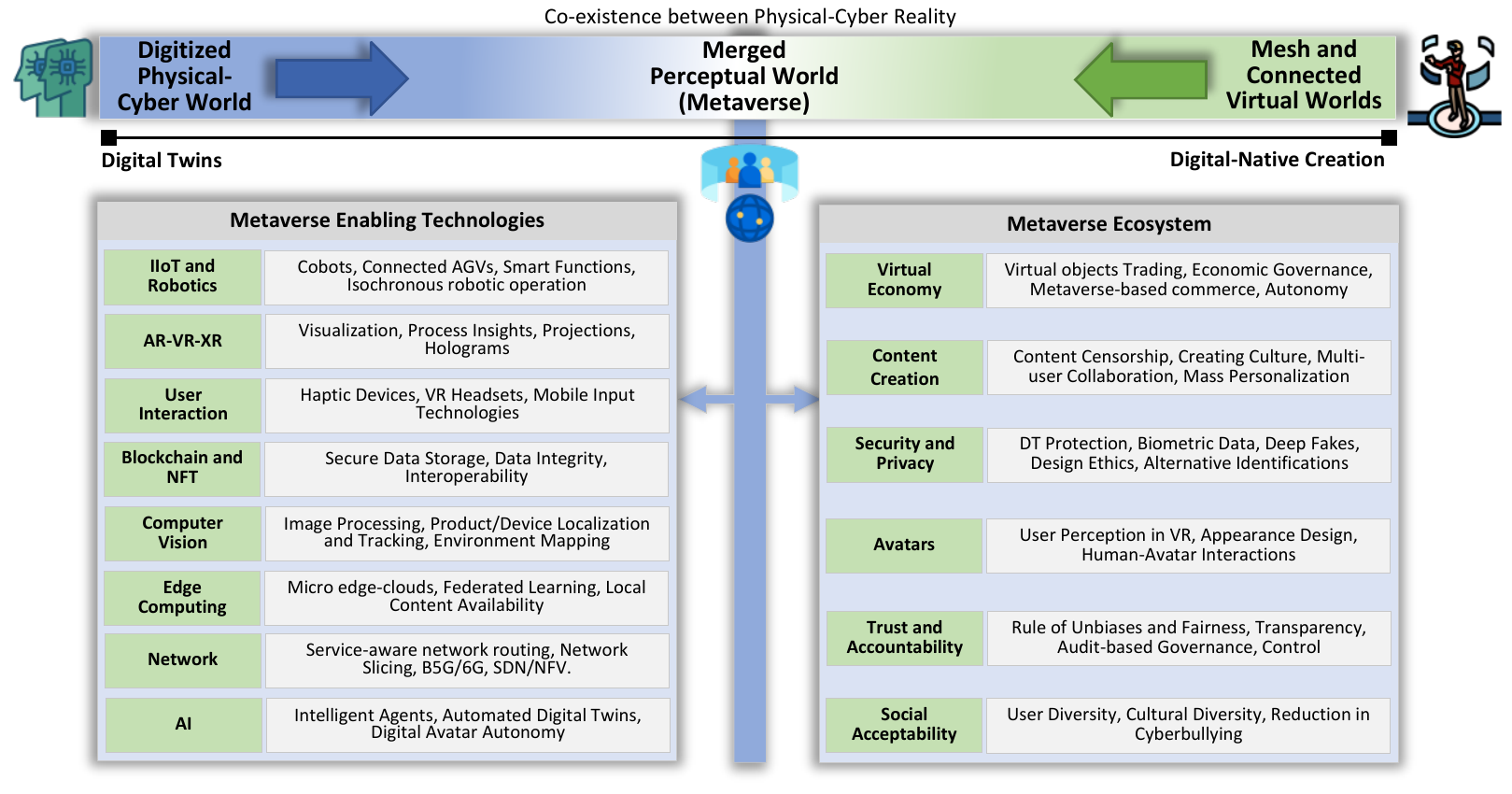}
	\caption{Linking the physical world (i.e., factories, devices, workers) to its counterpart digital twins, and shifting to the metaverse: (Left) the metaverse enbling technologies (i.e., blockchain, CV, distributed networks, pervasive computing, and ubiquitous interfaces); and (Right) the metaverse ecosystem features (i.e., avatars, content creation, data interoperability, security/privacy, trust, and others).}
	\label{fig:DTandMetaverse}
 	\vspace{-10pt}
\end{figure*}

The world has eventually reached a point where almost every technology is deeply embedded in our lives, leading to the blurriness among lines between business (factory production, big enterprise) and personal (factory workforce, employees)~\cite{CI1}.
\subsubsection{Digital Twin-as-a-Service (DTaaS)}
\label{sec:DTaaS}
In this regard, the emerging industrial digital twins bridge the physical and virtual space of factory assets and form a digital replica of the factory through IIoT and CPS-enabled cobots in nexus with CI-NGWNs as a hyperconnectivity medium that can optimize and interact with the physical environment autonomously~\cite{DT5,DT6,DT4}. Similarly, the symbiotic role of industrial DTs along with cognitive AI-based sensing paradigms, AR-VR-XR, and blockchain, accelerates towards realizing true digital transformation that can perform a wide range of on-demand functions and provide services, such as DT services for automated health services, industrial assets, optimizing B5G/6G networks, matrix productions. Especially the factory workforce such as disabled persons with blind or visually impaired disabilities needs to have the proper care in terms of stress management and trauma management, which forms an essential condition for the all-inclusive digital and sustainable future under Industry 5.0 vision. Moreover, DT technology can be utilized as standalone services, forming a \textit{DT-as-a-service (DTaaS)} paradigm to provide numerous digital services~\cite{aheleroff2021digital}, such as visualization and insights, status monitoring, process optimization, automated control, fault prediction, remote control maintenance, etc. Likewise, DT capabilities can enable true mass customization and personalization in matrix productions to achieve sustainable goals of zero-defect manufacturing. The two phases, i.e., DT training and implementation, of autonomous DT creation together with the deep learning method has been illustrated in Fig.~\ref{fig:DTaaSImplementation}.

\subsubsection{Metaverse}
\label{sec:metaverse}

Recently, the term “\textit{metaverse}”--- digital “big bang” of digital-native cyberspace, has been introduced to promote and streamline the digitizing of each physical reality~\cite{metaverse-1}, e.g., devices, people, processes, buildings, vehicles, etc. The essence of metaverse lies in the vision of shared digital-native realms of cyberspace, connected by the gigantic, immersive, unified, persistent, and sustainable Internet at its core, forming the mesh web of virtual worlds or digital twins which enables true digital transformation~\cite{metaverse-2,metaverse-3}. It is noted that true implementations of metaverse may seem futuristic at the moment, especially concerning its integration in the intelligent factory ecosystem. However, various emerging and enabling technologies are poised to catalyze metaverse formation. Fig.~\ref{fig:DTandMetaverse} shows the various enabling technologies and their critical functional role in the metaverse realization.

\fakeparagraph{3R's of Digital Twins and Metaverse.}
The interplay of working roles, relationships, and requirements (3R’s) between the digital twin and digital-native virtual world determines the actual realization and implementation of the metaverse vision~\cite{metaverse-4}. Fig.~\ref{fig:DTandMetaverse} shows the three stages for metaverse development, i.e., 1) \textit{digital twin implementation} of physical space, 2) creation of \textit{mesh-networked digital-native virtual worlds}, and 3) dual nexus of both digital twins and the digital-native virtual world towards the \textit{metaverse creation}. As a starting point, the digital twin digitizes and creates the virtual image of physical environments, providing the capability to project any physical world changes upon its virtual counterpart. For different environments and physical spaces/processes that encompass surroundings, digital twin technology creates many unique and distinct digital-native virtual worlds, which initially suffer from, 1)limited inter-connectivity among themselves and with physical worlds, and 2) information silos, i.e., communication between unrelated virtual images is difficult/impossible. Humans, intelligent agents, or devices can interact and work on different creations as different digital avatars within the build-up of digital-native images. The emerging trends towards the adoption of various enabling technologies and increased sharing/connection of many virtual worlds within a massive digital landscape have led to the convergence of the digitized real world and created meshed virtual world~\cite{metaverse-5}. With this convergence, the final stage of the interconnected reality of the physical and virtual world arises where the hyperconnectivity barrier has been reduced, and simultaneous co-existence of both physical-virtual entities entails much like surreality. It leads to the perpetually connected and 3D cyber-reality as the metaverse. The critical ingredients of the metaverse ecosystem and their significance have been identified in Fig.~\ref{fig:DTandMetaverse}.

\begin{figure}
\centering
\includegraphics[width=0.95\linewidth]{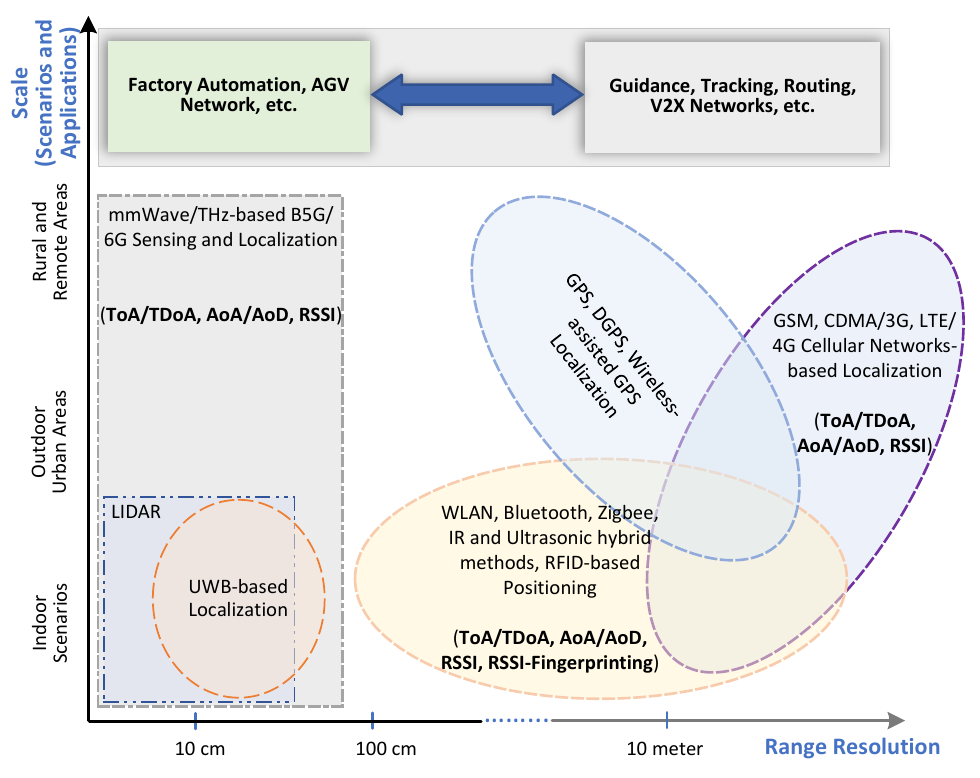}
	\caption{Illustration and classification of various localization systems and enabling technologies based on the scale (scenarios and applications) and range resolution (accuracies) }
	\label{fig:localization-techniques}
\vspace{-10pt}	
\end{figure}

\subsection{Navigation/Tracking-as-a-Service (NTaaS)}
\label{sec:NTaaS}
Navigation/Tracking-as-a-Service (NTaaS), together with the cloud-based infrastructure, can offer on-demand localization services to the retail and manufacturing sector, supply chain, transportation, healthcare, and logistic industries~\cite{TaaS-1,TaaS-5}. These services include navigationally tracking, maintaining real-time location information, and factory assets management, e.g., industrial cobots, AGVs, isochronous factory operations, environment sensing, and many other industrial applications. Moreover, the exponential rise in the adoption of ICT-based digital infrastructure across the factory ecosystem has brought forth the surge in the usage of NTaaS to drive enhanced productivity, revenue growth, and optimize financial efficiency, fulfilling the sustainability goals of the Industry 5.0 vision. The critical parts of NTaaS systems include~\cite{TaaS-3,TaaS-4,TaaS-2}, 1) suitable localization techniques, e.g., ToA, RSSI, etc., 2) localization system integration, e.g., optical and wireless systems, and 3) optimization and performance enhancement techniques, i.e., tracking algorithms, Kalman filters, neural networks, etc. 

\fakeparagraph{Enabling Localization Technologies.}
Numerous enabling technologies, such as light and wireless technologies, are used for localization and ranging purposes over time~\cite{TaaS-6}. The selection of the suitable localization technology for user-defined operations and applications generally depends on the performance metrics needed, e.g., accuracy, precision, complexity, scalability, robustness, and cost. LIDAR (light detection and ranging) is based on optical sensor technology which measures and senses the environment and distance range based on laser reflections~\cite{LIDAR}. They are typically used in AGVs applications (e.g., manufacturing warehouses, cell-based matrix productions) since they are compact, accurate, and more reliable than wireless-based localization technologies. However, they are primarily high costly, and their ranging performance drastically reduces with the atmospheric conditions, such as rain, snow, and fog, limiting LIDAR ranging performance in outdoor scenarios~\cite{liu2021research}. On the other hand, a range of wireless localization technologies is available, which are practically used for different scenarios and applications (c.f.~Fig.~\ref{fig:localization-techniques}). GPS-based localization is not suitable for indoor scenarios since the GPS signals get severely attenuated due to the rich multipath scatterer profile of the indoor factory environment~\cite{URLLCimpact}. On the other hand, other ranging solutions based on WiFi, RFID, ZigBee, LoRaWAN, and Bluetooth for indoor scenarios are developed over time, providing good accurate positioning services in indoor factory scenarios~\cite{aamir2022, Anjum_RSSI_LoRa}. However, ultra-wideband (UWB)-based localization solution, especially with a large bandwidth of 500~MHz and 1~GHz in 3.1–10.6~GHz band, provides much accurate localization and ranging performance over other available wireless technologies. Recent development in the use of mmWave and THz band with massive MIMO and beamforming technologies for high datarates communications in B5G and 6G increases the importance of mmWave/THz-based localization to provide integrated communication and ranging (ICAR) services~\cite{ICAR-1}.

\fakeparagraph{Localization and Ranging Techniques/Algorithms.}
Conventional wireless localization and ranging techniques worked on the measurements of received signal properties~\cite{obeidat2021review}, e.g., 1) received signal strength information is used in the RSSI-based localization schemes, 2) received signal angular information, i.e., angle-of-arrival (AoA) and angle-of-departure (AoD), are used on AoA/AoD-based localization schemes, and 3) received signal time-of-flight (ToF) information is used in time-of-arrival (ToA)-based localization schemes. 
ToA/TDoA-based localization algorithms generally performed better than the RSSI- and AoA/AoD-based algorithms in indoor factory scenarios since their performance is more affected due to thick-wall penetration losses and human blockages, machines and factory assets, and angular scatterers structures~\cite{ICAR-1}. However, multipath structure indeed induces erroneous errors in the positioning accuracies due to the absence of direct-path (DP) in line-of-sight (LOS) conditions. Hybrid algorithms---the combination of conventional localization techniques, e.g., RSSI-ToA, gives better performance than the standalone algorithms~\cite{TaaS-2}. Nevertheless, it increases the ranging and localization algorithm complexity. For URLLC and mMTC services in mission-critical industrial applications, timing synchronization among the industrial assets (wireless and wired) at the physical layer is very important~\cite{TSN-1}. At the 5G RAN level, 5G gNBs wirelessly disseminate the absolute reference time and physical propagation delay information for adjusting the timing clocks at the UE level~\cite{mmWave-ToA}. This increases the criticality of accurate positioning of industrial assets in a rich multipath indoor environment. Kalman filters and AI-enhanced localization algorithms are also developed over time to perform joint positioning and time synchronization and better UE navigational tracking in a realistic environment~\cite{kalman}. For M2M and D2D communication scenarios, a cooperative localization approach is adopted in the worst conditions (e.g., geographic positioning fails) where multiple wirelessly connected industrial sensors and assets communicate and collaborate to improve the accuracy of position estimates~\cite{D2D}. 

}

{\color{black}
\section{Open Standards, Frameworks, Project Demos, Groups}
\label{sec:Openstandards}
To the best of our knowledge, most of the working related to Industry 5.0 guidelines, standards, and frameworks have been carried out by European Union (EU). 

\subsection{EU Publications}

So far, the EU has published over 4 articles explicitly to advance the operations and realization of Industry 5.0. The first was the result from a workshop with Europe's Technology leaders where a discussion with funding agencies, technology, and research organizations was carried out~\cite{Muller2020}. Most of the focus was deviated towards the identification of enabling technologies that could realized the concept of Industry 5.0 along with the challenges. The publication considered individualized human-machine-interaction, bio-inspired technologies and smart materials, digital twins, emerging communication techniques for data transmission, data storage and analysis, AI, and technologies that can achieve autonomy, renewability, and energy efficiency. Furthermore, the challenges such as scalability, economic dimension, interdisciplinarity, governmental and political dimension, and social dimensions were also undertaken, accordingly.

The second publication in the series complemented Industry 4.0 paradigm and provided a pathway for its transition to fifth industrial iteration and first industrial evolution. Apparently, the focus was to make a transition from shareholder to stakeholder value for making industries resilient, human-centric, and sustainable~\cite{EC2021}. The publication discussed in detail the benefits for industries and workers in order to adapt Industry 5.0 guidelines. For instance, the guidelines for skilling, re-skilling, up-skilling, inclusive and safe work environment, and role of industry worker were defined. Similarly, to motivate industries, some benefits including increased resilience, resource efficiency for competitiveness and sustainability, and retaining as well as attracting talents were elaborated.

The most recent one considers the perspective of industries as well as recent challenges including the pandemic, biodiversity collapse, and climate change. The guidelines in this recent report mainly focus on the transformation rather than transition~\cite{EC2022}. The report consists of three segments, i.e., vision, governance, and action. The vision for Industry 5.0 was discussed to achieve Europe's 2030 goals by focusing on responsible ecosystem and value chain, regenerative circular economy design, adaptive and self-sustaining industries, decentralization to achieve sustainability and resilience, respecting planetary boundaries by limiting digitalization, and regulatory frameworks. The governance focuses on horizontal and vertical coherence with international standards and government levels, corporate governance 5.0, and role of government for Industry 5.0 transformation. Lastly, an action plan was proposed what needs to be achieved and who will strive to achieve it. The main course of action plan was to define the role of industries, economic orientation and approaches to improve industry performance, new design for supply chains, value chains, and business models, digital transformation while respecting planetary boundaries, policy making, and approaches and capabilities to innovation and research. Together these EU publications define most of the standards and provide guidelines for the transition and transformation of Industry 5.0.

\subsection{EU Infographs on Industry 5.0}
Next, the EU published an infographic that can be characterized by going beyond profit based services and goods production. The infographic emphasized more on societal contribution from the perspective of industries and suggests to place the health and well-being of the worker first~\cite{EU2021b}. The key elements of the infographic was to highlight the digitilization concerning Industry 4.0, its transition towards Industry 5.0, highlighting its importance in terms of reduced cost (resource efficiency), empowered workers (remaining in control), competitive industry (attracting talent), adaptation (training for evolving skills), competitive edge (new markets), and safety and well-being of the workers. It also highlights three core funding strategies through which Industry 5.0 can be uplifted, i.e. Beyond 4.0, HuMan Manufacturing, and KYKLOS 4.0. Beyond 4.0 mainly concerns the technological impact on welfare, business models, and jobs. HuMan Manufacturing focuses on the automation, factory workers, and their collaboration for increased safety, satisfaction, performance, quality, harmony, and productivity, respectively. KYKLOS 4.0 concerns about the circular manufacturing by focusing on AI-based methods, augmented reality, life-cycle assessment, product life-cycle management, and cyber-physical systems.

\begin{table*}[]
\centering
\caption{Ongoing projects that are closely related to Industry 5.0}
\label{tab:my-table}
\resizebox{\textwidth}{!}{%
\begin{tabular}{p{0.1\linewidth}|p{0.55\linewidth}|p{0.15\linewidth}|p{0.2\linewidth}}
\hline
Project Title & Project Sub-title                                                               & Project Length      & Project Website             \\ \hline
BEYOND 4.0 &
  Technological Inequality - Understanding the relation between recent technological innovations and social inequalities &
  Jan 2019 - Dec 2022 &
  https://beyond4-0.eu/ \\ \hline
EMPOWER       & European platforM to PromOte Wellbeing and hEalth in the woRkplace              & Jan 2020 - Dec 2023 & https://empower-project.eu/ \\ \hline
HR-Recycler   & Hybrid Human-Robot RECYcling plant for electriCal and eLEctRonic equipment      & Dec 2018 - Nov 2022 & https://www.hr-recycler.eu/ \\ \hline
H-WORK        & Multilevel Interventions to Promote Mental Health in SMEs and Public Workplaces & Jan 2020 - Jun 2023 & https://h-work.eu/          \\ \hline
KYKLOS 4.0 &
  An Advanced Circular and Agile Manufacturing Ecosystem based on rapid reconfigurable manufacturing process and individualized consumer preferences &
  Jan 2020 - Dec 2023 &
  https://kyklos40project.eu/ \\ \hline
MindBot       & Mental Health promotion of cobot workers in Industry 4.0                        & Jan 2020 - Dec 2022 & http://www.mindbot.eu/      \\ \hline
ROSSINI &
  Robot enhanced Sensing, Intelligence and actuation to improve job quality in manufacturing &
  Oct 2018 - Mar 2022 &
  https://www.rossini-project.com/ \\ \hline
SPIRE-SAIS &
  Skills Alliance for Industrial Symbiosis (SAIS) - A cross-sectoral Blueprint for a Sustainable Process Industry (SPIRE) &
  Jan 2020 - Dec 2023 &
  https://www.spire2030.eu/sais \\ \hline
\end{tabular}%
}
\label{table:Ongoing-Projects}
\end{table*}


\subsection{Projects and Demos}

Below, we highlight some of the completed and ongoing projects that either explicitly worked/working on Industry 5.0 or closely related to Industry 5.0 guidelines. 

\subsubsection{Incobotics}
One of the recent projects that was completed concerning Industry 5.0 was "\textit{Incobotics}" (2019 - 2021) that was funded by Erasmus+ programme. The funding united four EU partners, i.e. Spain, Greece, Italy, and France for the development of didactic materials that could be used to design collaborative robots (CoBots). The main objective of this project were to raise awareness concerning the evolution of Industry, enhance pool of skilled workers, and increase collaboration between educational centers and Industry. The modules designed by Incobotics are currently undergoing tests from the pilot groups in Italy, France, and Spain. Once the final products are reviewed and tested, they will be made freely available for downloading from their project website~\cite{incobots}.

\subsubsection{Industry 5.0: Inclusive design methodologies for digital manufacturing}
A project "\textit{Industry 5.0: Inclusive design methodologies for digital manufacturing}" (Sep 2021 - Feb 2025) is aligned with the Engineering and Physical Sciences Research Council's (EPSRC) project "Made Smarter Innovation: Center for People Led-Digitalization" is a collaboration between Nottingham, Bath, and Loughborough universities to overcome the digital skills gap, lack of senior management, employee resistance, and support for human capability in Industry 5.0. The project aims to achieve a sustainable digital technology adoption while keeping an emphasis on the well-being of a worker at the center of production process\footnote{https://www.lboro.ac.uk/study/postgraduate/research-degrees/phd-opportunities/industry-5-inclusive-design/}.

\subsubsection{AI REGIO}
In Europe, digital innovation hubs (DIHs) are being created to support small medium enterprises (SMEs) for their smooth transition to Industry 5.0. The project "\textit{AI REGIO}" was funded by EU to address business, technology, and policy barriers in making AI-focused DIHs a reality. In this regard, the project "\textit{Regions and DIH alliance for AI-driven digital transformation of European Manufacturing SMEs}" was funded (October 2020 - September 2023) by Horizon 2020 (H2020) with the co-ordination of Politecnico Di Milano, Italy. This project aims to address three major concerns that kind of create hurdles in implementing digital pathways between AI-driven DIHs and Manufacturing SMEs, i.e. European Union vs Regional Gap (policy driven concern), Innovation collaboration platform vs Digital Manufacturing Gap (Technology driven concern), and Industry 4.0 vs Industry 5.0 gap (Business driven concern)\footnote{https://cordis.europa.eu/project/id/952003}. 

The gap between EU and regional SMEs is about the scale at which AI innovations are applied and integrated. Furthermore, this gap also addresses the issue of accessibility to global marketplaces. A policy needs to be devised to provide equal opportunities to the SMEs in order to scale up the innovation as well as competition. Currently, the DIHs and digital manufacturing platforms (DMPs) fuel the digital European industry and digital single market concerning SMEs. So far, the related innovation, communities, and initiatives are working in an independent way, therefore, platform-related challenges and technology adoption measures need to be incorporated not only to increase the Socio-Business impact but also to bridge the gap between DIH and DMP. The business driven gap mainly concerns ICT innovation for manufacturing SMEs (I4MS) phase III platforms that pilot Industry 4.0 implementations through IoT, Motion and Analytics Data. It has been suggested in this project that I4MS does not properly addresses the synergy between AI revolution (autonomous systems) and Industry 5.0 (human-centricity) that creates the gap between Industry 4.0 and Industry 5.0. This projects aims to solve the aforementioned three concerns, accordingly.

\subsubsection{Collaborative Consortium on Industry 5.0}
A kind of collaborative consortium has recently been developed among the EU universities to collectively work on Industry 5.0, specifically with the focus on combining human craftsmanship and creativity with productivity, speed, and consistency of Robots\footnote{https://www.digitalfutures.manchester.ac.uk/themes/societal-challenges/industry-5-0/}. The consortium includes Henry Royce Insitute, The Thomas Ashton Institute, The Tyndall Centre, The Manchester Institute of Data Science and Artificial Intelligence, The Global Development Institute, The Manchester Urban Institute, and The Productivity Institute. The main idea of this consortium is to integrate human strength and AI for enabling mass personalization, improving products and productivity, deep understanding of customers, value-added creativity, and problem solving.
Some of the other projects that are indirectly related to Industry 5.0 are mentioned in the Table.~\ref{table:Ongoing-Projects}.

}

\section{Lessons Learned, Research Challenges and Future Directions}
\label{sec:lessonlearned}
With respect to the practical utilization of Industry 5.0 vision and the emerging CI-NGWNs, we discuss various important observations, recommendations, and open future research challenges/problems in this section.
{\color{black}
\subsection{Industry 5.0 Technological Enablers and CI-NGWNs }
In this subsection, we outlines the lessons learned, research challenges and research directions after in detail review of the Sec.~\ref{sec:Introduction}~and~\ref{sec:Industry5.0VisionandRequirements}.
\subsubsection{Lessons Learned}
In conclusion, the key practical lessons and recommendations derived from the systematic review can be summarized as follows:

    \fakeparagraph{1. Industry 5.0 Vision:}
    The concrete and universally-adopted definition of the Industry 5.0 vision, the next revolution to the Industry 4.0 paradigm, is yet to be defined as it is continuously evolving. It is observed from the various definitions of researchers and authors that Industry 5.0 vision will be based on human and robots co-working (c.f.~\ref{Sec:VisionTransition}). EU commission documents have reportedly defined the Industry 5.0 vision based on human-centric sustainable and resilient goals. More specifically, the vision comprises the critical technological enablers and digital systems that work in unison towards integrating the people-centric and people-empowered ecosystem with the goals of mass personalization and mass customization-based scalable manufacturing and productions. This can bring high and multifold positive impacts on various critical dimensions of the human ecosystem, e.g., digital society, government and political scenarios, economic and environmental perspectives. 
    
    \fakeparagraph{2. CI-NGWNs:}
    The nexus of computational intelligence as an enabling technology with NGWNs (6G) forms a CI-NGWNs, potentially a powerful integrated solution, which has seen a great deal of interest from academia and researchers (c.f.~Sec.~\ref{sec:Introduction}). Based on the systematic review, it is observed that CI-NGWNs can enable the wide range of scenarios in the Industry 5.0 vision by providing various critical network features, e.g., 1) inherent capabilities of CI technology (i.e., swarm AI, evolutionary AI, DRL algorithms, fuzzy systems, and others) to provide diverse, intelligent functions throughout the core network, 2) softwarized service provisioning and orchestration, 3) AI-native service architecture for ensuring diverse QoS and QoE requirements, 4) zero-touch network management and operations, 5) private intelligent networks, and 6) O-RAN architecture support.
    
    \fakeparagraph{3. Technological Enablers:}
    The technological enablers in Industry 5.0 vision bridge the physical and cyber world to allow seamless integration between human intelligence and cobots working in unison (c.f.~Fig.~\ref{fig:I5.0Enablers}). In this regard, six technological enablers and systems have been categorized and explored in Sec.~\ref{sec:I5.0Enablers} to identify the critical roles and requirements they incur. These categorizations are, 1) technologies that enable human-machine interactions in manufacturing and productions, e.g., CCPSS systems, AMR, augmented reality technologies for visualization insights, and cognitive services, 2) technologies and processes that enable virtual simulation and digital twins implementation, e.g., IIoT-based CPS systems, industrial process optimization, holograms, and digital 3D modeling, 3) AI-native smart systems to introduces secure, safe, energy-efficient and person-centric AI, 4) data infrastructure, sharing and analytics features to provide edge computing and micro clouds, communication interfaces, scalable and autonomy over multi-level IT infrastructure, and big data analytics-based management, 5) bio-inspired technologies, e.g., embedded bio-sensor networks, body area networks, and 6) energy-efficiency paradigms, e.g., integration of renewable energy sources and energy-autonomous systems-based on energy harvesting mechanism.
    
    \fakeparagraph{4. Network-in-Box Architecture:}
    The adoption of NIB architecture in CI-NGWNs enables the interoperable paradigm of intelligent network softwarization in tandem with AI-native SBA that is embedded throughout the entire network, providing reconfigurable integrated solutions and network resources for meeting the critical technical requirements of Industry 5.0 vision. In this regard, design objectives for the CI-NGWNs concerning Industry 5.0 vision are identified and discussed in Sec.~\ref{sec:CI-NGWNsupport}, which is followed by the NIBs architecture vision in CII-NGWNs that brings the numerous enabling functional features and softwarized service support using SDN/NFV-enabled networked HCIs. These enhanced CI-native features (c.f~.Fig.~\ref{fig:HyperarchitectureCI-NWGNs}) are distributed throughout all CI-NGWNs functional layers, e.g., intelligent application layer, intelligent network control layer, analytic layer, and sensing layer. It is observed and learned that NIBs architecture together with CI technology in NGWNs could efficiently cater to the critical incurred requirements of technological enablers in Industry 5.0 vision, identified and categorized in Sec.~\ref{sec:I5.0Enablers}.
    

\subsubsection{Research Challenges and Future Directions}
This discussion outlines the various foreseeable open problems that can be summarized as follows in light of the preceding analysis on lessons learned and recommendations:


\fakeparagraph{1. Challenges in Industry 5.0 Vision:}
The work on Industry 5.0 vision has recently started, and it has got much attention from researchers. However, particular challenges still exist on full human-robots collaboration realization and identification of other critical aspects of human-society impact on the vision since they are hard to quantify and measure~\cite{EU1}, such as 1) \textit{social challenges}, e.g., digitally skilled labor shortages, unemployments, corporate social responsibilities, customer value, and socio-centricity in technical training to learn complex ecosystem of technologies, 2) \textit{government and political challenges}, e.g., compliances with country-specific laws, adaptation with the speed of technological changes, protectionism, and digital system-oriented innovation policies, 3) \textit{economic challenges}, e.g., new business models, circular economy, productivity challenges, significant investments, and public-private partnerships, and 4) \textit{interdisciplinary approach} towards nexus of various research disciplines to address interrelations and cause-effects relations in system complexities, e.g., engineering and technology, mathematics, computer sciences, life and humanities, environmental and social sciences.

\fakeparagraph{2. CI-NGWNs Challenges:}
Some of the critical challenges in CI-NGWNs lie in the ultimate mobile experience, extreme global network coverage, and mission-critical low-latency communications compared to the 5G networks~\cite{CI-NGWNs1}. For extreme global network coverage, UAV-Satellite-based terrestrial networks have taken a great deal of promising research direction in providing communication and computation services where reliable coverage is not possible due to lack of infrastructure and environmental factors~\cite{AMR-1}.
Additionally, for URLLC-centric problems, ranging and positioning over the deployed NGWNs infrastructure at mmWave bands has gained importance since they require the nanosecond-level accuracy at the physical layer in ultra-dense wireless networks for timing synchronization procedures~\cite{ICAR-1}.
Similarly, to provide enhanced computing and storage service, there is a need to be smooth integration between cloud/edge networks and wireless access networks with critical enabling technologies, e.g., mmWave-THz bands, mMIMO technologies, advanced antenna arrays, NOMA and RIS metasurfaces. Furthermore, with the requirements of human intelligence integration, new developing performance metrics, such as Quality-of-Physical-Experience (QoPE), are needed since they can combine the required QoS and QoE with the human-centric perception and intelligence in new vertical applications~\cite{hu2020cellular}, e.g., AR-VR-XR in augmented reality technologies, holographic communication.

\fakeparagraph{3. Challenges in Technological Enablers:}
The core idea behind technological enablers for Industry 5.0 vision is to introduce and integrate the lost human-centric dimension (i.e., human values and needs) of the previous revolution~\cite{incobots}. In this regard, numerous complex systems and technologies are categorized and discussed in this reviewed work. However, technologies and systems assisting the human-machine collaboration to generate products and services, e.g., human-machine interfaces, integrating human-intelligence using person-centric AI, AMRs, cognitive CPSS systems, and others, are still in early development stages and possess critical research challenges before they are matured enough~\cite{fraga_lamas2021}. Similarly, the requirements and challenges differs with the type of Industry 5.0 scenario. For example, to ensure trust in the cloud/edge-based matrix production system, the critical private data exchange between humans (designer, collaborators) and machines at the factory physical floor for monitoring and control needs to be secured and not visible to any third-party system for potential cyber-attacks. Therefore, a blockchain and digital ledger transaction (DLT) in conjunction with federated learning-enabled quantum computing can provide a robust framework for Industry
5.0~\cite{Rupa2021}, enabling faster cognitive edge data processing for stringent time applications and data preserving secure transaction. \\ Simultaneously, the \textit{ethical and societal issues} with implementing AI needs to be addressed for seamless collaboration between the human resources and cobots since it can create negative societal impacts due to the notion of job displacements and humans replacements with AI-based cobots systems, while Industry 5.0 is poised to create more employment opportunities~\cite{Muller2020}. Moreover, data privacy in cobots needs to be preserved while doing the cognitive analysis in predictive maintenance applications with humans collaboration.
One of the essential requirements is \textit{regulatory compliance} in the co-production environment, where several issues relevant to technical policies and proper regulations may arise~\cite{xu2021industry}, e.g., 1) distinguishing between cognitive cobots and several other machines, i.e., drones, AMRs and 2) the inculcation use of critical AI in nexus with other complex digital systems, and 3) better predictions and control over more sophisticated mass-personalization goals.  

\fakeparagraph{4. NIB Architecture Challenges:}
Based on the critical systematic review and potential uses of NIB architecture in CI-NGWNs for supporting Industry 5.0 vision, we observed and categorized the relevant goal-based critical challenges into four groups, 1) \textit{hyper-intelligent network challenges}, i.e., integrating CI into the core network, 2) \textit{coordinated and harmonized network challenges}, i.e., co-existence of hyper-intelligent core network with other standalone networks and backward compatibility, e.g., 5G, terrestrial and non-terrestrial networks, visible light communication, 3) \textit{sustainable network challenges}, i.e., green computing-enabled energy-efficient networks, and 4) cyber security challenges. \\ Some of the hyper-intelligent network challenges includes~\cite{CI-NGWNs2}, 1) lack of standard and high-quality network datasets that has distinguished features, e.g., channel modeling, RAN technologies, device densities, 2) visibility of the underlying network operations since almost all proposed AI solutions are implemented in black-box manner, and 3) hybrid public-private network challenges in providing centralized and distributed AI-based solutions at the network edge. Similarly, the coordinated networks challenges and future directions includes~\cite{CI-NGWNs4}, 1) holistic approach towards designing frameworks that initiates the convergence between cyber, physical, biological, and social space through wirelessly connected biosensors and CCPSS systems using nano-networks, cell-free MIMO, autonomous D2D, above 100 GHz THz and optical spectrum, 2) global-level optimization scheme in coodinated networks to utilize the available communication and computation resources at different networks efficiently and collaboratively, and 3) trustworthy cooperation and protocols between the coordinated networks for seamless allocation and operation of AI-native functions.\\
Similarly, potential challenges in sustainable network goals include~\cite{CI-NGWNs5}, 1) energy-aware routing and resource management algorithms for ultra-dense wireless sensors network, 2) global-level network sustainability through integrating renewable energy systems and varying level of processing power in computing infrastructure, and 3) adaptive data modulation schemes for enabling data compression. Meanwhile, the cyber security challenges in the wake of AI inclusion include the design of new novel AI-based security architecture to provide basic security and privacy requirements in future networks~\cite{EAIS-2}, e.g., confidentiality, location privacy, integrity, identity privacy, availability and authenticity, data privacy and other meta-data privacy.

}

{\color{black}
\subsection{Computational Intelligence for Industry 5.0}
In this subsection, we outlines the lessons learned and recommendations, research
challenges and research directions after in detail review of the Sec.~\ref{sec:computationalintelligenceI5.0}.
\subsubsection{Lessons Learned}
Based on the systematic review presented in the previous sections, the key practical lessons and recommendations are as follows.
    
    \fakeparagraph{1. Diverse use of Computational Intelligence:}
    With such diversified characteristics of computational intelligence techniques, issues concerning resiliency, sustainability, and personalization are ought to be handled in Industry 5.0. For instance, variants of some supervised learning and federated learning can be widely used for resilient systems in Industry 5.0. Similarly, methods that are less computationally complex, such as self-supervised learning or soft computing techniques, can be used to achieve sustainability, and many intelligence techniques can be employed for achieving mass personalization as well. 
    
    \fakeparagraph{2. CI for Industrial Security:}
    The critical success of Industry 5.0 depends on the usage of CFE devices and computational intelligence. However, these devices are quite vulnerable to security and privacy attacks. Studies presented in the previous section are now moving toward implementation or adoption of computational intelligence techniques that not only provide smart services, but also preserve privacy and security of users' data specifically in Industry 5.0 setting. In this regard, the use of Federated learning and soft computing techniques have proven to be quite effective so far. As humans have a high stake in Industry 5.0, it is important to deal with such security and privacy threats in an efficient and effective manner. 
    
    
    \fakeparagraph{3. CI for Emerging Applications: }
    Since the advent of Industry 5.0, the use cases were mostly limited to manufacturing, supply chain, and human-robot collaboration. In recent years and with the popularity of the fifth industrial revolution, more applications are emerging and the use-cases are increasing. For instance, researchers have proposed the use of Industry 5.0 in industrial automation, data privacy and security, Society 5.0, and alternative energy sources. Researchers have also started to explore the use cases of Industry 5.0 in the domain of healthcare, emotional intelligence, and education. It is assumed that the increase in digital assets and technologies will pave the way for more applications that can be associated with Industry 5.0 to improve value creation and circular economy.  

\subsubsection{Research Challenges and Future Directions}
This section highlights the foreseeable open research problems concerning computational intelligence in Industry 5.0.

    \fakeparagraph{1. Exploring CI with Technological Frontiers:}
    The use of computational intelligence has been exploited for a variety of industrial and manufacturing use cases, but only a limited amount of attention has been diverted toward education, healthcare, and next generation network domains. Each of the aforementioned domain has its own corresponding challenges, for instance, personalization and resilience is the key to academic industry, acquisition of sensor data and its privacy are essential for health care data, and affordability is one of the keystones for Industry 5.0's integration with next generation networks \cite{Khowaja2021, dbfl2021}. Researchers must push the envelope to explore or extend the research in the aforementioned areas. 
    
    \fakeparagraph{2. Mental wellbeing and Affective Computing with CI:}
    Industry 5.0 is said to be an association with human workers while making the production process automated and user-friendly. However, existing studies have shown that working with machines causes emotional imbalance and stress \cite{Stress2020, Feri2021}. Considering that the fifth industrial revolution has a dedicated work-plan with machines, the monitoring of mental well-being and stress levels is an issues that needs to be undertaken. Computational intelligence techniques can play a vital role in monitoring and diagnosing stress or depressing state in a timely manner so that a necessary action can be executed for avoiding health issues. Currently, to the best of our knowledge, this issue has not been explored in the context of Industry 5.0. 
    
    \fakeparagraph{3. Seamless Integration of CI with Digital Spaces:}
    With the current rise in digital technologies such as virtual medicine, digital twins, and Metaverse, the concept of Industry 5.0 can be leveraged to make manufacturing and industrial spaces more sustainable, resilient, and human-centric \footnote{https://www.linkedin.com/pulse/industry-metaverse-next-big-thing-ribeiro-cybersecurity-msc-mba?trk=articles-directory}. However, in order to provide seamless integration, computational intelligence is of vital importance. For instance, digital avatars can be designed and simulated via digital twin to see how productive and sustainable the designed industry would be with certain characteristics. This will allow investors to not only foresee and plan budget issues, but also to identify the potential opportunities that could be exploited to generate more revenue \footnote{https://medium.com/@theo/digital-twins-iot-and-the-metaverse-b4efbfc01112}.  
}

{\color{black}
\subsection{Open and Intelligent 6G Network Architectures}
Here, we outline the lessons learned, research challenges, and future directions based on Sec.~\ref{sec:6G_Architecture}.

\subsubsection{Lessons Learned}
The following key lessons and recommendations are derived from the systematic review.

\fakeparagraph{1. SON and O-RAN Integration:} From Sec.~\ref{subsec:envision_6G_Arch}, it is learned that by merging self-organizing networks (SON) and O-RAN architectures, several advantages are gained, including the use of the same infrastructure nodes and monitoring databases, as well as cross-operations. In particular, the use of Management Data Analytic Service (MDAS) with Non-RT RIC is highly synergistic. Additionally, the O-RAN orchestrator provides services for SON. O-RAN communities are already implementing integration, which needs to be enforced in the long run.

\fakeparagraph{2. MEC, Network Slicing, and O-RAN Integration:} The support of MEC and Network Slicing (NS) in the O-RAN architecture will bring multi-fold features to enhance the performance of CI-NGWNs, e.g., 1)  MEC hosts (third party vendors) can be integrated with the near-RT RIC hosts of O-RAN, 2) O-RAN databases can integrate with MEC databases that can possibly include the critical information, e.g., UE locations, deployed cell performance, radio network information service (RNIS) and others, 3) mobile edge application orchestrator (MEAO) can be used for diverse type of xApps orchestration. e.g.,  SON functions, softwarized enterprise services, and 4) with NS support, an XApps slice can be include in the NR slice templates., which can be easily orchestrated by the O-RAN and/or MEC orchestrator on the separate and secure partition of O-RAN and/or MEC databases.
}

\subsubsection{Research Challenges and Future Directions}
This section reflects upon the imminent challenges in 6G architecture design and optimization, inclusive of network intelligence (NI). In particular:

\fakeparagraph{1. O-RAN Optimization:} 
Many challenges for implementing O-RAN-based  data-driven, open, programmable, and virtualized NextG architecture are still largely to be dealt with, where the main concerns are~\cite{ORAN_Learning_NextG}: functionalities and parameters of each network functions, distribution of network intelligence, optimization of data-driven control loop solutions, and RAN data access overhead for NI.

\fakeparagraph{2. E2E NI-native Architecture with Customized AI:}
The NextG architecture will require to host/integrate NI instances explicitly at NF levels and their coordination across entire network (i.e., macro-domains). Meanwhile, AI models should facilitate network automation and management, and thereby adapted (w.r.t. latency guarantees training and computational complexity) for NI-specific design aspects. Besides, the other two main challenges are~\cite{NI_ORAN}: AI feedback loop between the control and data planes not allowing the NI to monitor the system in a continuous manner, and the lack of interaction among NI instances at different levels in the current architectural trends.

{\color{black}
\subsection{Softwarized Service Provisioning and Potential Applications}
In this subsection, we outlines the lessons learned, research challenges and research directions after in detail review of the Sec.~\ref{sec:softwarizedservices}~and~\ref{sec:Softwarizedexamples}.

\subsubsection{Lessons Learned}
From the systematic review, we derive the following key lessons and recommendations:

\fakeparagraph{1. Multi-tenant Softwarized Service Provisioning:}
     Based on aformentioned discussion in Sec.~\ref{sec:softwarizedservices}, it is learned that the essential features of Industry 5.0 will be realized through softwarized components embedded throughout the hardware and cyber world in digital systems. For example, to achieve zero-defect manufacturing and mass-customization goals in matrix productions, softwarized components, like ACF functions, DT functions, and others, thus enabling fully self-automated control in production manufacturing and assembly. In this regard, a multi-tenant softwarized service provisioning framework based on CI-NGWNs is illustrated in Fig.~\ref{fig:softwarizedarchitecture}, aiming to provide the essential and on-demand softwarized services orchestration of required critical functions over many functional layers for on-demand applications in the Industry 5.0 vision. These functional layers are, 1) service layer providing multi-tenancy support and service mesh-enabled composite service blocks to orchestrate customized services for deployed multi-tenant industrial apps, 2) computational intelligence layer providing critical functions of AFL and DTFL, and 3) infrastructure layer providing functions of HCI resources, industrial access networks, autonomy and control over computing infrastructure and DTI functions. Together, all of these layers work in conjunction to enable the best practices and principles of SOA, microservices, and serverless architecture principles (c.f.~Fig.~\ref{fig:evolutionofsoftwarized} and Table~\ref{table:softarchitectures}) in a CI-enabled softwarized industrial framework to support a wide range of critical softwarized services, e.g., XaaS, AIaaS, DTaaS, NFaaS, and others, that were covered and discussed in Sec.~\ref{sec:Softwarizedexamples}.
    
    \fakeparagraph{2. Softwarized Communication and Industrial DevOps:}
    In Sec.~\ref{sec:Softwarizedexamples}, it is learned that the multi-tenant service provision architecture can orchestrate and deploy the softwarized XaaS paradigm over NIBs architecture in CI-NGWNs. For example, Fig.~\ref{fig:softwarizedcommunication} illustrates a hybrid NIBs architecture, i.e.,  aggregated CI-NGWNs and on-premise backhaul HCI industrial network, to provide enhanced services for critical applications, e.g., 1) intelligent control planes using CaaS, IaaS, AIaaS, SlaaS and NFVaaS for zero-touch networks, 2) customized DTaaS to enable and connect the machine/cell physical twins and virtual twins over the various placement of choices such as cloud, edge, and private databases, 3) hybrid cloud architectures to connect public and private cloud, 4) enabling time-sensitive networking to meet stringent time synchronization requirements for cobots operations in cells, and 5) NTaaS to provide critical localization and ranging services at physical manufacturing floors. Similarly, bi-directional softwarized communication (i.e., virtual links) is enabled over physical bi-directional links of hybrid NIBs architecture to provide real-time reliable and control communication between physical twins and DT placed at entire CI-NGWNs. Moreover, the integration of XaaS principles provides the infrastructure for enabling enhanced DevOps strategies designed to meet enterprise requirements. In this regard, it is learned from Table~\ref{table:softarchitectures} that various softwarized architecture provides various features and significance to adopt the suitable architecture for required DevOps and enabling strategies in cloud-native environments, which incurs the specific drawbacks/limitations according to the selected frameworks.  
    
    
   \fakeparagraph{3. Industrial DT and Metaverse:}
   From Sec.~\ref{sec:DTaaS}, we learned that DTaaS is poised to form a building block for realizing the different Industry 5.0 applications, e.g., predictive maintenance industries, zero-manufacturing, and others. It bridges and connects the created virtual image of industrial assets with counterpart physical space, thus efficiently increasing the hyperconnectivity to integrate human intelligence through augmented reality technologies. Moreover, CI-NGWNs architecture can also take services of DT implementation at the core network for creating the DT of the network to assist in the dynamic reconfiguration of communication and computation resource allocations. The autonomous DT creation using DL methods for any required applications can be realized through DTaaS implementation infrastructure, as illustrated in Fig.~\ref{fig:DTaaSImplementation}. Note that the mesh-connection and more superconvergence of created DT virtual images in cyberspace can lead to the new futuristic technology of tomorrow--- metaverse. In Sec.~\ref{sec:DTandMeta}, it is learned that the metaverse promotes the digitization of every physical reality comprised of humans, industrial assets and devices, vehicles (i.e., AMRs, AGVs), buildings (i.e., manufacturing floor, cells), and others to form a web of unified and ubiquitously connected virtual worlds. In this regard, three process steps from initial DT implementation to final metaverse realization have been explained. Moreover, the required metaverse enabling technology and resultant features of the metaverse-ecosystem has been illustrated in Fig.~\ref{fig:DTandMetaverse}, which indicates that DT enabling features in Industry 5.0 vision will evolve and transition towards the metaverse-based digitized industrial concepts to successfully realize the true fusion between physical and virtual space and goals of human-machine interaction.

    \fakeparagraph{4. Localization Services:}
    The requirement of centimeter-level accuracy in localization and ranging techniques has increased the importance of NTaaS services in mission-critical and location-aware Industry 5.0 applications like AMRS, AGVs, product and human tracking, and others. In this regard, it is learned from Sec.~\ref{sec:NTaaS} that various light-based and wireless-based enabling technologies exist for localization purposes, e.g., LIDAR, GPS, WiFi, RFID, ZigBee, LoRaWAN, UWB, and mmWave/THz. These technologies provide different performance metric-based solutions, e.g., cost, scalability, accuracy-precision, and implementation complexity. Similarly, various enabling technologies are reported in the literature to use different available localization techniques, e.g., ToA, RSSI, AOA/AoD, cooperative localization, and hybrid algorithms (i.e., RSSI-ToA-based frameworks). In, Fig.~\ref{fig:localization-techniques}, the classification of various enabling technologies and techniques based on their range-resolution performance in the diverse scenario has been illustrated, which shows that mmWave-THz-based localization is ideal and perfect to use since they can provide ICAR services by using the existing CI-NGWNs infrastructure.

\subsubsection{Research Challenges and Future Directions}
In light of lessons learned and recommendations in the preceding section, the following are some of the foreseeable open problems:

    \fakeparagraph{1. Multi-tenant Softwarized Service Provisioning Challenges:}
    It is imperative that unified service management in multi-tenant cloud/edge-native NIBs environments demand novel mechanisms and performance guarantees (e.g., agility and flexibility) for designing, instantiating, deploying, running, and optimizing composite services within softwarized HCI~\cite{SlaaS-1,CI-NGWNs2}. In this regard, the possible research directions are, 1) \textit{heterogeneous functions}, e.g., network, compute, and storage resources must be integrated in single infrastructure layer and managed holistically, even though they are in different tech-domains, 2) \textit{function design and system scalability}, e.g., the increase in the number of finer-grained service components (functions) and the communication overheads among them in microservices architecture make system scalability more challenging, 3) \textit{modelling unified abstraction for resources and services}, i.e., to overcome the heterogeneity challenge in convergence of enabling technologies in NIBs of CI-NGWNs, it is critical to design and regularize a standard set of models for unified abstractions of resources and services across these domains, and 4) \textit{new technologies for customized service management}, e.g., KubeEdge, Kubernetes, OpenShift, etc., can be used as automating software management, scaling, and deployment utilizing an open-source container orchestration system~\cite{AIS2}.

    
\fakeparagraph{2. Challenges in Softwarized Communication and Industrial DevOps:}
The complexity in NIBs architecture of CI-NGWNs is expected to unprecedently grow with the wide range of softwarized heterogeneous service requirements by highly diverse Industry 5.0 applications, e.g., cloud/edge functions often require lower latency, higher throughput, and greater reliability than the VNFs do~\cite{table_soft8,NFaaS-2}. Moreover, the resource limitations, inefficient deployed service application execution, communication protocols for data sharing, control signaling and resource allocations between heterogenous coordinated networks, and service mobility and connectivity form the performance bottleneck in heterogeneous cloud/edge-native environment~\cite{NFaaS-1}.
Thus, developing comprehensive solutions to deal with heterogeneity, scalability, flexibility, and performance in softwarized NIBs architecture of CI-NGWNs networks becomes an important research area. Moreover, acquired third-party service types dictate the DevOps strategies for enterprise softwarized solutions. For example, in third-party AIaaS solutions, the enterprise management and service development teams have no access to the underlying actual AI process since they only have the option of dealing with the inputs and outputs of the system, which changes the design guidelines of enabling software-oriented service architecture principles, affecting privacy and increasing security risks~\cite{AI-services-1}. 
    
    
    \fakeparagraph{3. Industrial DT and Metaverse Challenges:}
    Industrial DT and augmented reality technologies provide a smooth human-robotic experience in terms of visualization and interaction with the machine processes for monitoring, control, and optimization~\cite{CI-NGWNs1}. However, its performance can suffer from the typical KPIs of CI-NGWNs, e.g., network latency, bandwidth, throughput, and jitter, as they are critical and required features for a smooth user experience~\cite{metaverse-4}. Moreover, the mobility with multi-sensor fusion applications (e.g., AMRs, AGVs) can further complicate the critical machine tasks in remote operations through industrial DT~\cite{DT1}. Hence, for smooth DT operation, it is necessary to address industrial DT's stringent communication and computation requirements, eventually, the industrial metaverse-based applications, as they require a two-way reliable communication link over CI-NGWNs. Additionally, with more vendors, third parties, and service providers included in CI-NGWNs to offer MECs services, there is a high-security risk of multiple adversaries accessing, stealing, and tampering with the sensitive information and data from implemented edge DTs~\cite{metaverse-5}. Moreover, in the context of AI significance in Industrial DT and metaverse-based applications, there is a need to design privacy- and security-preserving AI models that can automate security and privacy preferences while creating virtual copies of physical assets and protect them from adversarial attacks~\cite{CI1}.
    
    \fakeparagraph{4. Localization Services Challenges:}
    Following are the possible future challenges and trends in indoor positioning systems that can aid the NTaaS paradignm.
    
    \begin{enumerate}[leftmargin=*]
    \item Indoor positioning and ranging coverage in mmWave and THz bands suffers from harsh multipath conditions due to severe NLOS blockages, e.g., human blockages, machine blockages, and scattering conditions, e.g., metallic surfaces, energy absorption. Moreover, the performance of ToA-based in terms of ranging and positioning accuracy degrades with indoor factory (InF) conditions~\cite{mmWave-ToA}. In this regard, RIS metasurfaces together with mMIMO technology operating at mmWave-THz bands can be used to provide enhanced ICAR services by~\cite{ICAR-1}, 1) aiding localization algorithms through intelligent reconfiguration of InF environments, 2) extending the ranging coverage by covering the NLOS blockages areas, and 2) reconfigured mMIMO channel gain matrix to increase sparsity for provision of the high capacity-based wireless connectivity. Similarly, the inclusion of smart beamforming at mmWave-bands can significantly increase the RSSI-based localization algorithms.  
    \item New hybrid localization algorithms are needed for the ultra-dense wireless sensor with the inherent features of low complexity, high precision, and high accuracy. In this regard, fusion-based localization algorithms can be explored that infer the estimated location using the fusion of location information obtained from multiple wireless technologies (WiFi, Bluetooth, LoRaWAN, and others) employing different wireless localization techniques (ToA, RSSI, fingerprinting, and others)~\cite{TaaS-5,alzubi2021multi}. One promising direction in fusion-based localization is multi-sensor fusion-based algorithms are promising for AGVs and AMR applications since they combine the wireless technologies with other technologies, e.g., optical (LIDAR), inertial, or ultrasonic~\cite{xu2021multi}.
    \end{enumerate}
        

}






\section{Conclusion}
\label{sec:conclusion}
\textcolor{black}{ In this paper, we reviewed the Industry 5.0 vision and the vision requirements for innovative uses in Industry 5.0 while primarily focusing on the role of connecting CI with the NIBs architecture of NGWNs, i.e., CI-native NGWNs, as a technological enabler to satisfy essential requirements of Industry 5.0 vision. In this regard, we first discussed the detailed overview of the Industry 5.0 vision in terms of three enabling goals, i.e., human-centric, sustainability, and resilience, and established the required technological enablers and critical requirements it incurs for the full realization of Industry 5.0. Second, we discussed the design objectives and vision of CI-NGWNs and the NIBs architecture as an enabler in CI-NGWNs design to fulfill the critical demands of identified Industry 5.0 technological requirements. Afterward, we identified and discussed the emerging critical features and technologies of CI-NGWNs, including the computational intelligence algorithms, tools, and frameworks, multi-tenant softwarized service provisioning framework for self-organizing industrial automation networks, virtualized O-RAN architecture, private mobile networks, and hybrid TSN-enabled NIBs architecture and potential enabling services, for Industry 5.0.  
In addition, we discussed the various technologies and services of NIBs architecture that CI-NGWNs could exploit to realize Industry 5.0. 
In the end, we reviewed our research findings and provided insights, recommendations, and future directions for research on Industry 5.0 and how CI and NGWNs shape it. In particular, the main challenges stem from the fulfillment of increasing human-machine interactions to provide human-centric solutions in Industry 5.0, where the support of scalable and flexible CI technology at the softwarized NIBs architecture of NGWNs is needed for hybrid and unified service deployments. Meanwhile, CI-NGWNs are expected to address the identified research challenges relevant to the O-RAN orchestration and deployments, ultimate user and machine experiences, hyper-intelligent network, coordinated and harmonized network, sustainable network, and cyber security.  
}

\ifCLASSOPTIONcaptionsoff
  \newpage
\fi
\bibliographystyle{IEEEtran}
\bibliography{Biblio}
\vspace{-0.5cm}
\begin{IEEEbiography}[{\includegraphics[width=1in,height=1.25in,clip,keepaspectratio]{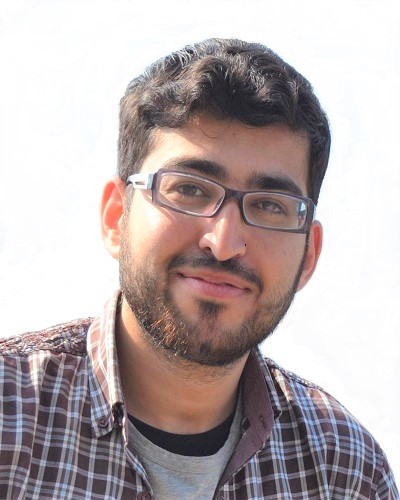}}]{Shah Zeb} (S'21) received the B.E. degree in Electrical Engineering (EE) from the University of Engineering and Technology, Peshawar, Pakistan, in 2016. He received the M.S. degree in Electrical Engineering from National University of Sciences and Technology (NUST), Pakistan, in 2019. 

He is currently pursuing Ph.D. degree in EE at NUST, Pakistan, where he works as a Research Associate with the Information Processing and Transmission (IPT) Lab, School of Electrical Engineering and Computer Science (SEECS), NUST, Pakistan. His research interests include emerging-enablers for Beyond-5G/6G wireless networks and industrial communication.
\end{IEEEbiography}


\begin{IEEEbiography}[{\includegraphics[width=1in,height=1.25in,clip,keepaspectratio]{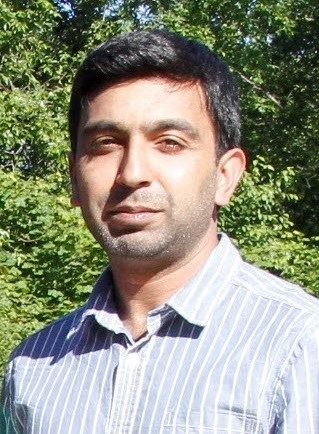}}]{Aamir Mahmood}
(M'18-SM’19) received the B.E. degree in electrical engineering from the National University of Sciences and Technology, Islamabad, Pakistan, in 2002, and the M.Sc. and D.Sc. degrees in communications engineering from the Aalto University School of Electrical Engineering, Espoo, Finland, in 2008 and 2014, respectively. Currently, he is an Assistant Professor with the Department of Information Systems and Technology, Mid Sweden University, Sweden. His research interests include intelligent time synchronization, positioning, RF coexistence, and radio resource management solutions for critical IoT networks. 
\end{IEEEbiography}


\begin{IEEEbiography}[{\includegraphics[width=1in,height=1.25in,clip,keepaspectratio]{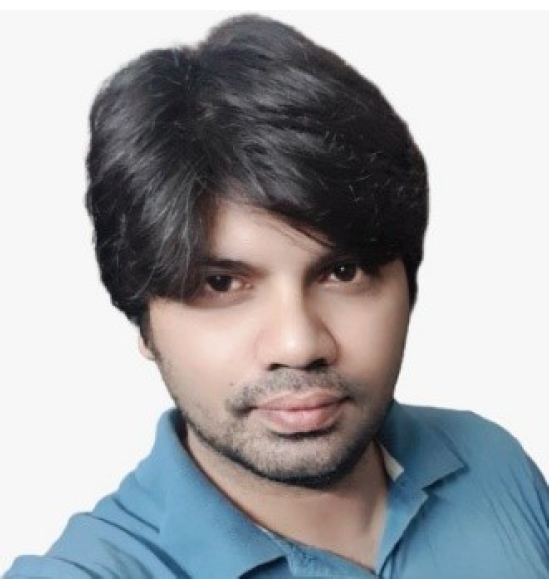}}]{Sunder Ali Khowaja}
received the Ph.D. degree in Industrial Information Systems Engineering from Hankuk University of Foreign Studies, South Korea. He is currently an Assistant Professor at Department of Telecommunication Engineering, University of Sindh, Pakistan. He is also a regular
reviewer of notable journals including IEEE Transactions, IET, Elsevier, and
Springer Journals. He has also served as a TPC member for workshops in A*
conferences such as Globecom and Mobicom. His research interests include
Data Analytics \& Machine Learning for Computer Vision applications. 
\end{IEEEbiography}


\begin{IEEEbiography}[{\includegraphics[width=1in,height=1.25in,clip,keepaspectratio]{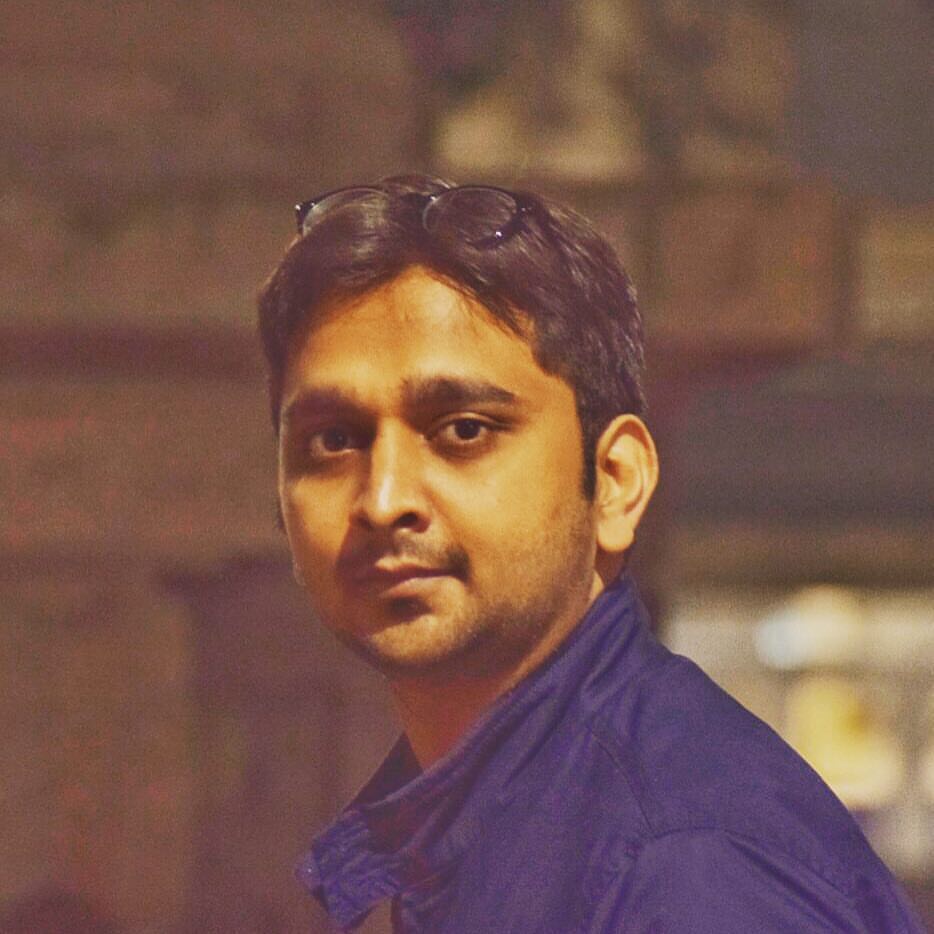}}]{Kapal Dev} is currently serving as Assistant Lecturer at Munster Technological University (MTU), Ireland and formerly he was senior researcher at same University. Previously, he was a Postdoctoral Research Fellow with the CONNECT Centre, Trinity College Dublin (TCD). He is founding chair of IEEE ComSoc special interested group titled as Industrial Communication Networks. He is serving as Associate Editor in NATURE, Scientific Reports, Springer WINE, HCIS, IET Quantum Communication, IET Networks, Area Editor in Elsevier PHYCOM. He performed duties as Guest Editor (GE) in IEEE Network, IEEE TII, IEEE TNSE, IEEE TGCN, IEEE STDCOMM. He has published over 40 research papers majorly in top IEEE Transactions, Magazines and Conferences. His research interests include Wireless Communication Networks, Blockchain and Artificial Intelligence.
\end{IEEEbiography}

\begin{IEEEbiography}[{\includegraphics[width=1in,height=1.25in,clip,keepaspectratio]{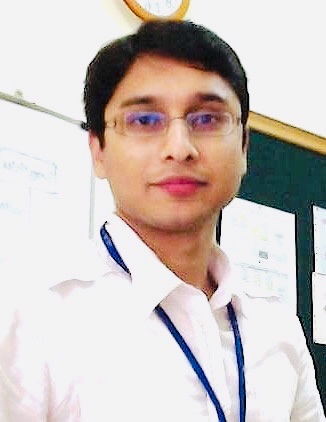}}]{Syed Ali Hassan}
(S'07--M'12--SM'17) received the B.E. degree in electrical engineering from the National University of Sciences \& Technology
(NUST), Islamabad, Pakistan, in 2004, the M.S. degree in mathematics from Georgia Tech in 2011, and the M.S. degree in electrical engineering from the University of Stuttgart, Germany, in 2007, and the Ph.D. degree in electrical engineering from the Georgia Institute of Technology, Atlanta, USA, in 2011. 
His research interests include signal processing for communications with a focus on cooperative communications for wireless networks, stochastic modeling, estimation and detection theory, and smart grid communications. He is currently working as an Associate Professor with the School of Electrical Engineering  and Computer Science (SEECS), NUST, where he is the Director of the Information Processing and Transmission (IPT) Lab, which focuses on various aspects of theoretical communications. 
He was a Visiting Professor with Georgia Tech in Fall 2017. He also held industry positions with Cisco Systems Inc., CA, USA, and with Center for Advanced Research in Engineering, Islamabad.
 \end{IEEEbiography}
\begin{IEEEbiography}[{\includegraphics[width=1in,height=1.25in,clip,keepaspectratio]{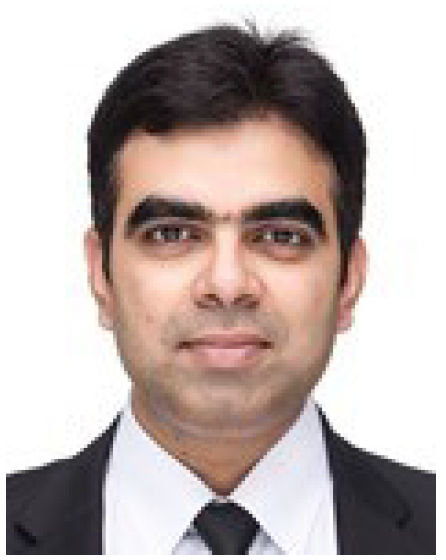}}]{Nawab Muhammad Faseeh Qureshi}
e is an Assistant Professor at Sungkyunkwan University, Seoul, South Korea.
He received Ph.D. in Computer Engineering from Sungkyunkwan University, South Korea, through
a SAMSUNG scholarship. He was awarded the 1st Superior Research Award from the College of
Information and Communication Engineering based on his research contributions and performance during
his studies. He has served as Guest Editor in more than 17 Journals. 
His research interests include big data analytics, machine learning, deep learning, context-aware data processing of the Internet of Things, and cloud computing.
\end{IEEEbiography}

\begin{IEEEbiography}[{\includegraphics[width=1in,height=1.25in,clip,keepaspectratio]{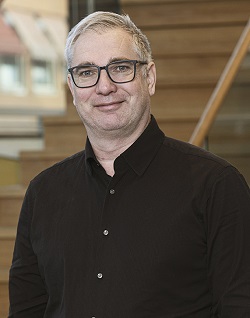}}]{Mikael Gidlund}
(M'98-SM'16) received the Licentiate of Engineering degree in radio communication systems from the KTH Royal Institute of Technology, Stockholm, Sweden, in 2004, and the Ph.D. degree in electrical engineering from Mid Sweden University, Sundsvall, Sweden, in 2005. From 2008 to 2015, he was a Senior Principal Scientist and Global Research Area Coordinator of Wireless Technologies with ABB Corporate Research, Västerås, Sweden. From 2007 to 2008, he was a Project Manager and a Senior Specialist with Nera Networks AS, Bergen, Norway. 
From 2006 to 2007, he was a Research Engineer and a Project Manager with Acreo AB, Hudiksvall, Sweden. 
Since 2015, he has been a Professor of Computer Engineering at Mid Sweden University. He holds more than 20 patents (granted and pending) in the area of wireless communication. His current research interests include wireless communication and networks, wireless sensor networks, access protocols, and security. Dr. Gidlund is an Associate Editor of the IEEE TRANSACTIONS ON INDUSTRIAL INFORMATICS.

 \end{IEEEbiography}
\begin{IEEEbiography}[{\includegraphics[width=1in,height=1.25in,clip,keepaspectratio]{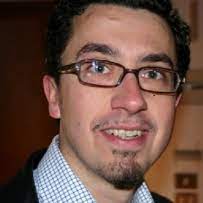}}]{Paolo Bellavista}
received the M.Sc. and Ph.D. degrees in computer science engineering from the University of
Bologna, Italy. He is currently a Full Professor of distributed and mobile systems with the University of Bologna. His research interests include from pervasive wireless computing to online big
data processing under quality constraints and from edge cloud computing to middleware for industry
4.0 applications. He serves on several editorial boards, including IEEE COMMUNICATIONS SURVEYS AND TUTORIALS (Associate EiC), ACM CSUR, JNCA (Elsevier), and PMC (Elsevier). He is the Scientific
Coordinator of the H2020 BigData Project IoTwins.
\end{IEEEbiography}
\end{document}